\documentclass[lettersize,journal]{IEEEtran}
\IEEEoverridecommandlockouts

\usepackage[ruled]{algorithm2e}
\usepackage{array}
\usepackage[caption=false,labelfont=sf]{subfig}
\usepackage{stfloats}
\usepackage{url}
\usepackage{verbatim}
\usepackage{bbding}
\usepackage{cite}
\usepackage{amsmath,amssymb,amsfonts}
\usepackage{algorithmic}
\usepackage{fontawesome}
\usepackage[colorlinks, linkcolor=blue]{hyperref}
\usepackage{orcidlink}
\usepackage{makecell}
\usepackage{colortbl}
\usepackage{multirow}
\usepackage{caption}
\usepackage{booktabs}
\usepackage{graphicx}
\usepackage{textcomp}
\usepackage[table,xcdraw]{xcolor}
\usepackage{arydshln}
\usepackage{enumitem}
\def\BibTeX{{\rm B\kern-.05em{\sc i\kern-.025em b}\kern-.08em
    T\kern-.1667em\lower.7ex\hbox{E}\kern-.125emX}}

\begin{document}

% The paper headers
\markboth{Journal of \LaTeX\ Class Files,~Vol.~14, No.~8, August~2025}%
{Shell \MakeLowercase{\textit{et al.}}: A Sample Article Using IEEEtran.cls for IEEE Journals}

\author{Feiyu Han~\orcidlink{0000-0002-7555-1232}, Mingtian Zhu, Dawei Yan~\orcidlink{0000-0002-9848-3017}, Weiwei Jiang~\orcidlink{0000-0003-4413-2497}, Yuanhao~Feng~\orcidlink{0000-0003-3969-9472}, Yanfei Zhang~\orcidlink{0009-0001-2244-613X}, and Panlong Yang~\orcidlink{0000-0003-1057-2793}~\IEEEmembership{Senior Member, IEEE}% <-this % stops a space

% \thanks{This work is partially supported by the National Natural Science Foundation of China with No. 62502214, the Basic Research Program of Jiangsu with No. BK20250727, the Jiangsu Province Higher Education Basic Science (Natural Science) Research Program with No. 25KJB520026, and the Startup Foundation for Introducing Talent of NUIST with No. 2025r068.}

\thanks{F. Han, M. Zhu, W. Jiang, and P. Yang is with School of Computer Science and School of Software, Nanjing University of Information Science and Technology, Nanjing, 210044, China. (E-mail: fyhan@nuist.edu.cn, 
zhumingtian29@gmail.com, weiwei.jiang@nuist.edu.cn, plyang@nuist.edu.cn)}
\thanks{D. Yan is now with the School of Cyber Security and Computer, Hebei University, Baoding, 071002, China. (E-mail: daweiyan@hbu.edu.cn)}
\thanks{Y. Feng is with University of Electro-Communications, Chofu, 182-8585, Japan. (E-mail: fengyuanhao@uec.ac.jp)}
\thanks{Y. Zhang is with Nanjing Chest Hospital, Affiliated Nanjing Brain Hospital, Nanjing Medical University, Nanjing, 210029, China. (E-mail: zhangyanfei0218@njmu.edu.cn)}
\thanks{This paper was produced by the IEEE Publication Technology Group. They are in Piscataway, NJ.}% <-this % stops a space
\thanks{Manuscript received 2025.}
}
% \title{\systemName: Enabling Multi-Level Diet Monitoring Based on Food Texture In-Ear Audio Response}
\title{Hear What You Eat: In-Ear Passive Acoustic Sensing for Multi-task Dietary Monitoring}
% \title{\systemName: Enabling Earable Dietary Logging with In-ear Audio Sensing}
% \title{\systemName: Enabling Earable-Based Multi-Level Diet Monitoring By Food Texture Sensing}
% \title{\systemName: Enabling Multi-Level Diet Monitoring on Earables with Food Texture Audio Profiles}
% Multi-Level Diet Monitoring on Earables with Food Texture Audio Profiles
% \title{\systemName: Enabling Earable-Based Multi-Level Diet Monitoring With Food Texture Audio Profiles}

\captionsetup{labelfont=bf}
\newcommand{\systemName}{\textsc{EarDiet}}
\newcommand{\etal}{\textit{et al}.}
\newcommand{\ie}{\textit{i}.\textit{e}.}
\newcommand{\eg}{\textit{e}.\textit{g}.}
\newcommand{\rmnum}[1]{\romannumeral #1}
\newcommand{\Rmnum}[1]{\expandafter\@slowromancap\romannumeral #1@}
\newcommand{\weiwei}[1]{\textcolor{red}{#1}}

\maketitle

\begin{abstract}

The increasing obesity health concerns are driving the advancement of continuous dietary monitoring. Yet, existing single-sensor solutions still face challenges of coarse-grained sensing resolution, while multi-sensor fusion strategies encounter significant engineering complexity or hardware requirements. In our work, we present a single modality-based multi-dimensional dietary monitoring and logging solution, named \systemName, which leverages inward‐facing microphones on earables to capture in-ear chewing sounds for fine-grained diet analysis. With the texture audio profiles derivation, the pipeline consisting of signal processing and lightweight deep learning, is carefully designed to achieve eating episode detection, food texture sensing, and food type recognition simultaneously. Comprehensive experiments involving 32 participants with diverse BMI profiles are conducted to validate the effectiveness of multi-dimensional dietary monitoring. Compared with the state-of-the-art solutions, \systemName~can derive multi-dimensional dietary information with only a single modality, outperforming by about 10\% in terms of eating episode detection. Furthermore, through real-world clinical validation, our solution can achieve masticatory disorder identification with an averaged F1-score of 85.97\%, demonstrating potential for healthcare in the future.

% A longitudinal study conducted in real-world settings further demonstrates the robustness of our system. 
% \systemName~paves the way for socially acceptable and multi-dimensional dietary monitoring through earables, advancing the development of wearable technology in individual dietary health assessment.

% with diverse body mass index (BMI) profiles

\end{abstract}


\section{\textbf{Introduction}}

% \textbf{Motivation.} According to the World Health Organization's obesity and overweight survey, about 2.5 billion adults suffer from overweight and obesity issues~\cite{world2024obese}. Overweight and obesity pose significant health risks that can affect individuals across various age groups~\cite{huttunen2013obesity,fock2013diet}, including cardiovascular diseases, type 2 diabetes, and osteoarthritis, therefore attracting widespread attention from society. In most cases, obesity is associated with multiple factors, including social environments, psychosocial factors, and genetic variation~\cite{world2024obese, huttunen2013obesity}. Essentially, overweight and obesity can be summarized as an imbalance between energy intake (diet) and energy expenditure (physical activity). Therefore, adopting a reasonable and healthy diet is the most effective way to manage weight, which contributes to long-term health and wellness.
\IEEEPARstart{O}{verweight} and obesity pose significant health risks that can affect individuals across various age groups~\cite{world2024obese,huttunen2013obesity,fock2013diet}, including cardiovascular diseases, type 2 diabetes, and osteoarthritis. Essentially, overweight and obesity are caused by the imbalance between energy intake and energy expenditure. Hence, maintaining a balanced and nutritious diet is crucial for weight management and long-term health~\cite{world2024obese}. However, accurately recording food intake remains a significant challenge, as it requires consistent, precise tracking of dietary habits, which can be influenced by numerous factors, including social environments, psychological aspects, and personal biases~\cite{world2024obese, huttunen2013obesity}. This makes monitoring and managing nutrition effectively a complex task that is essential for addressing obesity and promoting overall wellness.

\textit{Limitation of Prior Work.} Effective dietary management fundamentally relies on capturing multi-dimensional dietary information, as it provides a holistic basis for accurate nutritional assessment and long-term health optimization. The traditional manual food intake logging presents several disadvantages, including time consumption and easy forgetfulness, causing missing or inaccurate data collection. To address this issue, the research community has explored the potential of wearable devices for diet monitoring. However, most of these solutions are inconvenient to wear (like wearing them on the neck or chin~\cite{bi2015autodietary, san2020eating}) or difficult to integrate into existing commercial devices (like requiring dedicated sensors or multi-sensor/multi-device fusion~\cite{mirtchouk2017recognizing, min2025multimodal, mirtchouk2016automated, bedri2020fitbyte}). More importantly, most of them only provide single-dimensional dietary sensing, like only identifying eating activities~\cite{shin2022mydj, thomaz2015practical, zhang2022eatingtrak}, only identifying food types~\cite{papapanagiotou2021recognition}, or only estimating food weight~\cite{papapanagiotou2021bite}. It is difficult to provide sufficient fine-grained information for dietary health analysis and evaluation.

% Until now, many efforts have been conducted to achieve diet monitoring. The traditional manual food intake logging presents several disadvantages, including time consumption and easy forgetfulness, causing missing or inaccurate data collection.%which may hinder effective dietary monitoring and management. 

\begin{figure}
    \centering
    \includegraphics[width=1\linewidth]{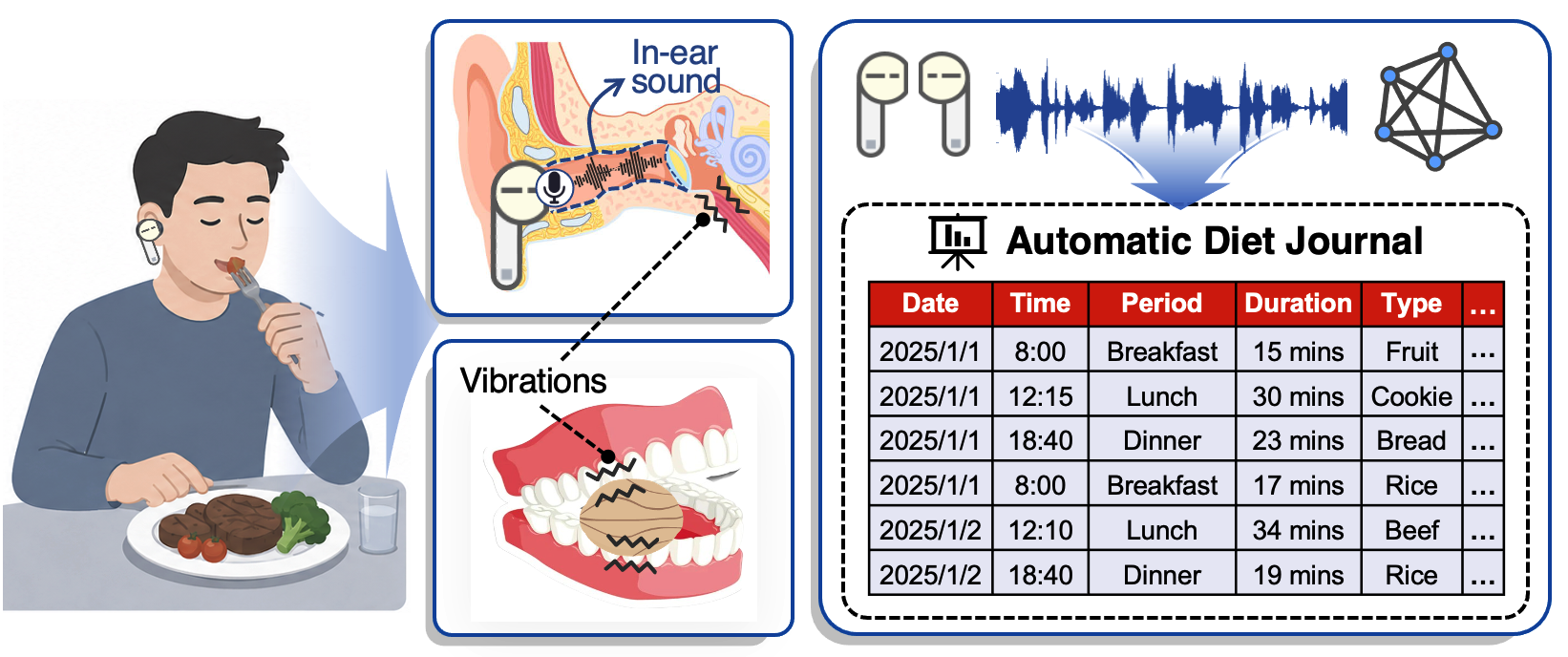}
    \caption{\textbf{Based on the social behavior change that more and more people (especially young people) tend to eat with earables, we design \systemName~that leverages inward-facing microphones on earables to capture chewing-induced in-ear sounds for automatic diet logging. }}
    \label{fig:fig_1}
    \vspace{-0.5cm}
\end{figure}

\textit{Potential and Opportunity of Earables}. In light of these challenges, earables have emerged as a promising alternative. Recent studies show novel applications in multiple areas such as physiological activity monitoring~\cite{han2024earsleep, hu2024breathpro}, human-computer interaction~\cite{han2024earspeech, 10843381}, and user authentication~\cite{han2023breathsign, hu2024lr, 10663927, 10566858}. Compared with other wearable devices, ear-worn devices have several significant advantages in terms of diet monitoring (see \S \ref{sec:motivation_advantages}), including \textit{(\rmnum{1}) proximity to the mouth,} 
\textit{(\rmnum{2}) high social acceptability,} and 
\textit{(\rmnum{3}) wearing comfort}. 
Earable diet monitoring solutions based on a single sensor (like an IMU sensor~\cite{ketmalasiri2024imchew}) only capture coarse-grained dietary information. Recently, a work~\cite{chugh2025bitesense} leverages the IMU sensor on earables to sense eating activities and recognize food types, while it still lacks high accuracy and robustness due to narrow-bandwidth inertial signals and sensitivity to motion artifacts.
To capture fine-grained and multi-dimensional dietary information, multiple sensors are often required to work together~\cite{bedri2020fitbyte, bedri2020fitbyte, chun2018detecting, zhang2020necksense}, resulting in a high hardware burden for limited-resource earables. Thus, we raise a problem: 
$\diamondsuit$ \textit{Can we leverage a single sensing modality to capture sufficient fine-grained and multi-dimensional dietary information?}

% That is because the low-sampling IMU sensor due to limited sampling rate. For instance, 

% This is because the IMU sensor with a low sampling rate has low perception accuracy.

% leveraging an IMU sensor can only identify whether the eating activity occurs, but it cannot identify what the user has eaten. Recently, 

Here, we focus on leveraging chewing sound to capture users' dietary data. In principle, during chewing, the collision and friction between teeth and food will produce sounds. Prior solutions use microphones to capture airborne chewing sounds for diet monitoring~\cite{bi2015autodietary, bi2018auracle}. However, airborne chewing sounds are extremely susceptible to interference from external environmental noises. Chewing sounds can be transmitted through two channels: the air channel and the body channel. Compared with airborne sounds, bone-conducted sounds are more resistant to external noise. Chewing sounds propagate along bone conduction and arrive at the ear canal, where they are eventually picked up by the inward-facing microphones. Prior efforts have been conducted to use chewing-induced in-ear sounds for single-task diet sensing, such as only chewing activity recognition or food type classification. In addition, most of them are studied in constrained environments, lacking robustness in real-world scenarios. Different from these existing solutions, in our work, we aim to explore the feasibility and effectiveness of leveraging chewing-induced in-ear sounds for multi-task diet monitoring.

\textit{Proposed Solution.} In this work, we design \systemName~that leverages inward-facing microphones on earables to capture chewing-induced in-ear sounds to enable automatic diet logging, including meal time, duration, chewing frequency, and intake food texture, which could provide fine-grained dietary information for weight management, as shown in Fig.~\ref{fig:fig_1}. 
The novelty lies in that \systemName~recognizes intake food based on texture attributes rather than the regular food types. The variety of foods available is vast, reaching into the hundreds or even thousands, making it challenging to enumerate them all. However, different types of foods may exhibit similar texture attributes, such as hardness, crispness, adhesiveness, and juiciness. Food texture attributes play a crucial role in daily diet management, including nutritional assessment, portion control, and food quality~\cite{nishinari2018perception}. Thus, we believe that food texture recognition could provide valuable insight into designing personalized nutritional strategies and balanced dietary structures.

\textit{Challenges and Strategies.} To achieve the aforementioned goals, several technical challenges are addressed. 

\begin{itemize}
    \item \textit{(\rmnum{1}) Recognize continuous chewing activities under complex interferences during the long-term eating process.} Compared with daily body movements, the amplitude of chewing activities is smaller, resulting in motion artifacts mixing with target audio signals or even drowning out target audio signals. In addition, the diversity of chewing activities brings challenges to continuous recognition. To address that, we design an audio stream processing pipeline to enhance the SNR of chewing in-ear sounds. Based on the physical characteristics of chewing activities, we extract key temporal and spectral features and design a mechanics-aware eating activity detection approach to derive accurate diet episodes from continuous chewing-induced in-ear audio streams.

    \item \textit{(\rmnum{2}) Unclear relationship between food texture and audio profiles makes it difficult to derive interpretable features.} Although prior solutions have explored the feasibility of leveraging statistical features of chewing sounds for food texture classification, the relationship between food texture and audio profiles has not been well studied yet, resulting in significant challenges in terms of effectiveness and accuracy. To address that, we investigate the mechanical properties of dietary foods and establish the semantic relationship between food texture and audio profiles. Based on that, we conduct a comprehensive analysis to demonstrate the unique response of different food texture attributes and derive interpretable CWT-based audio profiles.

    \item \textit{(\rmnum{3}) The negative impact of individual variability on fine-grained and generalized diet monitoring.} The eating patterns, oral structures, and skull geometry, etc., vary from person to person, which affects the generation and propagation of chewing sounds. Such individual variability affects the generalization and robustness of the system in real scenarios. To address that, we carefully design a multi-task learning network with dual feature extraction branches to capture generalized feature maps for food type and food texture recognition. Additionally, we adopt the channel attention and spatial attention mechanisms to adaptively emphasize important components and suppress less relevant ones.

\end{itemize}

\textit{Evaluation Summary.}
Comprehensive experiments in real-world scenarios validate the effectiveness and robustness of our solution. During long-term monitoring, \systemName~still yields excellent performance in eating episode detection with 96.08\% of the averaged F1-score (10\% improvement over the state-of-the-art solution). In the four daily scenarios, \systemName~achieves an averaged accuracy of 78\% in terms of food type recognition, an averaged of 88.39\% in terms of macro evaluation, and an averaged of F1-score of 74.95\% in terms of micro evaluation. Most importantly, through the real-world clinical validation with 51 patients, our solution can identify masticatory disorders with an averaged F1-score of 85.97\%. \systemName~paves the way for socially acceptable and multi-dimensional dietary monitoring through earables, advancing the development of wearable technology in individual dietary health assessment.

% demonstrates great potential in terms of masticatory disorders

% We believe our technique could promise enhanced accuracy, real-time feedback, and personalized dietary insights, ultimately promoting healthier lifestyles and improved nutrition management.
% The future of earable-based dietary logging with in-ear audio sensing envisions seamless integration of advanced AI algorithms, allowing users to effortlessly track their eating habits. This technology promises enhanced accuracy, real-time feedback, and personalized dietary insights, ultimately promoting healthier lifestyles and improved nutrition management.

\section{\textbf{Related Work}}

\begin{table*}[]
\caption{\label{tab:comparison_related_work}  \textbf{Comparison with related works. Qualitative analysis indicators: \faBatteryEmpty~-~Low, \faBatteryHalf~-~Medium, \faBatteryFull~-~High.}}
\renewcommand{\arraystretch}{1.1}
\resizebox{1\linewidth}{!}{
\begin{tabular}{c|c|c|ccc|c|c|cc}

\Xhline{1.2pt}
\multicolumn{1}{c|}{\multirow{2}{*}{\raisebox{-1\height}{\textbf{No}}}} & \multicolumn{1}{c|}{\multirow{2}{*}{\raisebox{-1\height}{\textbf{Hardware}}}} & \multicolumn{1}{c|}{\multirow{2}{*}{\raisebox{-1\height}{\textbf{Sensor}}}} & \multicolumn{3}{c|}{\textbf{Sensing Tasks}}                                        & \multicolumn{1}{c|}{\multirow{2}{*}{\raisebox{-1\height}{\begin{tabular}[c]{@{}c@{}}\textbf{User} \\ \textbf{Scale}\end{tabular}}}} & \multicolumn{1}{c|}{\multirow{2}{*}{\raisebox{-1\height}{\textbf{Performance}}}} & \multicolumn{2}{c}{\textbf{Robustness}} \\ \cline{4-6} \cline{9-10}

\multicolumn{1}{c|}{}                    & \multicolumn{1}{c|}{}                          & \multicolumn{1}{c|}{}                        & \multicolumn{1}{c}{\begin{tabular}[c]{@{}c@{}}\textbf{Eating} \\ \textbf{episode}\end{tabular}} & \multicolumn{1}{c}{\begin{tabular}[c]{@{}c@{}}\textbf{Food} \\ \textbf{type}\end{tabular}} & \multicolumn{1}{c|}{\begin{tabular}[c]{@{}c@{}}\textbf{Food} \\ \textbf{texture}\end{tabular}} & \multicolumn{1}{c|}{}  & \multicolumn{1}{c|}{}   & \multicolumn{1}{c}{\begin{tabular}[c]{@{}c@{}}\textbf{Motion} \\ \textbf{artifacts}\end{tabular}} & \multicolumn{1}{c}{\begin{tabular}[c]{@{}c@{}}\textbf{Ambient} \\ \textbf{noise}\end{tabular}} \\ \Xhline{1.2pt}

\textbf{EarBit}~\cite{bedri2017earbit}              & Wearables    & \begin{tabular}[c]{@{}c@{}}IMU + Proximity\\ + Microphone\end{tabular}         & \Checkmark     & \XSolidBrush  & \XSolidBrush    & 20   & Eating activity (F1-score of 90.9\%)   & \faBatteryFull   & \faBatteryHalf         \\ \hline

\textbf{IMChew}~\cite{ketmalasiri2024imchew}     & \includegraphics[width=0.45cm]{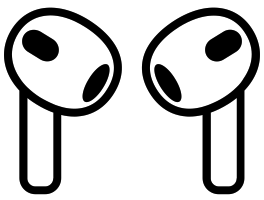} Earbuds  & IMU    & \Checkmark       & \XSolidBrush    & \XSolidBrush        & 8      & Eating activity (F1-score of 91.0\%)   & \faBatteryEmpty    & \faBatteryFull          \\ \hline

\textbf{MyDJ}~\cite{shin2022mydj}     & \raisebox{-0.3\height}{\includegraphics[width=0.4cm]{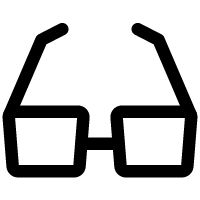}} Eyeglasses    & \begin{tabular}[c]{@{}c@{}}Accelerometer +\\ Piezoelectric\end{tabular} & \Checkmark       & \XSolidBrush    & \XSolidBrush       & 30   & Eating activity (F1-score of 91.9\%) & \faBatteryEmpty    & \faBatteryFull          \\ \hline

\textbf{EatingTrak}~\cite{zhang2022eatingtrak}          & \raisebox{-0.4\height}{\includegraphics[width=0.45cm]{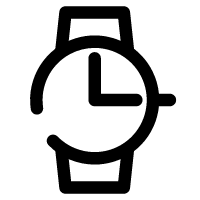}} \begin{tabular}[c]{@{}c@{}}Wrist-worn \\ device\end{tabular} & IMU   & \Checkmark       & \XSolidBrush    & \XSolidBrush     & 9   & Eating activity (F1-score of 83.8\%)  & \faBatteryEmpty              & \faBatteryFull          \\ \hline

\textbf{FitByte}~\cite{bedri2020fitbyte}             & \raisebox{-0.3\height}{\includegraphics[width=0.4cm]{./Figures/three-d-glasses.png}} Eyeglasses        & \begin{tabular}[c]{@{}c@{}}Camera + IMU\\ + Proximity\end{tabular} & \Checkmark       & \XSolidBrush    & \XSolidBrush       & 23   & Eating activity (F1-score of 89\%)  & \faBatteryFull             & \faBatteryFull          \\ \hline

\textbf{NeckSense}~\cite{zhang2020necksense}           & \raisebox{-0.3\height}{\includegraphics[width=0.5cm]{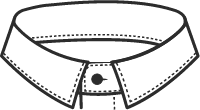}} Necklace          & \begin{tabular}[c]{@{}c@{}}Proximity +\\ Light + IMU\end{tabular}  & \Checkmark       & \XSolidBrush    & \XSolidBrush    & 18   & Eating activity (F1-score of 81.6\%)  & \faBatteryEmpty              & \faBatteryFull          \\ \hline

\textbf{BiteSense}~\cite{chugh2025bitesense}           & \raisebox{-0.3\height}{\includegraphics[width=0.45cm]{./Figures/earbud.png}} Earbuds   &  IMU  & \Checkmark       & \Checkmark    & \XSolidBrush   & 38  & \begin{tabular}[c]{@{}c@{}}Food type (Accuracy of 75.7\%), \\ Eating activity (F1-score of 86\%)\end{tabular}   & \faBatteryEmpty              & \faBatteryFull          \\ \hline

\textbf{CNN-based~\cite{papapanagiotou2017chewing}}  & \raisebox{-0.3\height}{\includegraphics[width=0.45cm]{./Figures/earbud.png}} Earbuds   &  \begin{tabular}[c]{@{}c@{}}In-ear \\ microphone\end{tabular}  & \Checkmark       & \XSolidBrush    & \XSolidBrush   & 14  & Eating activity (F1-score of 88.3\%)   & \faBatteryFull     & \faBatteryFull         \\ \hline

\textbf{SSL-based~\cite{papapanagiotou2021self}}  & \raisebox{-0.3\height}{\includegraphics[width=0.45cm]{./Figures/earbud.png}} Earbuds   &  \begin{tabular}[c]{@{}c@{}}In-ear \\ microphone\end{tabular}  & \Checkmark       & \XSolidBrush    & \XSolidBrush   & 14  & Eating activity (F1-score of 86\%)   & \faBatteryFull     & \faBatteryFull         \\ \hline

\begin{tabular}[c]{@{}c@{}}\textbf{\systemName} \\ \textbf{(Ours)}\end{tabular} & \raisebox{-0.3\height}{\includegraphics[width=0.45cm]{./Figures/earbud.png}} Earbuds  & \begin{tabular}[c]{@{}c@{}}In-ear \\ microphone\end{tabular}    & \Checkmark       & \Checkmark    & \Checkmark    & 32   & \begin{tabular}[c]{@{}c@{}}Food type (Accuracy of 78.2\%), \\ Food texture (F1-score of 88.3\%), \\ Eating activity (F1-score of 96.1\%)\end{tabular} & \faBatteryFull     & \faBatteryFull          \\ \Xhline{1.2pt}

\end{tabular}
}
\end{table*}

Diet monitoring contains a wide variety of research topics~\cite{min2019survey, yang2022review}, including eating episode/activity detection, food type recognition, and food intake weight estimation. 
Compared with non-contact diet monitoring like mmWave-based~\cite{wang2023eat} and vision-based~\cite{min2023large}, wearable-based diet monitoring has shown great advantages in terms of accuracy, privacy preserving, and anti-interference. Next, we mainly introduce wearable solutions in the context of dietary monitoring. 

% \subsection{\textbf{Wearable Diet Monitoring}}

\subsection{\textbf{Wearable Diet Monitoring}}
Based on the sensing platform, we divide contact diet monitoring solutions into the following categories.

\subsubsection{\textbf{Wrist-worn based}} 

Arm, wrist, and hand movements show significant patterns during eating episodes. Therefore, eating activity recognition based on wrist-worn devices has been widely studied. For example, Thomaz~\etal~\cite{thomaz2015practical}explored the feasibility of eating moment recognition based on inertial sensing. To improve the robustness in the wild, Zhang~\etal~\cite{zhang2022eatingtrak} designed a robust eating moment recognition approach with high time resolution by estimating arm postures. The above solutions can only identify whether eating activities occur, which is insufficient to provide fine-grained information for healthy diet management. Therefore, prior literature studied the feasibility of fluid/food intake weight estimation using inertial sensors on wrist-worn devices~\cite{hamatani2018fluidmeter, levi2025estimation}.

% \cite{hamatani2018fluidmeter}
% \cite{liu2024ieat}

\subsubsection{\textbf{Head/Ear-worn based}}
Compared with wrist-worn devices, head-worn devices (like smart glasses or earbuds) are positioned close proximity to the mouth, enabling a promising sensing platform for diet monitoring~\cite{min2019survey}. Nowadays, eyeglasses have become socially acceptable wearable devices, promoting the development of eye-worn eating activity recognition~\cite{shin2022mydj, mahmud2024munchsonic, parikh2024echoguide}. Bedri~\etal~\cite{bedri2020fitbyte} equipped eyeglasses with inertial sensors, a proximity sensor, and a camera to achieve multi-modal diet monitoring, including food intake event recognition and food type identification. As another mainstream head-worn device, ear-worn devices also provide promising sensing modalities for diet monitoring. For example, prior literature leveraged the in-ear microphone on earphones to capture chewing-induced sounds for chewing side preference analysis~\cite{kim2022earchew}, food-texture attribute identification~\cite{papapanagiotou2021recognition}, and bite-weight estimation~\cite{papapanagiotou2021bite}. Bi~\etal~\cite{bi2018auracle} used a contact-skin microphone embedded on earpieces to detect eating episodes. IMU sensors on earables are widely used to detect eating-related activities~\cite{chugh2025bitesense, ketmalasiri2024imchew}.
Moreover, \textit{EarBit} was designed to fuse IMU sensors, a microphone, and a proximity sensor to enable robust eating episode detection on the head-worn device~\cite{bedri2017earbit}. Papapanagiotou~\etal~\cite{papapanagiotou2017chewing,papapanagiotou2021self} firstly explore the feasibility of leveraging in-ear microphones on earphones to capture chewing-induced in-ear sounds for eating activity detection. Although promising, they still only focus on a single eating activity detection task.

\subsubsection{\textbf{Neck-worn based}}

In addition to the above wearable devices, neck-worn devices could sense chewing and swallowing activities, becoming a potential sensing platform to capture diet-related information. By attaching a microphone to the neck, Bi~\etal~\cite{bi2015autodietary} extracted unique acoustic responses of chewing and swallowing sounds to recognize food intakes. San~\etal~\cite{san2020eating} leveraged an IMU sensor to sense jaw movements during eating. 
To overcome the limitations of single-modal sensing, some significant solutions have integrated multiple sensors to achieve food intake-related activity recognition. For instance, a proximity sensor, a motion sensor, and a light sensor were integrated on a necklace to detect several chewing activities\cite{chun2018detecting, zhang2020necksense}. Nguyen~\etal~\cite{nguyen2017swallownet} explored the feasibility of fusing piezoelectric sensors and a motion sensor to enable passive swallow counting.

\subsubsection{\textbf{Multi-device based}}
As the wearable device ecosystem is being completed, multi-device collaborative diet monitoring has attracted wide attention~\cite{mirtchouk2017recognizing, min2025multimodal}. Mirtchouk~\etal~\cite{mirtchouk2016automated} leveraged sensors on earbuds and the smartwatch to achieve food type and eating activity recognition. Bedri~\etal~\cite{bedri2017earbit} built a hardware prototype with ear-worn and neck-worn components to detect eating moments.

\subsubsection{\textbf{Review Summary}} From the above literature review, prior significant works have leveraged various sensing modalities or sensing platforms for diet-related monitoring. Furthermore, we highlight the key differences between our solution and prior solutions in Tab.~\ref{tab:comparison_related_work}.
Prior solutions still face the following limitations: \underline{\textbf{(1)}} Multi-sensor collaborative sensing suffers from engineering complexity, hardware costs, and potential synchronization or calibration issues. \underline{\textbf{(2)}} Special wearing position requirements lead to inconvenience for users, decreasing acceptability, comfort, and portability; \underline{\textbf{(3)}} Staying at the single eating activity sensing task, like identifying whether the eating activity occurs, insufficient to provide fine-grained information for diet health assessment. 

% To address the limitations of prior solutions, we leverage ubiquitous in-ear microphones on wearing-comfort earbuds to derive multi-dimensional dietary information, including eating episode, food texture, and food type.

\subsection{\textbf{In-ear Audio-based Sensing on Earables}}

In-ear audio-based sensing refers to leveraging built-in microphones on earbuds to capture activity-induced bone-conducted sounds for various passive sensing tasks~\cite{roddiger2022sensing}. Compared with bioelectric signals, the acquisition of in-ear audio does not rely on dedicated biosensors, but only on ubiquitous microphones. In addition, in-ear audio signals have a wider bandwidth compared with inertial signals, enabling them to have more information. As a promising modality, in-ear audio signals have enabled various applications, including user authentication~\cite{10663927, han2023breathsign, hu2024lr}, human-computer interaction~\cite{10843381, han2024earspeech}, vital sign detection\cite{butkow2023heart, hu2024breathpro}, and health assessment~\cite{han2024earsleep}.

% However, most of them only stay at single eating activity sensing tasks, like identifying whether the eating activity occurs, insufficient to provide fine-grained information for diet health assessment. 

% However, most of them currently only stay at the coarse-grained eating activity detection, such as identifying whether it is eating behavior or food identification.

% fine-grained diet monitoring which includes eating episode recognition, food type recognition, and food intake weight estimation.

\section{\textbf{Background and Preliminary}}

\begin{figure*}[t]
\centering
    \begin{minipage}[t]{1\linewidth}
       \subfloat[\textbf{Scenario preference.}]{\includegraphics[width=0.22\textwidth]{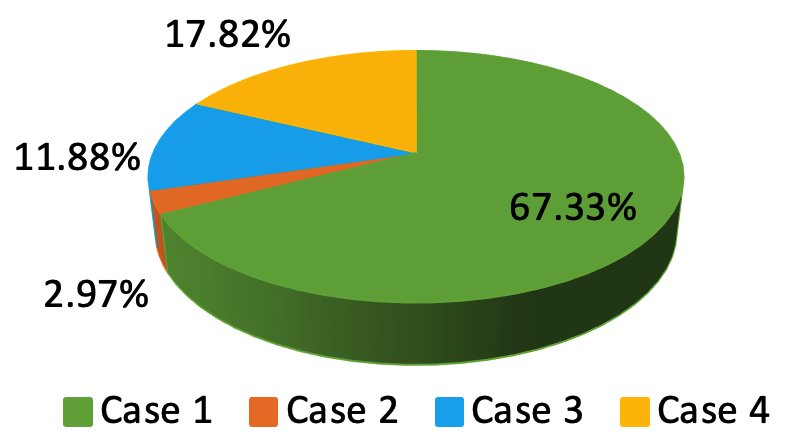}
        \label{fig:survey_scenarios}} \hfill
        \subfloat[\textbf{Wearing frequency.}]{\includegraphics[width=0.22\textwidth]{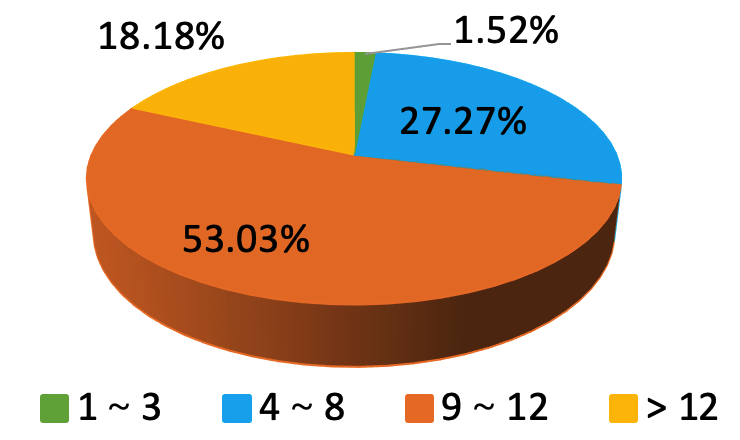}
    \label{fig:survey_frequency}} \hfill
        \subfloat[\textbf{Age distribution.}]{\includegraphics[width=0.22\textwidth]{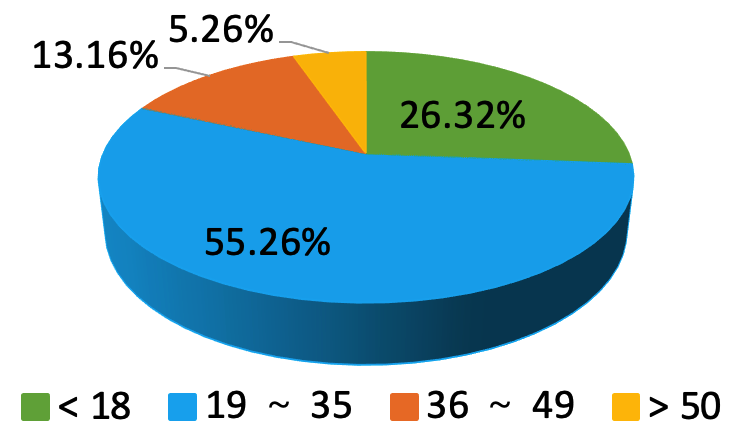}
    \label{fig:survey_age}} \hfill
     \subfloat[\textbf{Willing survey.}]{\includegraphics[width=0.23\textwidth]{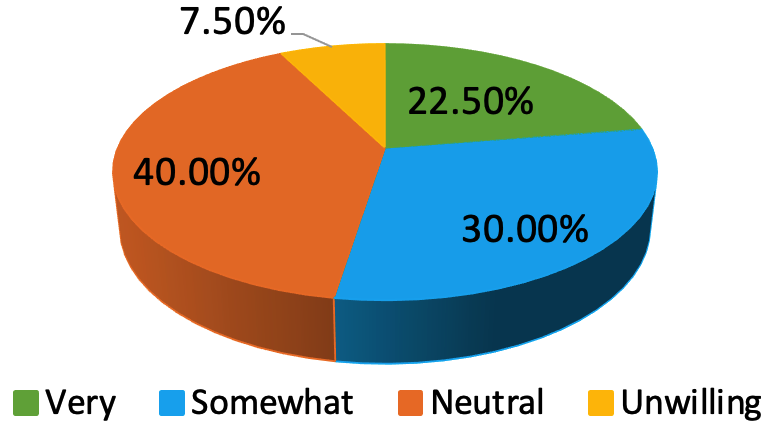}
    \label{fig:survey_willing}}
    \caption{\label{fig:user_survey}\textbf{User survey on eating with earables. }} 
    \end{minipage}
    \vspace{-2mm}
\end{figure*}

\subsection{\textbf{Socially Acceptable Diet Monitoring with Earables}} \label{sec:motivation_advantages}
% Earables have become a novel sensing platform~\cite{roddiger2022sensing}. 
Nowadays, eating while wearing earables has become socially acceptable due to the widespread integration of technology into daily life~\cite{kim2022earchew, papapanagiotou2021recognition, papapanagiotou2021bite, bedri2017earbit, chugh2025bitesense}, especially among young people. Compared with other wearable devices, earables are positioned closer to the mouth, enabling accurate sensing of mandibular and palatal movements during eating. Modern earables are ergonomically designed with soft earplugs, providing comfortable long-term wear without imposing additional pressure on users.

To understand the prevalence and social acceptance of wearing earables during mealtime, we conducted a user study among 108 individuals ranging in age from 15 to 55 on our campus. Each individual takes about 1 minute to complete the survey questionnaire. 
Fig.~\ref{fig:user_survey}\subref{fig:survey_scenarios}~shows a survey on the acceptance of eating with earables in terms of different scenarios, including only in single-person dining scenarios (Case 1), only in multi-person dining scenarios (Case 2), both acceptable (Case 3), and neither acceptable (Case 4). We can observe that nearly 85\% of respondents are willing to wear earables while eating. The above results indicate that users largely regard wearing earables while eating as a socially acceptable practice. Fig.~\ref{fig:user_survey}\subref{fig:survey_frequency}~demonstrates that the majority of respondents reported wearing headphones while eating between 9 and 12 times per week, followed by 27.27\% who did so 4 to 8 times per week, indicating that regular use during meals is common among participants. Among respondents who accepted wearing headphones while eating, we can find that the largest proportion were aged under 35 years old, suggesting that acceptance of this behavior is highest among younger demographics, as shown in Fig.~\ref{fig:user_survey}\subref{fig:survey_age}. Lastly, Fig.~\ref{fig:user_survey}\subref{fig:survey_willing}demonstrates that about half of the respondents exhibit interest in integrating a dietary monitoring function on earables. Based on the questionnaire results, the domain of earable-based dietary monitoring exhibits considerable potential for future development.

\begin{figure}[t]
    % \begin{minipage}[b]{0.38\linewidth}
    %     \centering
    %     \includegraphics[width=0.9\linewidth]{./Figures/teeth_food_contact_fig.png} \caption{\label{fig:tooth_food_tooth_contact}
    %     \textbf{In-ear microphones on earbuds can capture sounds induced by chewing activities.}}
    % \end{minipage}\hfill
    \begin{minipage}[b]{1\linewidth}
        \centering
        \includegraphics[width=1\linewidth]{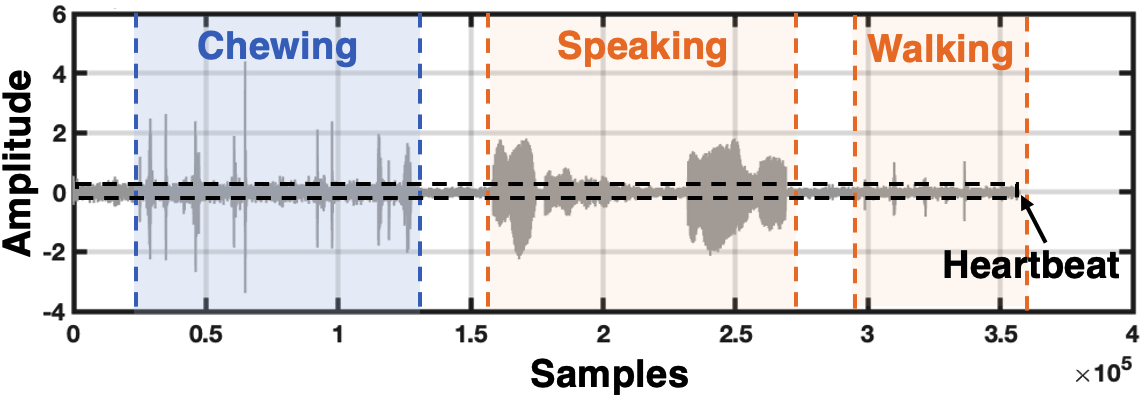}
    \caption{\textbf{Sound waveform comparison between chewing and artifacts (like speaking, walking, and heartbeat).}}
    \label{fig:time_domain_analysis}
    \end{minipage}
    \vspace{-0.5cm}
\end{figure}

\begin{figure}[t]
    \begin{minipage}[t]{1\linewidth}
       \subfloat[\textbf{Same individual}]{\includegraphics[width=0.46\textwidth]{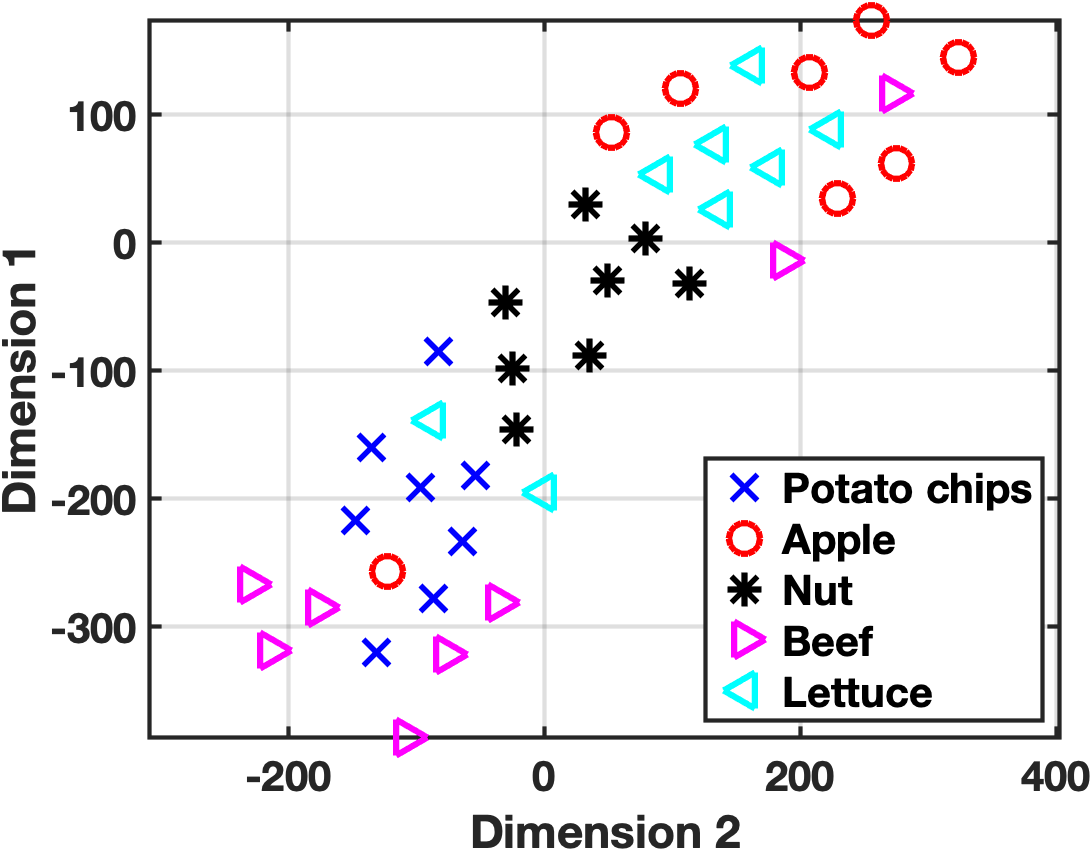}
        \label{fig:same_person_different_food_types}} \quad
        \subfloat[\textbf{Different individuals.}]{\includegraphics[width=0.46\textwidth]{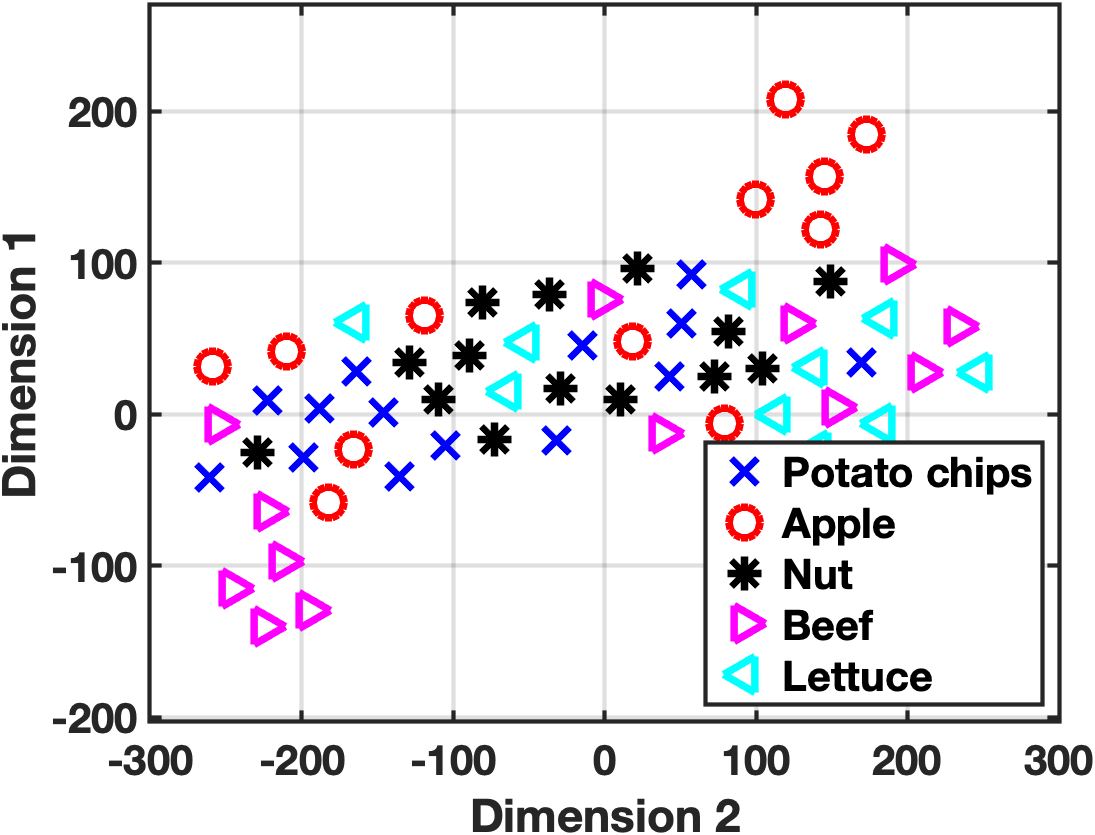}
    \label{fig:different_person_different_food_types}} 
    \caption{\label{fig:person_food_types}\textbf{Temporal feature visualization of five food types.}} %within the same individual and between different individuals
    \end{minipage}
    \vspace{-2mm}
\end{figure}

\subsection{\textbf{Eating/Chewing-induced In-ear Sounds}} \label{sec:chewing_induced_inear_sounds}

When teeth break down food, vibrations occur, producing various sounds like crunching, squelching, or grinding. These eating-induced sounds travel along the skull to the ear canal and are further captured by in-ear microphones that are built into earables. Benefiting from the occlusion effect, the low-frequency components ($\leq$1000 Hz) of in-ear sounds are significantly enhanced ~\cite{han2023breathsign, han2024earsleep, han2025earoe}, which enables the feasibility of detecting eating-induced sounds on earables. Fig.~\ref{fig:time_domain_analysis} demonstrates the sound waveform comparison between chewing and other activity artifacts like speaking, walking, and heartbeat. We can observe that the target chewing sounds present a lower signal-to-noise ratio compared with activity artifacts. In addition, other body movements may occur while eating, which poses a challenge to the continuous detection and segmentation of food intake events. 

% As mentioned in prior literature~\cite{han2023breathsign, han2025earoe}, chewing-induced in-ear sounds denoted by $s_{ie}$ can be modeled as follows: 
% \begin{equation}
%     s_{ie} = H_{ear} \ast  H_{bone} \ast  s_{o}
% \end{equation}
% where $H_{ear}$, $H_{bone}$, and $s_{o}$ represent the gain of occlusion effect, the gain of bone conduction, and original vibrations generated by the interaction between teeth and food, respectively.
% $ s_{ie} = H_{ear} \ast  H_{bone} \ast  s_{o}$. 

\subsection{\textbf{Understanding Mechanical Properties of Food Textures}} \label{sec:understanding_mechanical_food_texture}

Food texture attributes play an important role in daily diet monitoring. By understanding the texture attributes of foods, individuals can make more informed choices that align with their health goals. Furthermore, it can aid healthcare professionals in developing tailored dietary recommendations that consider individual preferences and needs, ultimately promoting better adherence to nutritional guidelines. Additionally, recognizing food textures can help design personalized nutritional strategies and balanced dietary structures. As for regular food intake, the food textures can be divided into the following categories~\cite{nishinari2018perception}:
\begin{itemize}
    \item \textit{Hardness} refers to the resistance of food to deformation when pressure is applied, distinguishing between soft and hard foods. Hard foods require more force to chew and generate distinct sounds that can be more pronounced.
    \item \textit{Crispness} is associated with the freshness and crispness of food. Crunchy items often create sharp, audible sounds when bitten into.
    \item \textit{Adhesiveness} indicates the property of food that causes it to stick to surfaces, such as teeth or gums. Foods with high adhesiveness require more effort to chew, leading to unique sounds and prolonged chewing.
    \item \textit{Juiciness} pertains to the moisture content of food, as well as the food's ability to release liquid when bitten or chewed. Foods with high moisture can produce additional sounds related to liquid release during chewing.
\end{itemize}

In a broad sense, food texture profiles include mechanical profiles, geometrical profiles, and nutritional profiles~\cite{nishinari2018perception}. In our work, we mainly focus on the impact of mechanical profiles on sound responses. When teeth exert force on the food surface, foods with various textures have different mechanical responses, which have been widely studied in feeding biomechanics~\cite{nishinari2018perception, papapanagiotou2021recognition}. Next, we first leverage the critical stress intensity ($K_{ic}$) to describe how much force a material can withstand before cracking or breaking, helping us understand food texture attributes such as hardness and crispness. Based on the criteria in~\cite{berthaume2016food}, the critical stress intensity can be expressed as follows:

\begin{equation}
    \begin{aligned}
      K_{ic} = C_1 \sigma_{F} \sqrt{\alpha}, \quad \sigma_{F} = C_2\sqrt{\frac{ER} \alpha}
    \end{aligned}
\end{equation}
where $\sigma_{F}$, $\alpha$, $E$, and $R$ represent fracture stress, the depth of teeth inserted in the food, the Young’s modulus of the food, and the toughness of the food, respectively. In general, $C_1$ and $C_2$ can be considered as constant values. Afterwards, $K_{ic}$ can be defined as follows:
\begin{equation}\label{eq:K_ic}
    \begin{aligned}
      K_{ic} &= C_1 C_2(\sqrt{\frac{ER} \alpha}) \sqrt{\alpha} = C\sqrt{(ER)}
    \end{aligned}
\end{equation}
Based on Eq.~\ref{eq:K_ic}, different food textures determine the ability of the food to resist deformation during elastic deformation. As for hard foods, more forces are applied to break them down.
When food is broken, the force from outside will generate a large acceleration between the teeth and the food, and between the food fragments, causing frequency-varying vibrations.

\begin{figure}[h]
\centering
    \begin{minipage}[b]{1\linewidth}
        \centering
        \includegraphics[width=0.8\linewidth]{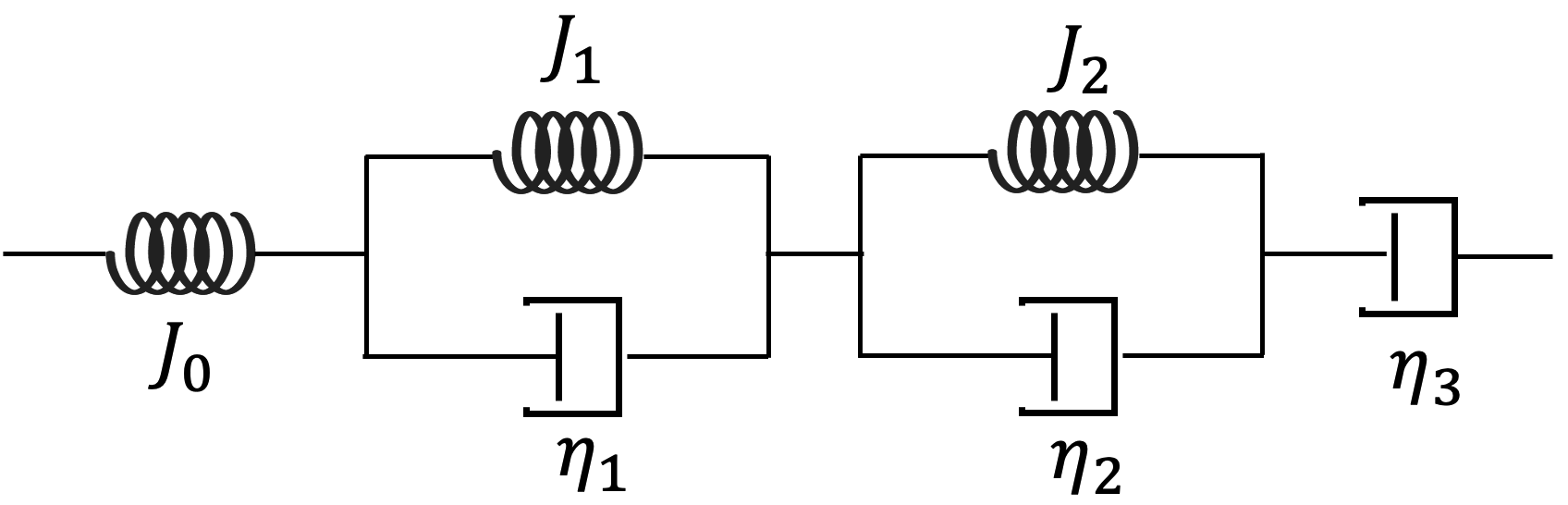} \caption{\label{fig:Kelvin_Voigt_Material_Model}
        \textbf{The mechanical model describing the adhesive characteristics of foods.}}
    \end{minipage}
    \vspace{-0.5cm}
\end{figure}

Next, we also introduce a Kelvin-Voigt Material model (shown in Fig.~\ref{fig:Kelvin_Voigt_Material_Model}) to represent the creep compliance parameter that describes the adhesive characteristics of foods. Based on~\cite{alzamora2008exploring}, the formula is simplified as follows:
\begin{equation}\label{eq:creep_compliance_parameter}
    \begin{aligned}
      J(t, \sigma_c) = J_1(1-e^{-t/\lambda_1}) + J_2(1-e^{-t/\lambda_2}) + \frac{t}{\eta}
    \end{aligned}
\end{equation}
where $J$, $t$, $\sigma_c$, $\lambda$, and $\eta$ represent the creep compliance, time, constant stress, retardation times, and the coefficient of viscosity, respectively. Specifically, Eq.~\ref{eq:creep_compliance_parameter} can be used to quantify a food's tendency to deform under a constant stress applied over time. 
As for foods with various viscosity coefficients, they require different levels of stress and time to exhibit significant deformation, resulting in unique characteristics in both time and frequency domains.

\subsection{\textbf{Audio-Mechanical Analysis}} \label{audio_mechanical_analysis}

The mechanical profiles of food have varying vibrations and resonance levels, directly affecting sound intensity, frequency, and duration during the chewing process. To understand the impact of food texture on in-ear sound responses, we collected the chewing-induced in-ear sounds of five food types (\ie, Photo chips, Apple, Nut, Beef, and Lettuce) with different food textures. Fig.~\ref{fig:person_food_types} demonstrates statistical features of sound waveforms of different food types. From Fig.~\ref{fig:person_food_types}\subref{fig:same_person_different_food_types}, as for the same individual, temporal features present a distinct clustering pattern, although some overlap still exists among the groups. Fig.~\ref{fig:person_food_types}\subref{fig:different_person_different_food_types} shows the feature visualization of food types across different individuals. We can observe that the data points are more dispersed and exhibit a higher degree of overlap between the food types. It indicates that inter-individual variability interferes with the distribution of temporal features across the same food categories. That is because chewing patterns vary from person to person. In addition, each individual has unique oral and skull structures, bringing different in-ear sound attenuation during bone conduction~\cite{han2023breathsign, han2024earsleep}. 

% Characteristics of chewing-induced in-ear sounds are influenced by food texture attributes~\cite{nishinari2018perception, papapanagiotou2021recognition}. 

% contributes to a broader distribution of temporal features across the same food categories. 

% shows the temporal features when the food types are consumed by various individuals. Here, the data points are more dispersed and exhibit a higher degree of overlap between the food types. This suggests that inter-individual variability contributes to a broader distribution of temporal features across the same food categories.

% different food types show 

% From Fig.~\ref{fig:person_food_types}\subref{fig:same_person_different_food_types}, 

% select five food types with different food textures to conduct a preliminary study.

\begin{figure*}
\centering
    \begin{minipage}[t]{1\linewidth}
        \centering
        \includegraphics[width=0.95\linewidth]{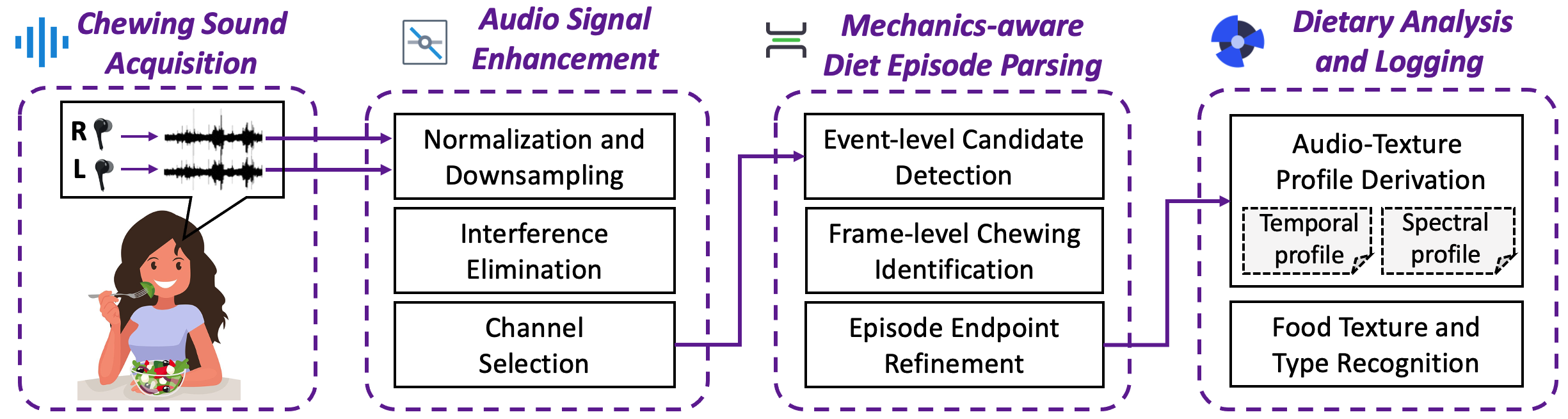}
        \caption{\textbf{System overview of \systemName, which can provide multi-level information for diet monitoring, such as eating episode detection, food texture recognition, and food type recognition.}}
        \label{fig:eardiet_system_overview}
    \end{minipage}
\end{figure*}

\section{\textbf{System Design}}

\subsection{\textbf{Overview}}
Fig.~\ref{fig:eardiet_system_overview} demonstrates the design framework of \systemName. When the user wears earbuds to eat, the built-in microphones on the earbuds automatically collect two channels (left and right channels) of in-ear audio streams. Nowadays, most noise-canceling earphones are equipped with inward-facing microphones, which are compatible with our solution. The SNR of in-ear audio signals is improved by the signal enhancement module. Then, the diet episode parsing module continuously identifies eating activity from high-SNR in-ear audio signals. Diet events are fed into the dietary analysis and logging module to identify food types and food textures simultaneously. 

% The built-in microphones on the earbuds automatically collect two channels (left and right channels) of in-ear audio streams. 

% food texture recognition module to identify the intake food types. Based on the recognized parameters like food textures, eating durations, and chewing frequency, the food weight estimation module will calculate the weight of food intake.

% \begin{figure}[t]
%     \begin{minipage}[t]{0.45\linewidth}
%         \centering
%         \includegraphics[width=0.75\linewidth]{./Figures/earphone_structure.png} \caption{\label{fig:earphone_structure}
%         \textbf{Hardware prototype structure.}}
%     \end{minipage} \quad
%     % \begin{minipage}[t]{0.3\linewidth}
%     %     \centering
%     %     \includegraphics[width=1\linewidth]{./Figures/3D_print.png} \caption{\label{fig:3D_print}
%     %     \textbf{3D-printed cavity.}}
%     % \end{minipage}\quad
%     \begin{minipage}[t]{0.45\linewidth}
%         \centering
%         \includegraphics[width=1\linewidth]{./Figures/hardware_prototype.png} \caption{\label{fig:hardware_prototype}
%         \textbf{Portable prototype.}}
%     \end{minipage}
    
% \end{figure}

% \subsection{\textbf{Hardware Prototype: Chewing Sound Acquisition}}\label{sec:chewing_sound_acquisition}

% A hardware prototype is implemented to collect in-ear sounds. As shown in Fig.~\ref{fig:earphone_structure}, a small-sized and low-cost in-ear microphone sensor is embedded into the earbud. We make a 3-D printed earphone cavity  

% \textcolor{red}{To be added!}

\subsection{\textbf{Audio Signal Enhancement}}\label{sec:signal_enhancement}

\subsubsection{\textbf{Normalization and Downsampling}} 
In-ear microphones embedded into earbuds collect in-ear audio signals at a sampling rate of 48 kHz. Based on frequency-domain analysis, chewing-induced in-ear sounds are predominantly distributed below 1 kHz, as high-frequency components are attenuated by bone conduction while low-frequency components are amplified due to the occlusion effect. To enhance computational efficiency, the collected in-ear audio signals are first normalized to remove amplitude variations across recordings, ensuring consistent feature scaling. Subsequently, the signals are downsampled to 8 kHz, which preserves the relevant frequency content while significantly reducing data size and processing demands.

\subsubsection{\textbf{Interference Elimination}} 
\begin{figure}[t]
    \begin{minipage}[t]{0.48\linewidth}
        \centering
        \includegraphics[width=0.99\linewidth]{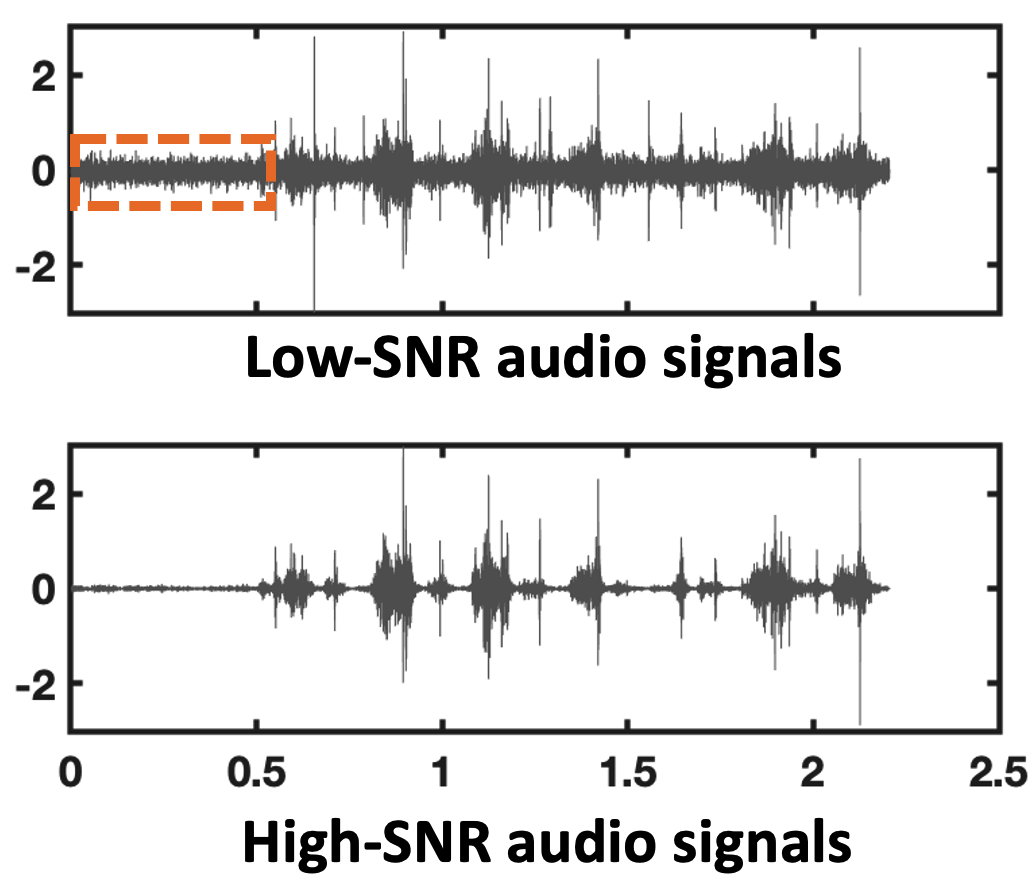} \caption{\label{fig:noisy_en_signals}
        \textbf{The effectiveness of interference elimination.}}
    \end{minipage} \hfill
    \begin{minipage}[t]{0.485\linewidth}
        \centering
        \includegraphics[width=1\linewidth]{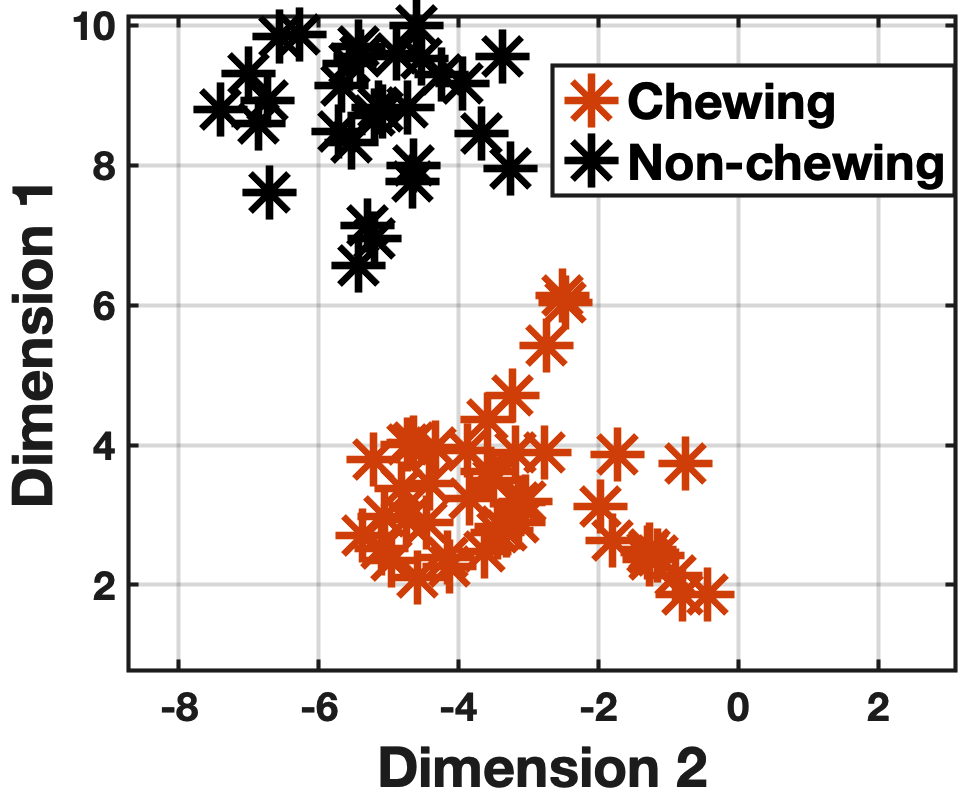} \caption{\label{fig:event_feature_effectiveness}
        \textbf{The effectiveness of event feature extraction.}}
    \end{minipage}
    \vspace{-0.5cm}
\end{figure}

As analyzed in Sec.~\ref{sec:chewing_induced_inear_sounds}, other activities may occur simultaneously during the eating process, bringing artifacts to chewing-induced in-ear sounds. 
It is noted that individuals typically do not chew when they are speaking because chewing requires the jaw to move in a repetitive motion, which can interfere with the precise movements needed for articulate speech. Prior literature has explored that the frequency components of these artifacts mainly distribute below 50 Hz. Thus, a high-pass filter is used to remove these artifacts.
After filtering-based artifact removal, residual noises still interfere with the target signals, as shown in the orange box of Fig.~\ref{fig:noisy_en_signals}. Residual noises mainly consist of blood flow-induced noises and ambient noises that enter the ear canals through earbuds. Ambient noises yield little impact on chewing-induced in-ear sounds, because the ear canal is blocked by the earphone, making it difficult for most environmental noise to enter the ear canal.~\cite{han2024earspeech,han2025earoe}. 
The distribution of blood flow-induced noises is close to a normal distribution. Thus, we adopt a Wiener filtering algorithm with a priori SNR estimation to improve the SNR further. From the comparison in Fig.~\ref{fig:noisy_en_signals}, the carefully designed signal processing approach improves the SNR of chewing-induced in-ear sounds significantly.

% \textcolor{red}{Music Echo}
% In our work, we first analyze the characteristics of residual noises and then design a corresponding denoising method. As shown in Fig.~\ref{}, ambient noises that are collect 

\subsubsection{\textbf{Channel Selection}}
Considering the impact of hardware limitations, earable wearing position, and other factors on recording quality, channel selection is required to determine which channel can be used for subsequent processing. To ensure effectiveness while keeping computational efficiency, the asymmetric variance ratio ($R$) between right-channel signals ($s_{r}$) and left-channel signals ($s_{l}$) is calculated. We perform asymmetric variance ratio calculations on the RMS envelope of the original waveform, which can eliminate the negative impact of outliers (such as spikes or noise). 
\begin{equation}
    R = \frac{\text{std}(E_{r})-\text{std}(E_{l})}{\text{std}(E_{r})+\text{std}(E_{l})} 
\end{equation}
where $E_{r}$ and $E_{l}$ denote the RMS envelope of the right-channel waveform and the left-channel waveform, respectively.
Ideally, the asymmetric variance ratio should be 0. Therefore, after channel selection, the output in-ear audio signals $s_{en}$ can be expressed as follows: 
\begin{equation}
    s_{en} = \left\{\begin{matrix}
 s_{r}  & \beta < R \\ 
 (s_{r} + s_{l})/2  & -\beta \leq R \leq \beta \\ 
 s_{r} & R < -\beta
\end{matrix}\right.
\end{equation}
We set a threshold $\beta$. If $R$ is larger than $\beta$, it means that the right channel contributes more than the left channel. If $-\beta \leq R \leq \beta$, it means that the left channel and the right channel contribute equally to the sensing task. In our work,  $\beta$ is set to 0.1.

\subsection{\textbf{Mechanics-aware Continuous Diet Episode Parsing}}\label{sec:mechanics_aware_diet_episode_parsing}
During the eating process, other oral activities (like talking, laughing, coughing, and sneezing) interfere with chewing activity recognition. Thus, \systemName~needs to recognize continuous chewing activities and isolate them from audio signals.
Existing chewing activity recognition solutions~\cite{chugh2025bitesense,ketmalasiri2024imchew, papapanagiotou2017chewing, papapanagiotou2021self} generally formulate diet event detection as an independent frame-level classification problem. However, chewing is inherently a temporally structured biomechanical process rather than a collection of isolated acoustic events. As introduced in Sec.~\ref{audio_mechanical_analysis}, during food mastication, repetitive jaw movements generate a sequence of chewing cycles interleaved with short pauses, swallowing transitions, and occasional speech interruptions. Thus, we reformulate chewing activity detection as a continuous temporal parsing problem and propose a mechanics-aware diet event parsing framework. The entire workflow is shown in Alg.~\ref{algo_diet_episode_parsing}.

\begin{algorithm}[h]
\small
\caption{\label{algo_diet_episode_parsing} \textbf{Mechanics-aware Continuous Diet Episode Parsing}} 

\LinesNumbered
\KwIn{$s$: the input chewing-induced in-ear sounds. }
\KwOut{$\mathcal{C}$: output episode set.}

$E_{rms} = \text{Smooth}(\text{Envelope}(s))$ \;
$\tau_{E} = \mu_E + k\sigma_E$ \; 
$ M(t)= \begin{cases} 1, & E_{rms}>\tau_{E}\\ 0, & \text{otherwise} \end{cases} $ \;
Identify all contiguous active regions in $M(t)$\;
Obtain candidate episode set: $\mathcal{C} = \{c_1,c_2,\ldots,c_{N}\}$\;

\For{$i \gets 1$ \KwTo $\operatorname{Length}(\mathcal{C})$}{
    \If{$\operatorname{Interval}(c_{i+1}, c_{i}) < 0.5$}{
            Merge $c_{i+1}$ and $c_{i}$ \;  
    }
    \If{$\operatorname{Duration}(c_{i}) < 1$}{
            Remove $c_{i}$ from $\mathcal{C}$ \; 
    }
}

\For{$i \gets 1$ \KwTo $\operatorname{Length}(\mathcal{C})$}{
    
    $\mathcal{B} = \{b_1,b_2,\ldots,b_n\} = \text{Findpeaks}(c_i)$\;
    Calculate $\mathcal{I} = \{Itv_1,Itv_2,\ldots, Itv_{n-1}\}$, where $Itv_{j} = b_{j}-b_{j-1}$\; 
    \For{$j \gets 1$ \KwTo $\operatorname{Length}(\mathcal{I})$}{
        Calculate the rhythm consistency score: $R_j = \exp \left( -\frac{(Itv_j-\mu_{ICI})^2}{2\sigma_{ICI}^2}\right)$\;
        \If{$R_j > \tau_{rhy}$}{
            Remove $b_{i+1}$ from $\mathcal{B}$\;           
        }
    } 
    Refine $c_i$ based on $\mathcal{B}$\;
}

\For{$i \gets 1$ \KwTo $\operatorname{Length}(\mathcal{C})$}{
   $\mathcal{V}= \text{Enframe}(c_i)$\;
   \For{$j \gets 1$ \KwTo $\operatorname{Length}(\mathcal{V})$}{
        $\{f_{zcr}, f_{ste}, f_{scd}, f_{ssd}, f_{mf}\} = \text{FeatureExtract}(\mathcal{V}_j)$\;
        $score = \text{Classifier}(\{f_{zcr}, f_{ste}, f_{scd}, f_{ssd}, f_{mf}\})$\;
        \If{$score > \tau_{frame}$}{
            $\mathcal{V}_{j}$ is identified as a chewing frame\;      
        }\Else{
            $\mathcal{V}_{j}$ is identified as a non-chewing frame\; 
            Remove it from $\mathcal{V}$;   
        }
   }
    Refine $c_i$ based on $\mathcal{V}$;
}

\end{algorithm}

% This framework first estimates frame-level chewing evidence and subsequently infers temporally consistent chewing episodes through rhythm-aware sequence modeling and adaptive boundary inference. 

\subsubsection{\textbf{Event-level Candidate Detection}}
Chewing-induced in-ear sounds originate from repetitive tooth-food and food-food mechanical interactions. Unlike speech and environmental noise, chewing typically generates quasi-periodic acoustic bursts with concentrated low-frequency energy and characteristic temporal envelopes. Given the enhanced in-ear audio signal ($s$), we first derive the Root Mean Square (RMS) envelope ($E_{rms}$) to emphasize chewing-related acoustic responses while suppressing transient disturbances. Then, a dynamic threshold is employed:
\begin{equation}
\tau=\mu_E+k\sigma_E,
\end{equation}
where $\mu_E$ and $\sigma_E$ denote the mean and standard deviation of the entire envelope, respectively, and $k=3$ is an empirically selected coefficient. The candidate episode set denoted by $\mathcal{C}=\{c_1, c_2, \ldots,c_N\}$ consists of envelope portions exceeding the threshold. Furthermore, if the duration of an episode is less than 1 second, we will remove it from $\mathcal{C}$; if the interval between two consecutive episodes is less than 0.5 seconds, we will merge them together.
% Envelope portions exceeding the threshold are considered candidate chewing episodes. 

Furthermore, a key observation is that chewing is fundamentally a biomechanical process consisting of repetitive jaw movements. Therefore, chewing events do not occur independently but exhibit strong temporal dependencies. Then, in each candidate chewing episode, a peak detection approach is leveraged to denote the chewing bursts that are denoted by $\mathcal{B} = \{b_1,b_2,\ldots,b_n\}$. Then, the inter-chew interval (ICI) is defined as $Itv_i=b_{i}-b_{i-1}$. Based on observations from natural eating behaviors, the intervals between two consecutive chewing bursts are approximately 0.4 to 1 s. Detected peaks violating this constraint are discarded as non-chewing events. Within a chewing episode, the ICIs generally fluctuate around an individual-specific chewing rhythm. Based on this observation, we define a Gaussian rhythm consistency score as:
\begin{equation}
R_i=\exp \left( -\frac{(Itv_i-\mu_{ICI})^2}{2\sigma_{ICI}^2}\right),
\end{equation}
where $\mu_{ICI}$ and $\sigma_{ICI}$ denote the running mean and standard deviation of previously observed ICIs, respectively. A higher rhythm consistency score indicates that the current chewing burst is more likely generated by mastication rather than accidental oral activities such as coughing, laughing, or transient speech. By incorporating rhythm consistency into the episode-level candidate detection, \systemName~can better distinguish genuine chewing events from spurious acoustic peaks and achieve more robust chewing identification.

\subsubsection{\textbf{Frame-level Chewing Identification Refinement}}
\label{sec:event_feature_extraction}
Based on the above coarse-grained candidate chewing episodes, a fine-grained frame-level chewing identification is conducted. Within each candidate episode, the audio signals are segmented into multiple frames by a sliding window with a length of 1 second and a step of 0.5 seconds. Compared with other oral activities, chewing activities present unique behavioral characteristics. Thus, \systemName~extracts behavioral acoustic responses from each frame to distinguish chewing activities from other oral activities. Firstly, chewing involves mechanical actions of the jaw, tongue, and teeth, which create intermittent and often rhythmic sounds. As the jaw moves up and down and side to side, it generates a series of noise bursts that can cause rapid fluctuations in the audio signal. Thus, chewing-induced sounds are typically periodic and rhythmic and can be characterized by Zero Crossing Rate ($f_{zcr}$) due to rapid changes in the signal as the jaw moves. Secondly, the interaction between teeth, food, tongue, etc., produces short bursts of sound with varying energy levels during the chewing process. However, as for talking, laughing, coughing, etc., they involve the vocal cords and the respiratory tract, resulting in different levels of intensity. Thus, the short-time energy ($f_{ste}$) is calculated to represent the intensity level. Furthermore, they also present distinct resonant frequencies that are related to an individual’s vocal tract shape. Since chewing activities do not involve the vibrations of the vocal cords and the vocal tract, they generate complex sounds that may have certain resonant frequencies depending on the food being chewed, the structure of the mouth, and the physical dynamics of the jaw. Thus, we extract spectral features including Spectral Centroid ($f_{scd}$), Spectral Spread ($f_{ssd}$), and Mel Frequency Cepstral Coefficient (MFCC, $f_{mf}$) with 12 coefficients to describe the resonant characteristics of chewing-induced in-ear sounds. To form a characteristic representation, we compute the statistical mean values of all extracted features across each frame and standardize based on the standard deviation of the original data. This results in a 16-dimensional feature matrix for each frame, where each dimension represents the average value of a domain feature. Fig.~\ref{fig:event_feature_effectiveness} demonstrates the effectiveness of extracted features in terms of chewing activity detection.

\begin{figure*}[t]
\centering
    \begin{minipage}[t]{1\linewidth}
       \subfloat[\textbf{Hardness analysis}]{\includegraphics[width=0.232\textwidth]{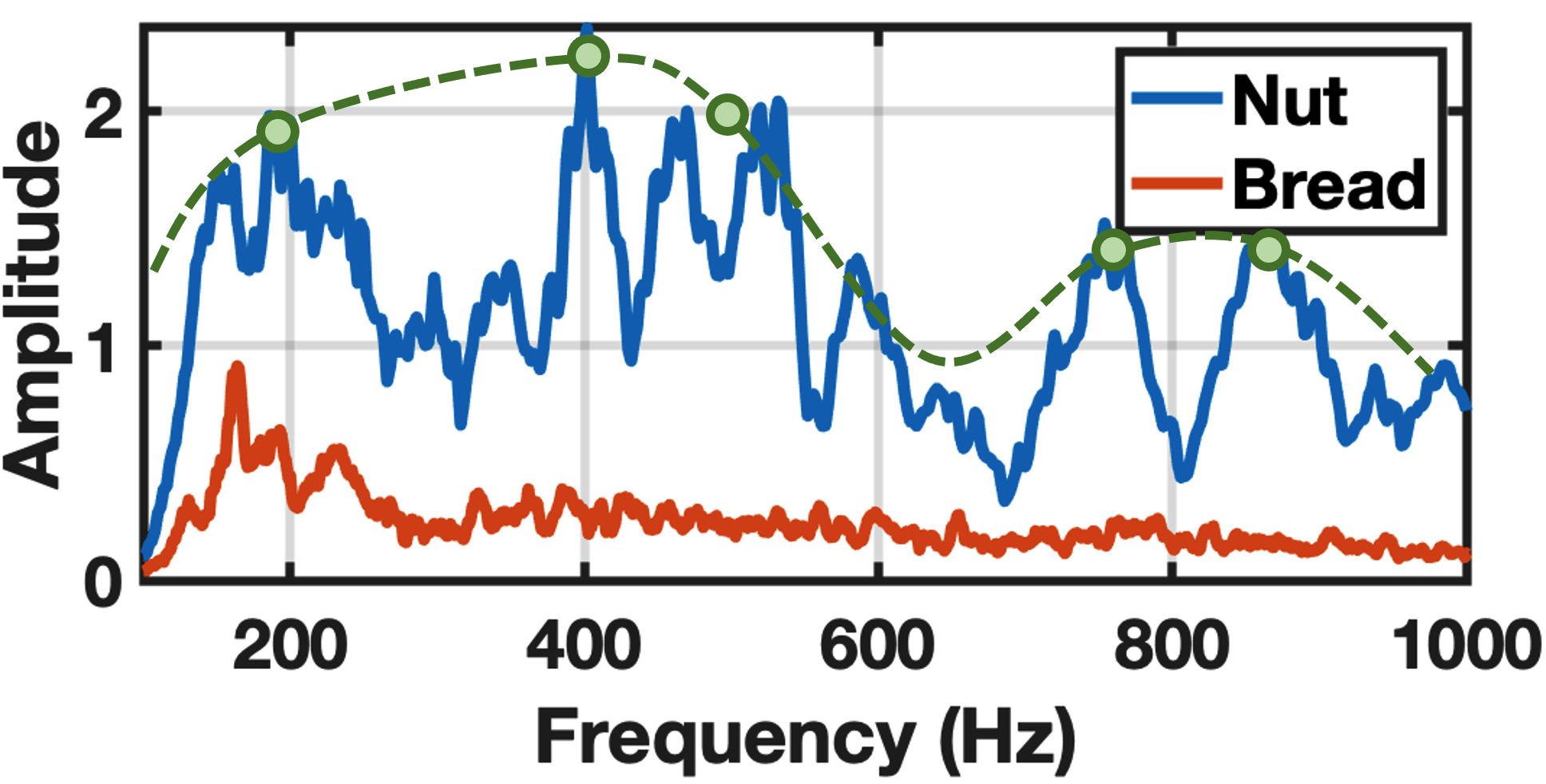}
        \label{fig:Hardness_analysis}} \hfill
        \subfloat[\textbf{Crispness analysis.}]{\includegraphics[width=0.23\textwidth]{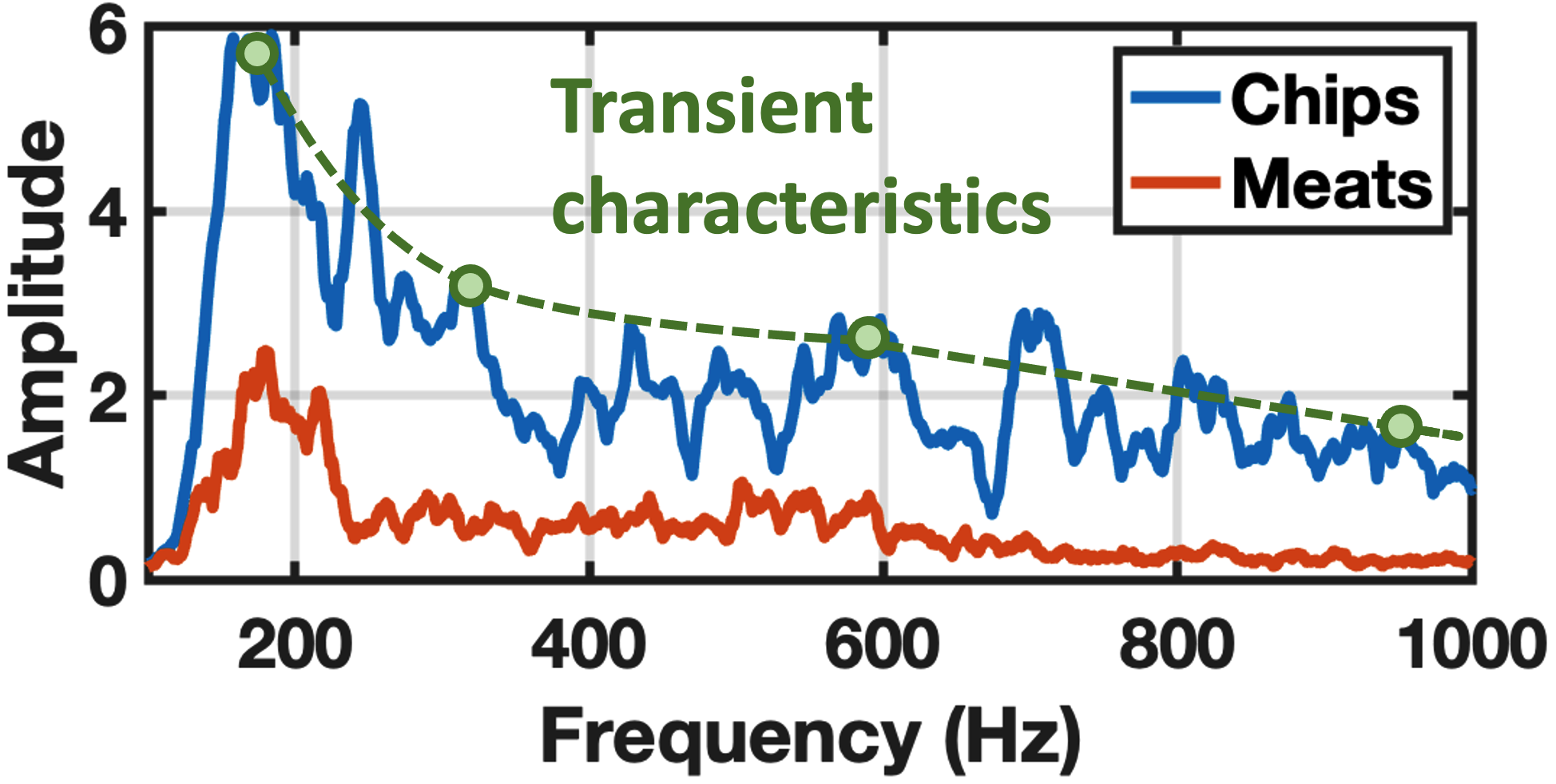}
    \label{fig:Cripness_analysis}} \hfill
        \subfloat[\textbf{Juiciness analysis.}]{\includegraphics[width=0.23\textwidth]{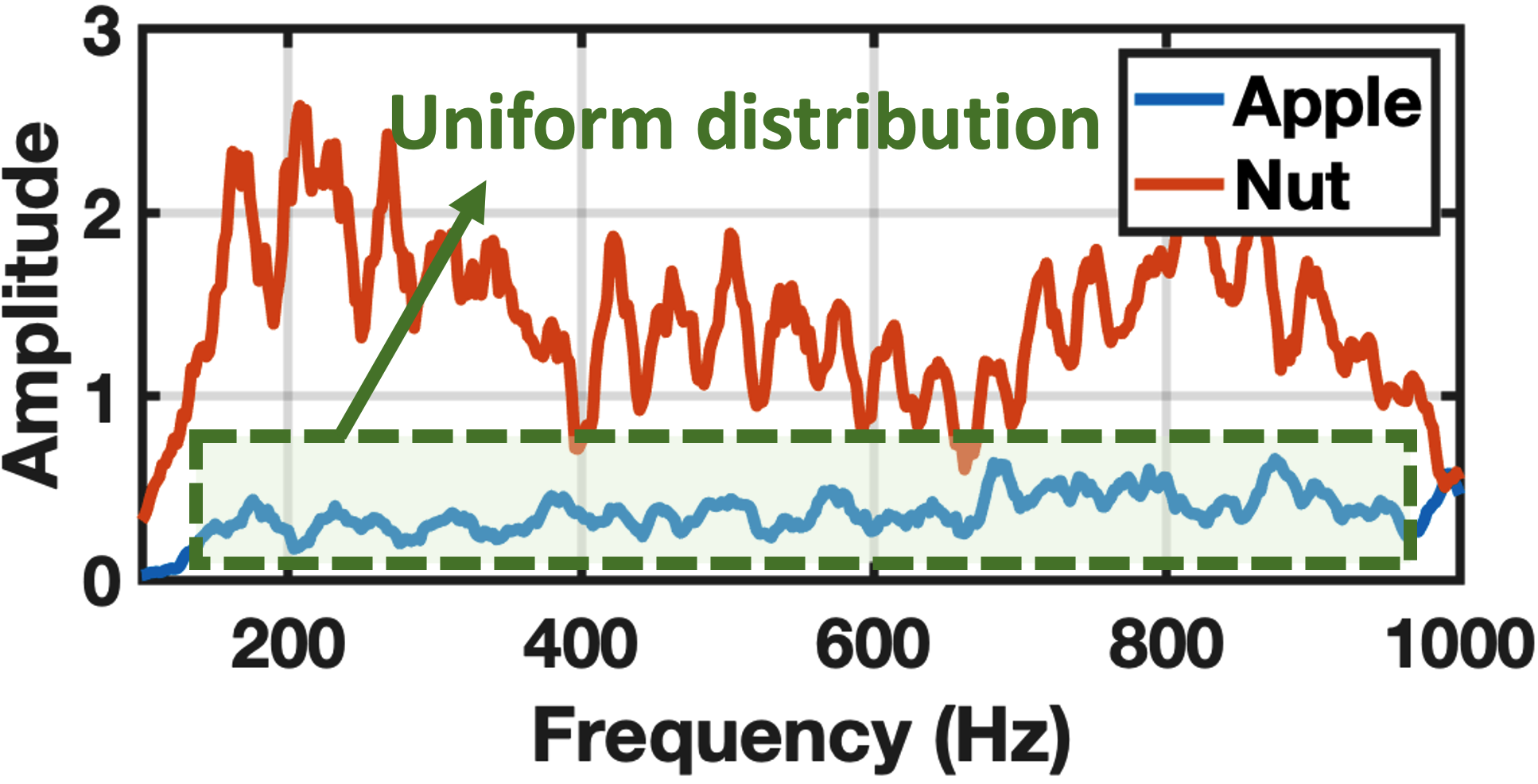}
    \label{fig:Juiciness_analysis}} \hfill
    \subfloat[\textbf{Adhesiveness analysis.}]{\includegraphics[width=0.23\textwidth]{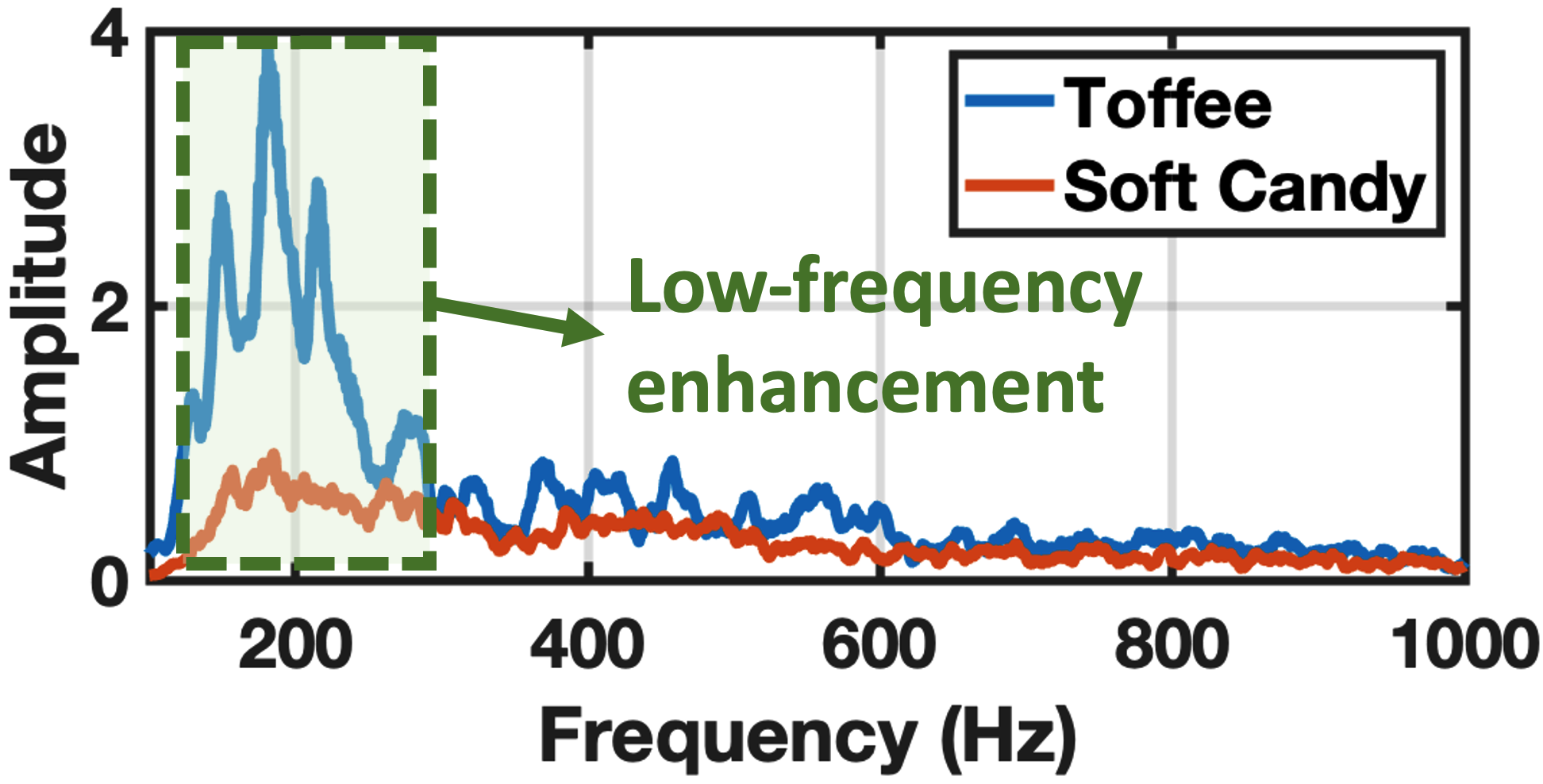}
    \label{fig:Adhesiveness_analysis}} 
    \caption{\label{fig:food_texture_analysis}\textbf{Food texture audio spectral profile analysis.}} 
    \end{minipage}
    % \vspace{-2mm}
\end{figure*}

To enable frame-level chewing identification, we employ a lightweight ensemble classifier consisting of Support Vector Machine (SVM), Random Forest (RF), and Decision Tree (DT) models. Each classifier independently predicts the chewing probability of a given frame, and their outputs are subsequently fused through a fully connected layer followed by a sigmoid activation function. During training, a frame is labeled as chewing if chewing activity occupies more than half of the window duration; otherwise, it is labeled as non-chewing. During inference, the estimated chewing probability is compared against a predefined decision threshold, where frames exceeding the threshold are classified as chewing, while the remaining frames are classified as non-chewing. 

% Fig.~\ref{fig:event_segmentation_demo} shows the effectiveness of the chewing episode detection and segmentation. Compared to the hard threshold-based approach, our approach can accurately identify eating episodes.

\subsection{\textbf{Dietary Analysis and Logging}}
\label{sec:dietary_analysis_logging}

\subsubsection{\textbf{Texture-Audio Analysis}}
Food texture attributes contribute to tooth-food and food-food contact vibrations, resulting in unique temporal and spectral characteristics. Next, we aim to explore the relationship between food texture and acoustic responses. Common texture attributes of food consumed daily can be divided into four categories, \ie, hardness, crispness, juiciness, and adhesiveness. Although prior solutions have explored the feasibility of leveraging chewing sounds for food texture recognition~\cite{papapanagiotou2021recognition}, they still lack the interpretation of audio profiles for texture attributes, resulting in low accuracy and weak robustness.

\textit{(\rmnum{1}) Spectral profile analysis.} 
In our work, we first conduct a detailed spectral analysis of the texture audio profiles in Fig.~\ref{fig:food_texture_analysis}. To validate the impact of a specific attribute on audio profiles, we keep other attributes the same across different food types. 
\begin{itemize}
    \item \textit{Hardness analysis.} Hard foods such as nuts exhibit pronounced high-frequency and high-power spectral components due to the strong mechanical interaction between teeth and the food. As shown in Fig.~\ref{fig:food_texture_analysis}\subref{fig:Hardness_analysis}, nuts with a hardness attribute present multiple peaks between 200–900 Hz, reflecting repetitive fracture events of rigid structures. In contrast, softer foods like bread show a markedly attenuated spectrum across all frequencies, indicating reduced fracturing energy during mastication.
    
    % Since hard foods need more force to chew, the interaction between teeth and food and the interaction of food structures cause high-frequency and high-power sounds, as shown in Fig.~\ref{fig:food_texture_analysis}\subref{fig:Hardness_analysis}. 
    \item \textit{Crispness analysis.} Foods with high crispness, such as chips, produce transient, high-intensity low-frequency bursts (around 200 Hz) that decay gradually as frequency increases, as observed in Fig.~\ref{fig:food_texture_analysis}\subref{fig:Cripness_analysis}. This decay pattern reflects the rapid rupture of brittle structures, where the spectral power is initially concentrated in lower harmonics and dissipates in higher bands. Meats, with lower crispness, maintain a consistently low power across the spectrum, indicating the absence of such transient fracture events.

    % The rupture of foods with high crispness generates transient low-frequency sounds with high intensity. As the frequency increases, the spectral power decreases gradually, as shown in Fig.~\ref{fig:food_texture_analysis}\subref{fig:Cripness_analysis}.

    \item \textit{Juiciness analysis.} High-moisture foods like apples generate a relatively uniform spectral power distribution across 200–1000 Hz, as shown in Fig.~\ref{fig:food_texture_analysis}\subref{fig:Juiciness_analysis}.
    This uniformity of spectral energy arises from the combined effects of food tissue rupture and liquid release. In contrast, low-moisture foods such as nuts exhibit uneven spectral profiles dominated by mid-frequency components, reflecting the absence of fluid-mediated acoustic scattering.

    % When the foods with high moisture content are broken, the released liquid will produce a unique sound compared to solid materials, presenting a uniform spectral power distribution as shown in . 

    \item \textit{Adhesiveness analysis.} Sticky foods introduce more resistance when chewing, resulting in the significant enhancement of low-frequency components ($<$ 300 Hz), as shown in Fig.~\ref{fig:food_texture_analysis}\subref{fig:Adhesiveness_analysis}. This enhancement is due to prolonged contact and resistance between the food and teeth during mastication, generating sustained, low-frequency frictional sounds. Yet, soft candies with lower adhesiveness yield a comparatively flatter and lower-amplitude spectral profile, indicating reduced resistance and shortened chewing contact time.
    
    % Sticky foods introduce more resistance when chewing, resulting in the significant enhancement of low-frequency components. 
    % Different from the crispness attribute, the spectral power of the adhesiveness audio profile presents a decay pattern, as shown in .    
\end{itemize}

% \begin{table}[t]
% \centering
% \caption{Summary of spectral characteristics for different food texture attributes.}
% \label{tab:spectral_summary}
% \begin{tabular}{p{2.0cm}  p{6cm}}
% \toprule
% \textbf{Texture}  & \textbf{Spectral Pattern Description} \\
% \midrule
% \textbf{Hardness} &  Strong high-frequency and high-power bursts caused by rigid structure fracture \\
% \textbf{Crispness} &   Transient fracture events; energy concentrated at low frequencies with decay towards high frequencies \\
% \textbf{Juiciness} &   Broadband energy distribution from tissue rupture and liquid release \\
% \textbf{Adhesiveness} &  Sustained low-frequency components due to prolonged chewing resistance \\
% \bottomrule
% \end{tabular}
% \end{table}

\begin{figure}[t]
    \centering
    \includegraphics[width=0.9\linewidth]{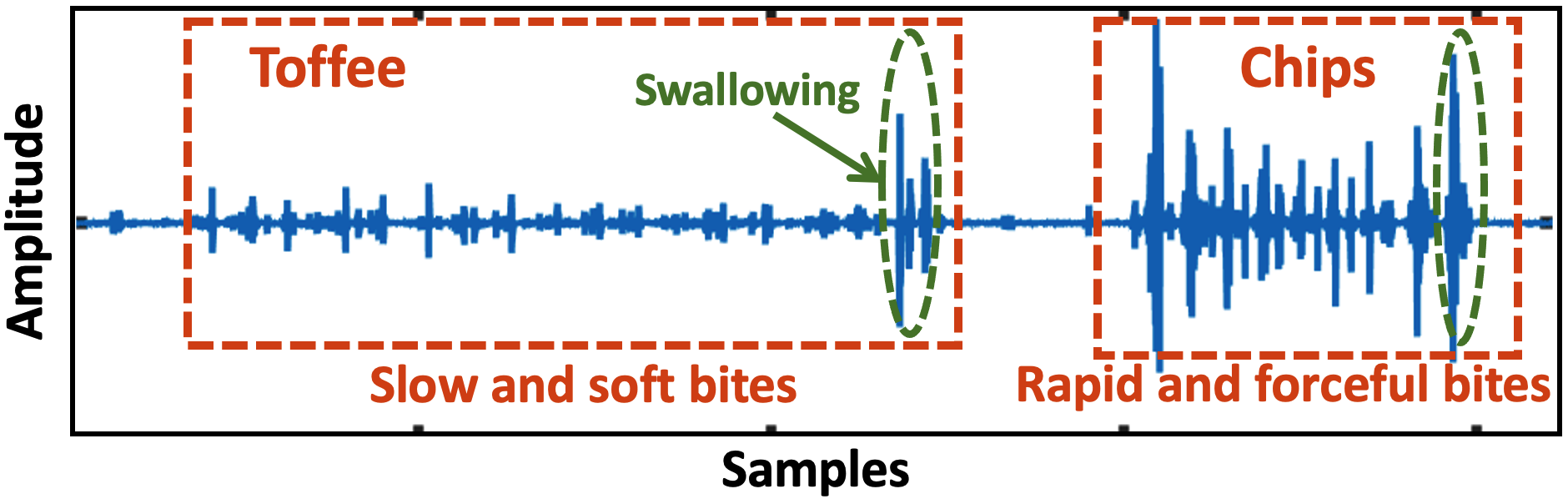}
    \caption{\textbf{Food texture audio temporal profile analysis.}}
    \label{fig:Texture_time_analysis}
    % \vspace{-0.5cm}
\end{figure}

\textit{(\rmnum{2}) Temporal profile analysis.} The in-ear sounds of chewing different foods not only present different patterns in the frequency domain but also present unique patterns in the time domain, as shown in Fig.~\ref{fig:Texture_time_analysis}. For example, adhesive foods require more effort to chew and may lead to longer mastication times or slower chewing rates. The chewing pattern for crisp foods often includes rapid and repetitive bites. In addition, when the crisp foods are first bitten into, they produce a loud sound due to the rapid fracturing of their structure. As the food is chewed into pieces, the amplitude of chewing sounds gradually decreases. Thus, we can conclude that the temporal and spectral characteristics of chewing in-ear sounds both contain a wealth of information on food texture.

% As you continue to chew, the food breaks into smaller pieces, which generally produce softer sounds.

% Foods that are highly adhesive require more effort to chew and may lead to longer mastication times. During the chewing process, the food may cling to the teeth, causing the individual to make more lateral and vertical movements to dislodge and break it down. This can result in a slower chewing rate initially, as the individual focuses on managing the stickiness.

\begin{figure}[t]
\centering
    \begin{minipage}[t]{1\linewidth}
    \centering
        \subfloat[\textbf{STFT representation.}]{\includegraphics[width=0.85\textwidth]{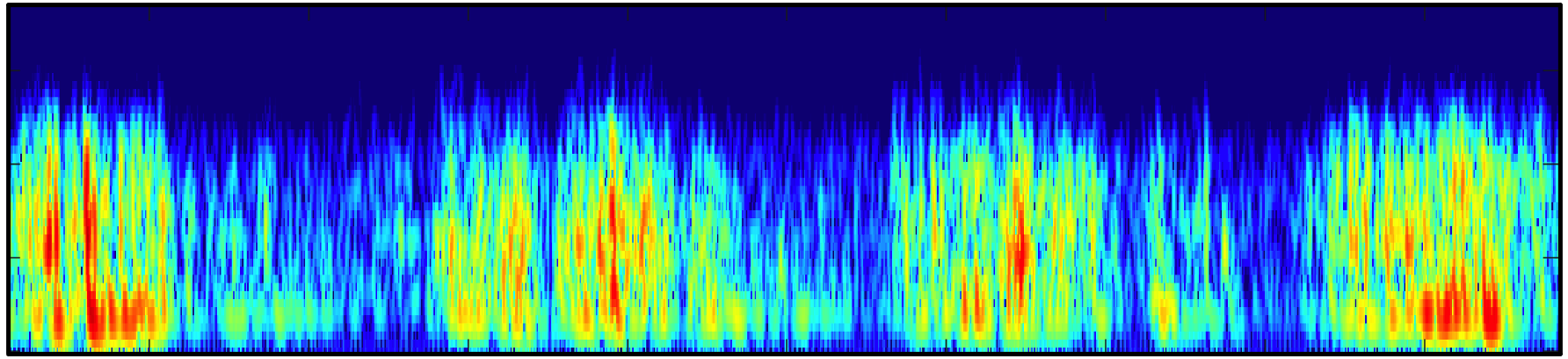}
        \label{fig:stft}} \\
        \subfloat[\textbf{CWT representation.}]{\includegraphics[width=0.85\textwidth]{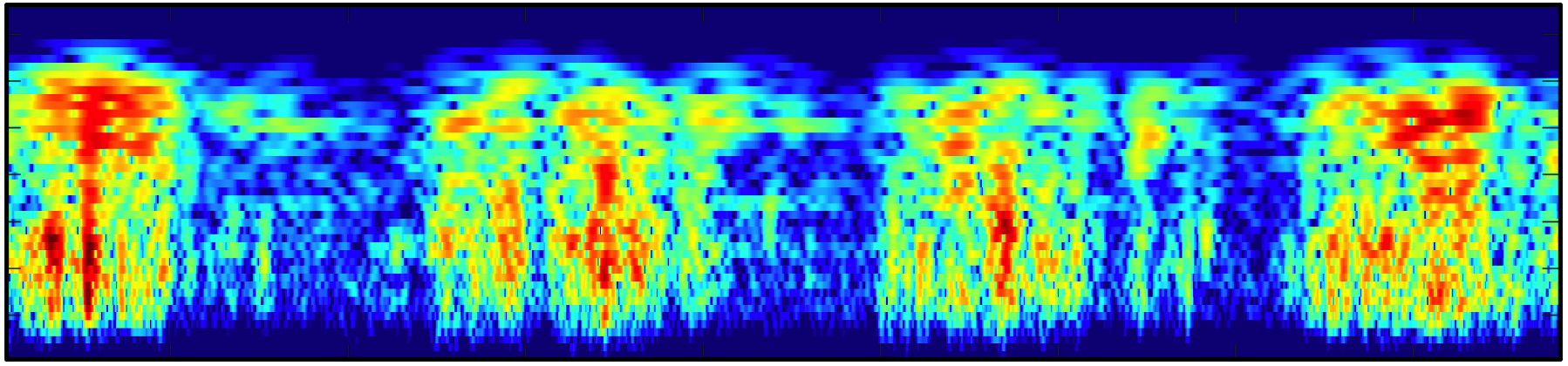}
        \label{fig:cwt}} 
    \caption{\label{fig:stft_cwt_comparison}\textbf{Comparison between STFT and CWT representations.}} 
    \end{minipage}
    \vspace{-0.5cm}
\end{figure}

\subsubsection{\textbf{CWT-based Profile Derivation}} 
Short-time Fourier Transform (STFT) is a common technique to analyze the frequency variations of signals that change over time~\cite{han2024earspeech, shin2022mydj, roddiger2022sensing}. However, the poor frequency resolution and high noise sensitivity of STFT make it unsuitable for our work. In our work, we adopt Continuous Wavelet Transform (CWT) to convert audio signals into time-frequency representations. CWT could provide high resolution for low-frequency components, especially suitable for narrow-bandwidth in-ear sounds. As shown in Fig.~\ref{fig:stft_cwt_comparison}, the CWT-based representation demonstrates fine-grained time–frequency structure, which can offer discriminative features more effectively for different food textures. 

Next, we elaborate on the CWT representation derivation pipeline. The continuous wavelet transform (CWT) analyzes $s(t)$ via scaled and shifted versions of a mother wavelet $\psi(t)$. To capture good
time-frequency localization, the complex Morlet function is selected as a mother wavelet, which is defined as follows:
\begin{equation}
    \psi(t) = \pi^{-\frac{1}{4}} \, e^{j 2 \pi f_c t} \, e^{-\frac{t^2}{2\sigma^2}}
    \label{eq:morlet}
\end{equation}
where $f_c$ is the center frequency, $\sigma$ controls the temporal spread, and $j=\sqrt{-1}$. 

For a sampled signal $s[n]=s(nT_s)$, $n=0,\dots,N-1$, with sampling period $T_s=1/f_s$, a discrete approximation of CWT is denoted as follows
\begin{equation}
    CWT_s(f,b) \approx \frac{1}{\sqrt{|a|}}
    \sum_{n=0}^{N-1} s[n]\,
    \psi^{*}\!\left(\frac{nT_s-b}{a}\right).
\label{eq:discrete_cwt}
\end{equation}
where $f = \frac{f_c}{a}$, $a$ is the frequency scale parameter, $b$ is the time shift parameter, and $\psi^{*}{(\cdot)}$ denotes the complex conjugate of the mother wavelet. Due to the occlusion effect, the effective frequency components of in-ear sound are concentrated below 1000~Hz, so we only retain the low-frequency components (below 1000~Hz) of the CWT representation subsequently. Furthermore, we conduct pixel-level processing to highlight salient time–frequency components and eliminate insignificant noise with Otsu’s thresholding method.

% to suppress background noise while retaining salient time–frequency structures. 

% in which high-energy ridges and harmonics were preserved while diffuse low-energy noise was suppressed. To further improve structural continuity, a light morphological opening/closing was optionally applied to the binary mask. This parameter-free procedure adapts to recording-level intensity variations and enhances class separability by emphasizing discriminative time–frequency features for subsequent classification tasks.

% \textcolor{red}{To be added!}

% \textit{(\rmnum{1}) Normalization}
% \textit{(\rmnum{1}) Thresholding}

\begin{figure}[t]
\centering
    \begin{minipage}[t]{1\linewidth}
        \centering
        \includegraphics[width=1\linewidth]{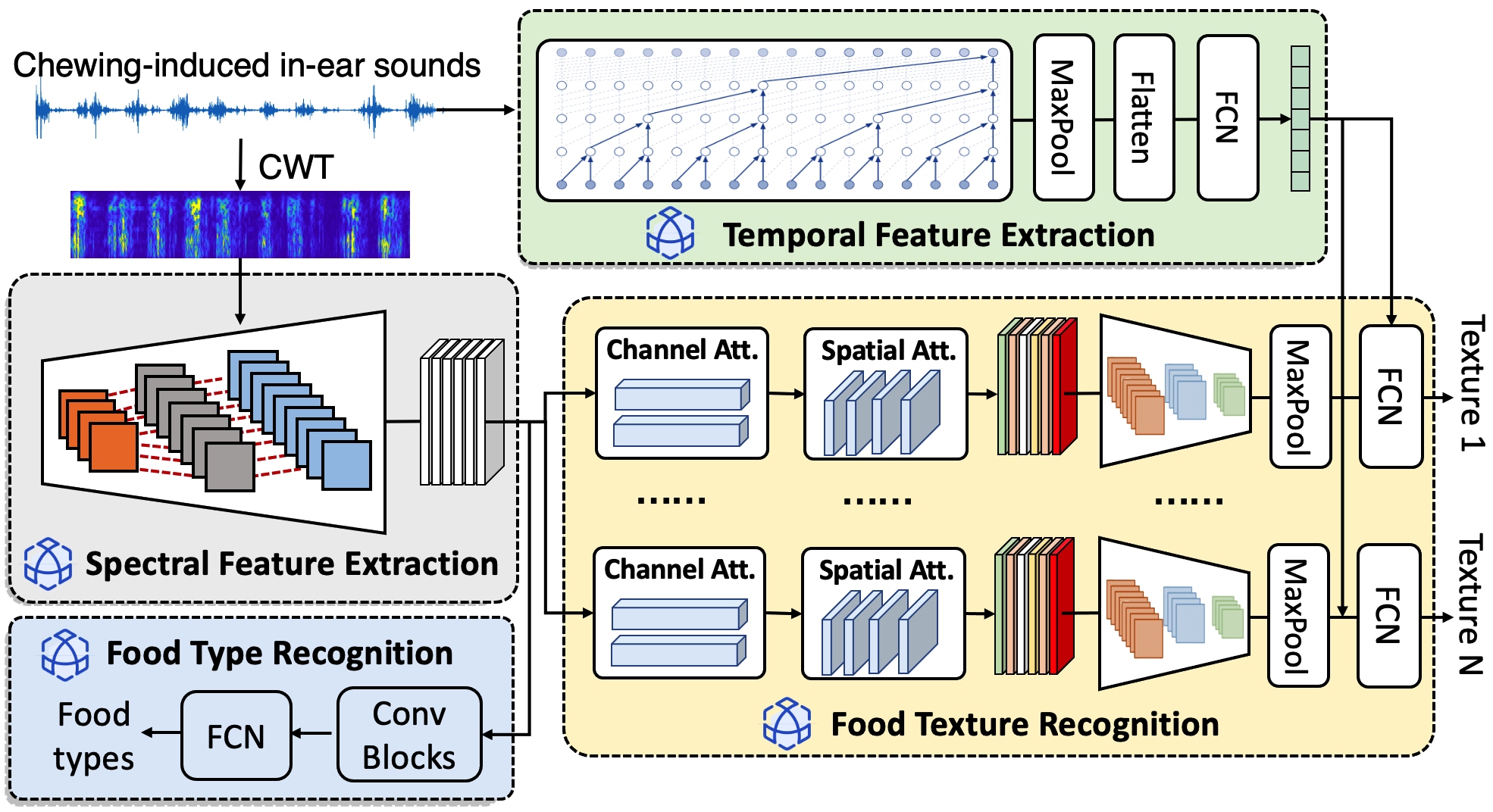} \caption{\label{fig:model_structure_1}
        \textbf{The model structure of food texture and type recognition.}}
    \end{minipage}
    \vspace{-0.5cm}
\end{figure}

\subsubsection{\textbf{Multi-branch Deep Learning Model for Food Texture and Type Recognition}}\label{sec:food_texture_type_model}
We design a multi-branch learning framework for simultaneously identifying food types and food textures. The detailed structure of the food texture recognition network is in Fig.~\ref{fig:model_structure_1}.  

\textit{(\rmnum{1}) Spectral Feature Extraction Module.} The pre-processed in-ear audio signals are transformed into a CWT spectrogram denoted by $S_{cwt}^{T  \times F \times 1}$, where $F$ denotes the number of frequency bins and $T$ denotes the number of time frames. 
Then, $S_{cwt}$ is pumped into the spectral feature extraction module that is designed to learn discriminative texture-related time-frequency representations. 
Specifically, the spectral feature extraction backbone consists of six convolutional blocks. The first convolutional block consists of three consecutive convolutional layers with the kernel size of $3\times1$, $1\times3$, and $3\times3$, respectively. This asymmetric design decomposes two-dimensional time-frequency feature extraction into frequency-aware and time-aware operations, thereby jointly modeling local time-frequency patterns. Then, five encoder convolutional blocks are employed to progressively refine the spectral representation. In each encoder convolutional block, an encoder convolutional layer is followed by batch normalization and ReLU activation. The detailed parameters of these convolution blocks are shown in Tab.~\ref{tab:shared_backbone}. Since all convolutional layers use appropriate zero-padding and no pooling operation is applied, the temporal and frequency resolutions of the feature maps are preserved throughout this module. The final output of the spectral feature extraction module is a high-level spectral representation, \ie, $\mathbf{F}_{spec} \in \mathbb{R}^{T  \times F \times C}$, where $C=64$. To provide a unified spectral representation of food-related acoustic responses, the extracted spectral representation is shared by subsequent task-specific branches, including food texture recognition and food type classification.

\begin{table}[t]
\centering
\caption{\textbf{Configuration of the spectral feature extraction module.}}
\label{tab:shared_backbone}
\begin{tabular}{cccc}
\toprule
Block & Kernel Size & Output Channels & Operation \\
\hline

Encoder-1-1 & $3\times1$ & 16 & Conv + BN + ReLU \\
Encoder-1-2 & $1\times3$ & 16 & Conv + BN + ReLU \\
Encoder-1-3 & $3\times3$ & 16 & Conv + BN + ReLU \\
Encoder-2 & $3\times3$ & 16 & Conv + BN + ReLU \\
Encoder-3 & $3\times3$ & 32 & Conv + BN + ReLU \\
Encoder-4 & $5\times5$ & 32 & Conv + BN + ReLU \\
Encoder-5 & $5\times5$ & 64 & Conv + BN + ReLU \\
Encoder-6 & $5\times5$ & 64 & Conv + BN + ReLU \\
\bottomrule
\end{tabular}
\end{table}

\textit{(\rmnum{2}) Temporal Feature Extraction Module.} 
Based on the analysis in Sec.~\ref{sec:understanding_mechanical_food_texture}, in addition to spectral representations, chewing-induced in-ear sounds also contain rich temporal dynamics, such as chewing rhythm, burst evolution, and short-term intensity variations. To capture these sequential patterns, we design a temporal feature extraction module that directly operates on the one-dimensional sound waveform. First, the input in-ear audio signals are fed into a Temporal Convolutional Network (TCN)~\cite{lea2017temporal}, which contains four dilated and causal 1-D convolutional layers with channel dimensions configured as $[1,3,3,1]$. Each temporal convolution uses a kernel size of 5 and a dropout rate of 0.2. Through stacked one-dimensional convolutions, the TCN can effectively capture local and mid-range temporal dependencies of chewing-induced acoustic responses.
The output of the TCN is then passed through a one-dimensional max-pooling layer with a kernel size of 5, a stride of 2, and a padding of 2. Then, the pooled feature map is subsequently flattened into a one-dimensional vector and projected into a compact latent representation through a fully connected layer. Lastly, a 1-dimensional feature vector $F_{temp}^{1\times 256}$ is output from the temporal feature extraction module. This compact temporal vector provides complementary temporal information to the spectral representations in the subsequent food texture recognition stage.

% and is subsequently fused with spectral features for food texture and food type recognition.

% the spectral feature extraction network, and the T-domain representation is fed into the temporal feature extraction network.
% In the spectral feature extraction network, we leverage a DenseNet-30 network~\cite{huang2017densely} as the backbone. After processing, the input CWT representation $S_{cwt}^{T\times F\times 1}$ is transformed as $F_{spec}^{T\times F\times C}$, where $C$ is the number of feature map channels. 

\textit{(\rmnum{3}) Food Texture Recognition Module.} The output of the spectral feature module (\ie, $F_{spec}^{T\times F\times C}$) is fed into four food texture recognition modules with the same structure. Each food texture recognition network consists of a channel attention module and a spatial attention module~\cite{woo2018cbam} to weight the channel dimension and the spatial dimension of the incoming spectral feature map. 
The refined feature map is processed by three successive convolutional blocks that progressively compress the channel dimension from 32 to 1. Each block consists of a $3\times3$ convolution, batch normalization, and ReLU activation. The resulting feature map is further downsampled using a $5\times5$ max-pooling operation to obtain a compact texture representation. Then, the max-pooled feature map is flattened and projected into a one-dimensional latent embedding $F_{text}^{1\times 256}$ through a fully connected layer. Lastly, 
$F_{temp}^{1\times 256}$ is concatenated with $F_{text}^{1\times 256}$ together and then fed into the two fully connected layers for food texture recognition. 

% Lastly, the maxpooled feature map is flattened and projected into a one-dimensional latent embedding through two fully connected layers. The obtained embedding is subsequently used by the corresponding texture prediction head for recognizing food texture attributes.

\textit{(\rmnum{4}) Food Type Recognition Module.} 
The food type recognition module is designed to predict the food category from the shared spectral representation. Thus, a food type recognition module is designed to process the spectral feature map $F_{spec}^{T\times F\times C}$) and outputs the food types. The structure of convolutional blocks is the same as that of the food texture recognition module. After the subsequent processing of convolutional layers and fully connected layers, $F_{type}^{T\times F \times C}$ is transformed to a feature vector $F_{out}^{1 \times K}$, where $K$ represents the number of food types.

\begin{figure*}[!t]
    \begin{minipage}[t]{0.3\linewidth}
        \centering
        \includegraphics[width=1\linewidth]{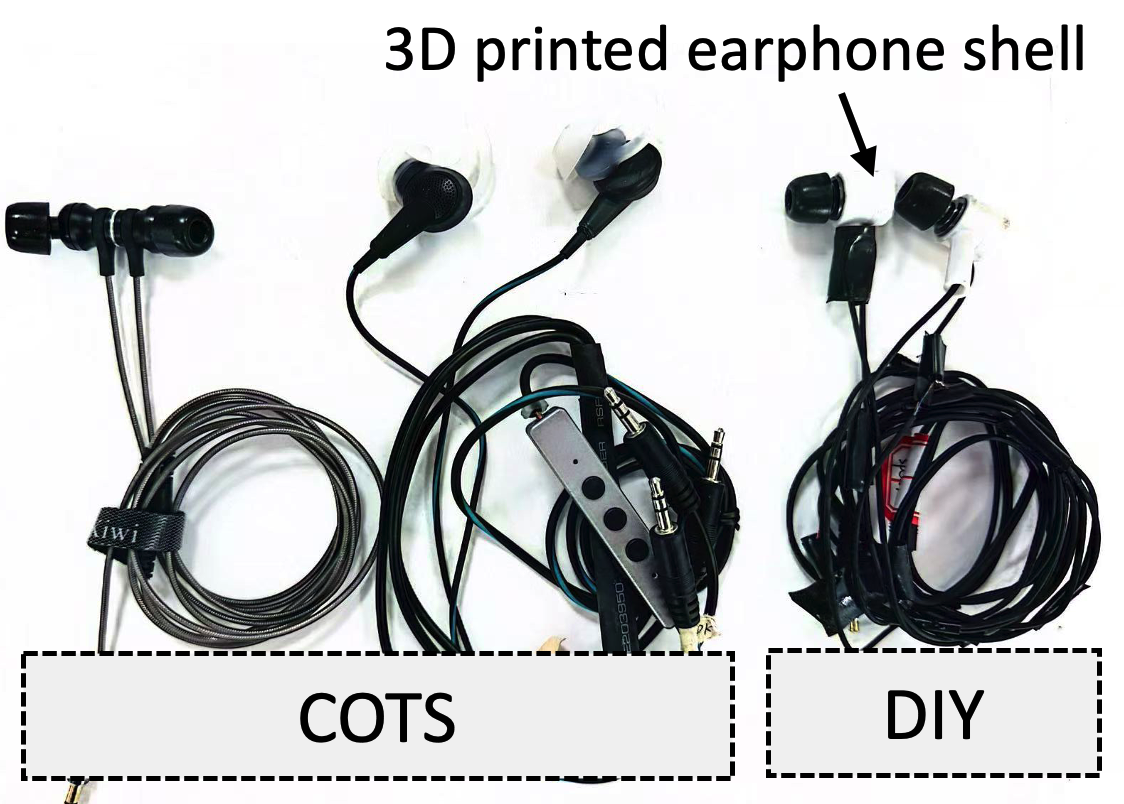} \caption{\label{fig:earphones_figure}
        \textbf{Three types of earphones.}}
    \end{minipage} \quad
    \begin{minipage}[t]{0.3\linewidth}
        \centering
        \includegraphics[width=0.95\linewidth]{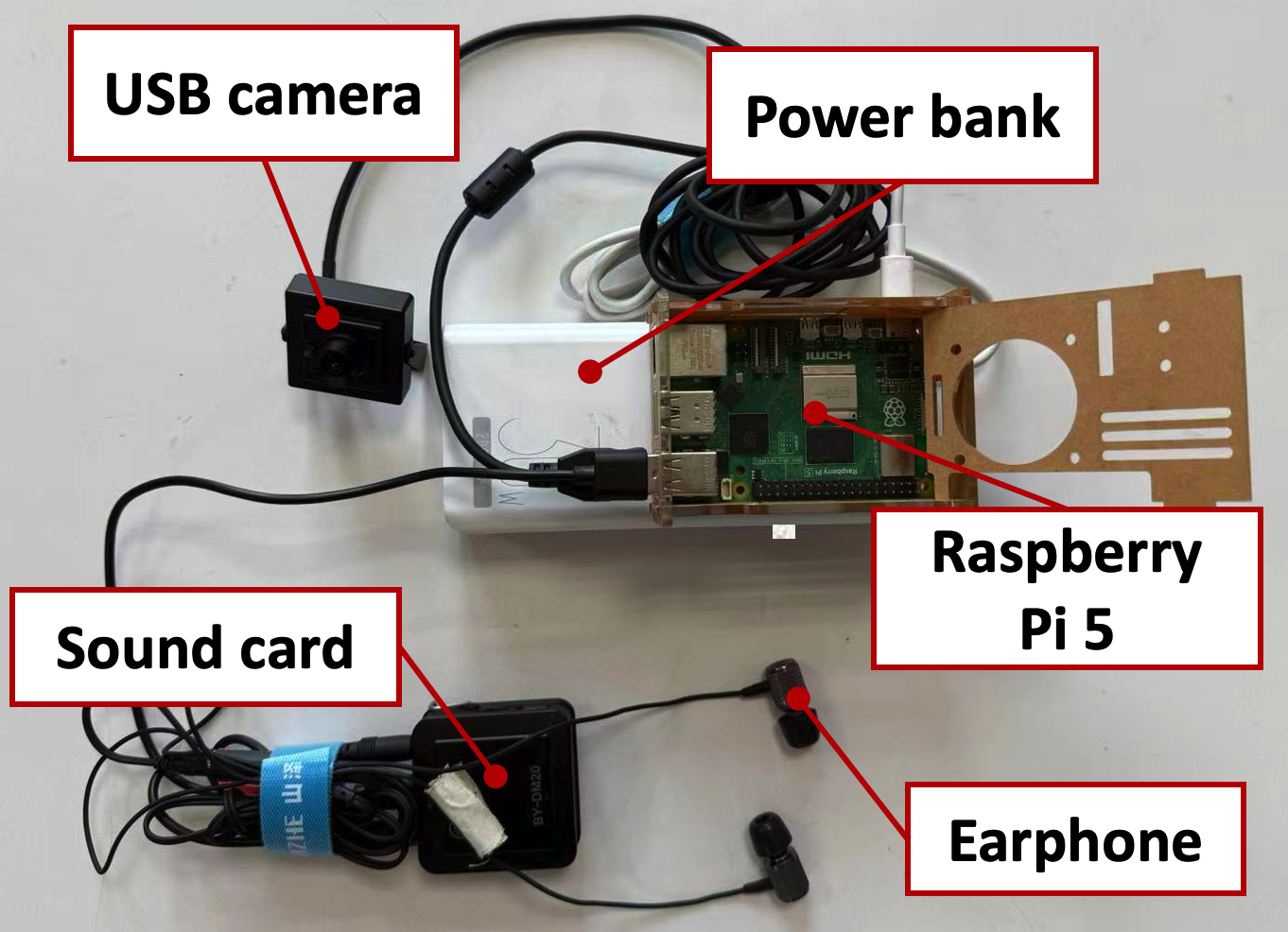} \caption{\label{fig:data_collection}
        \textbf{Data collection setup.}}
    \end{minipage} \quad
    \begin{minipage}[t]{0.3\linewidth}
        \centering
        \includegraphics[width=1\linewidth]{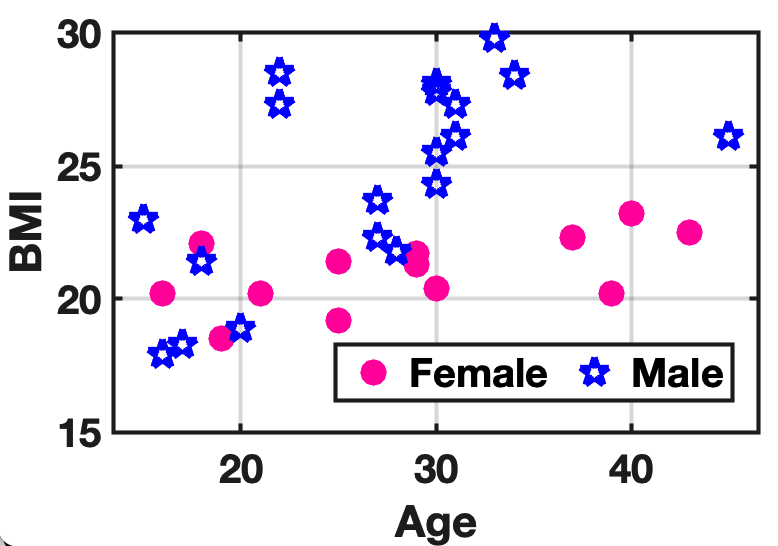} \caption{\label{fig:volunteer_list}
        \textbf{Age and BMI distribution.}}
    \end{minipage}
\end{figure*}

\textit{(\rmnum{5}) Customize Loss Design.} 
For food texture recognition networks, we leverage Binary Cross-Entropy loss functions ($L_{text}^{hard}, L_{text}^{crip}, L_{text}^{adhe}, L_{text}^{juic}, L_{type}$) to quantify the difference between the predicted outputs and the actual values. For the food type recognition network, we calculate the Cross Entropy loss function $L_{type}$ to guide model optimization and training. Considering the task balance, we combine the above loss functions as follows:
\begin{equation}
\footnotesize 
    L_{joint} = L_{type} + (L_{text}^{hard}+ L_{text}^{crip}+L_{text}^{adhe}+L_{text}^{juic})/4
\end{equation}

\section{\textbf{Evaluation Analysis}}

\subsection{\textbf{Experimental Setup}}\label{sec:hardware_setup}
\subsubsection{\textbf{Hardware Prototype}} Three types of earphones with different in-ear microphone types are involved in data collection. As shown in Fig.~\ref{fig:earphones_figure}, we use two pairs of commercial earphones and one pair of 3D printed earphones that are equipped with AS-B6027AL30-RC electret microphones. Specifically, we developed a portable data collection platform, as shown in Fig.~\ref{fig:data_collection}. The earphones are connected to a Raspberry Pi 5 device (4$\times$Cortex-A76@2.4GHz, 8GB memory) via a 3.5 mm audio adapter that supports two-channel audio stream recording. In addition, a USB camera is also connected to the Raspberry Pi 5 device. During collection sessions, audio and video streams are recorded simultaneously. 

\subsubsection{\textbf{Data Collection}} 
\begin{figure}[h]
\centering
    \begin{minipage}[t]{1\linewidth}
        \centering
        \includegraphics[width=0.9\linewidth]{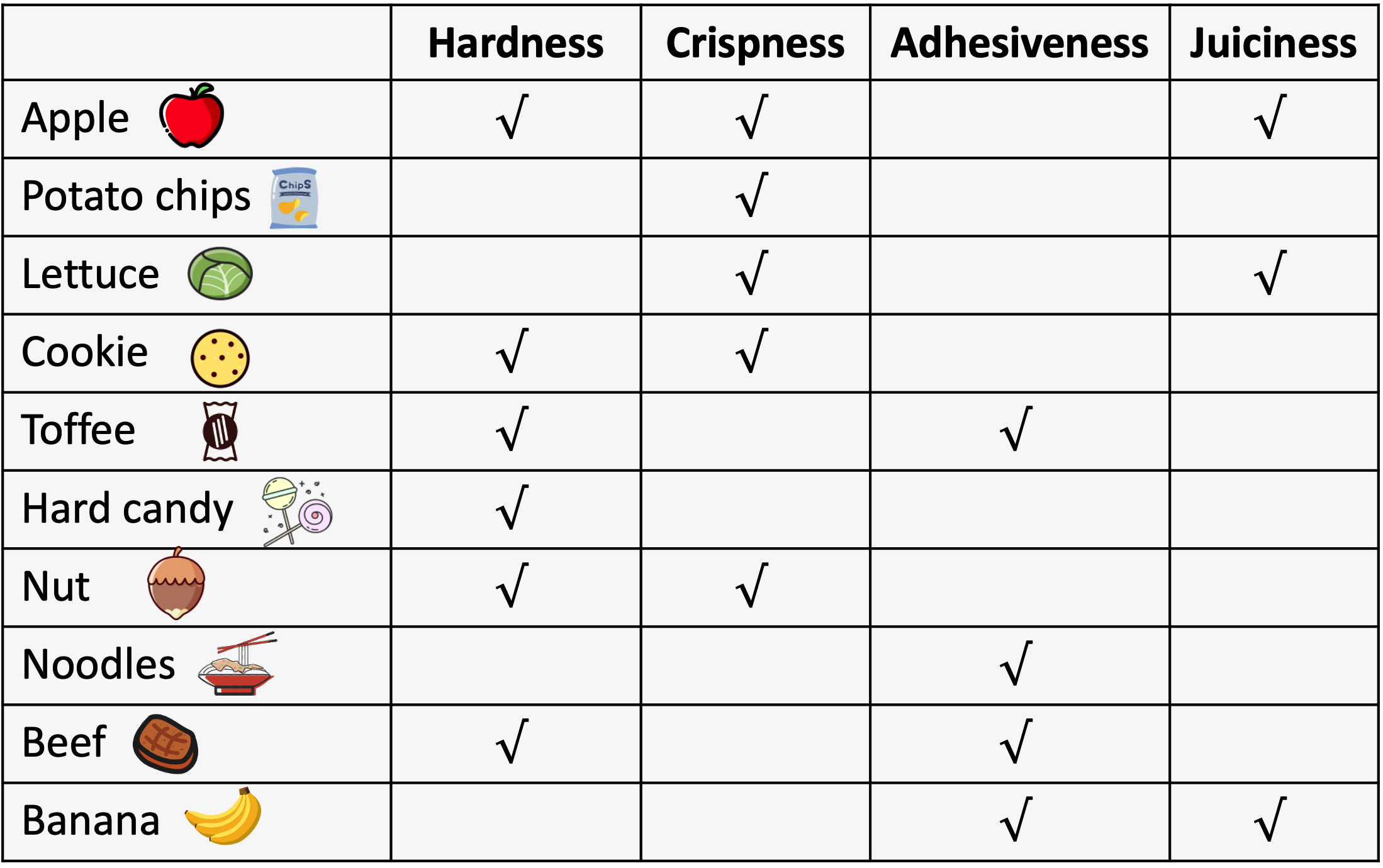} \caption{\label{fig:food_list}
        \textbf{Food types and food texture attributes.}}
    \end{minipage}
\end{figure}

We recruit 32 participants aged between 15 and 50 to wear our earphone prototype in four common dining scenarios. Among these participants, there are 18 students, 3 cafeteria staff, 2 security staff, 4 research staff, and 5 administrative staff. The personal information of participants is shown in Fig.~\ref{fig:volunteer_list}. During the data collection process, participants are free to select the type of earphones and wear them in the way that is comfortable for them. 
We select 10 common snacks, sweets, staple foods, vegetables, fruits, etc., from people's daily diet. Each participant is required to chew several food types randomly selected from Fig.~\ref{fig:food_list}. The USB camera is used to record eating moments to provide the ground truth. To synchronize the recorded audio and video streams, each participant needs to manually control the device to record before eating. All recorded audio and video streams are exported to local storage. It is noted that the data collection process has been approved by the local Institutional Review Board. In total, the dataset contains approximately 192 hours of in-ear audio signals. Specifically, approximately 136 hours of data are manually labeled as chewing signals, and the remaining data are labeled as non-chewing signals.

\subsubsection{\textbf{Data Annotation}} 
The six authors of this paper annotate the audio data based on the recorded video data, including the eating episode, food type, and food texture. 
Specifically, we follow the standardized definitions of food texture attributes reported in the food science literature~\cite{nishinari2018perception}. Based on these definitions, each food item is manually annotated according to its corresponding texture. Specifically, six participants independently assign texture labels to all food items following the same annotation guidelines. The annotations are performed independently without discussion, and the resulting inter-rater agreement has a Fleiss' $\kappa$ score of 0.88.

% In addition, we also use a camera to record eating moments to provide the ground truth. It is noted that the data collection process has been approved by the local Institutional Review Board. In total, the dataset contains approximately 192 hours of in-ear audio signals. 

% for the PC platform and the mobile platform, respectively. As for the PC platform, the earphones are connected to the laptop with a 3.5 mm jack adapter. 

% The data collection software is designed using MATLAB 2021B. As for the mobile platform, an application is developed using Android Studio Iguana (2023.2.1 Ver). 

\subsubsection{\textbf{Model Training}}
We fully shuffle the dataset and then select 70\% of the dataset for model training and the remaining 30\% of the dataset for testing. The deep learning model in Sec.~\ref{sec:food_texture_type_model} is designed with the Pytorch framework and is trained using the Adam optimizer with a learning rate of 0.0001, batch size of 24, and for 64 epochs. 
All training processes are completed on a server that is equipped with 144 CPUs of Intel(R) Xeon(R) Platinum 8352V @ 2.10GHz and 3 NVIDIA RTX A6000 graphics cards.

\begin{table*}[t]
\caption{\textbf{Performance evaluation of eating episode detection among different scenarios.}}
\label{tab:eating_episode_detection} 
% \normalsize
\renewcommand{\arraystretch}{1.2}
\resizebox{0.99\linewidth}{!}{
    \begin{tabular}{lcccccccccccc}
    \toprule
    
    \multirow{2}{*}{} & \multicolumn{3}{c}{ \textbf{Scenarios 1}}      & \multicolumn{3}{c}{ \textbf{Scenarios 2}}       & \multicolumn{3}{c}{ \textbf{Scenarios 3}}       & \multicolumn{3}{c}{ \textbf{Scenarios 4}} \\ 
                      & \textbf{Precision} & \textbf{Recall} & \textbf{F1-score} & \textbf{Precision} & \textbf{Recall} & \textbf{F1-score} & \textbf{Precision} & \textbf{Recall} & \textbf{F1-score} & \textbf{Precision}     & \textbf{Recall}     & \textbf{F1-score}    \\ \midrule
    \textbf{1D CNN-based~\cite{papapanagiotou2017chewing}} & 86.66\%  & 81.12\%  & 83.80\%  & 86.18\%  & 83.31\%  & 84.72\%   & 85.41\%  & 82.06\%  & 83.70\% &  84.69\%  &  80.05\% & 82.30\% \\
    
    \textbf{SSL-based~\cite{papapanagiotou2021self}}  & 84.91\%  & 86.27\%  & 85.58\% & 85.02\%  & 82.86\%  & 83.93\%   & 85.58\%  & 83.92\%  & 84.74\% &  84.07\%  &  81.22\% & 82.62\%   \\

    \rowcolor[HTML]{E6E6FA} \textbf{\systemName} & 97.19\% & 95.76\% & 96.47\%    & 97.31\% & 94.85\% &   96.06\%   & 96.75\% & 94.91\% &  95.82\%   & 96.21\% &95.73\% & 95.97\%    \\

    \textbf{- w/o Event-level} & 92.23\% & 90.51\% & 91.36\%   & 92.54\% & 90.67\% & 91.60\%  & 93.42\% & 91.46\% &   92.43\%     & 90.93\% & 88.16\% & 89.52\%      \\

    \textbf{- w/o Frame-level} & 88.57\% & 86.64\% & 87.59\%    & 89.83\% & 87.31\% & 88.55\%  & 87.52\% & 86.21\% &    86.86\%    & 88.53\% & 86.20\% & 87.35\%    \\

    \bottomrule
    \end{tabular}
    }
\end{table*}

\subsection{\textbf{Performance of Eating Episode Detection}}

Audio signals are collected in four scenarios: a single-person home environment, a multi-person home environment, a single-person public canteen, and a multi-person public canteen. The averaged noise pressure levels of the above scenarios are 36.45 dB, 43.76 dB, 54.61 dB, and 63.79 dB, respectively. 
A detected eating episode is considered a true positive if its temporal overlap with a ground-truth episode exceeds 75\%. This setup is similar to the metrics in previous solutions~\cite{papapanagiotou2017chewing, papapanagiotou2021self}. Episodes without matched ground-truth annotations are counted as false positives, while missed ground-truth episodes are regarded as false negatives.

\subsubsection{\textbf{Baseline Comparison.}} 
We incorporate two prior significant solutions~\cite{papapanagiotou2017chewing, papapanagiotou2021self} into experiments as baselines. Specifically, these two baselines leverage 1D convolutional networks (\ie, 1D CNN-based) and 1D convolutional networks combined with self-supervised feature learning (\ie, SSL-based), respectively, to achieve chewing event detection. Since the baselines directly process sound waveforms, we split the entire sound waveform into multiple two-second windows for baseline training and testing. Tab.~\ref{tab:eating_episode_detection} shows the precision, recall, and F1-score in terms of eating episode detection under four different scenarios. 
Overall, \systemName~consistently achieves the best performance across all evaluation metrics, significantly outperforming both the CNN-based and SSL-based baselines. For instance, compared with the CNN-based approach, \systemName~improves the F1-score by 12.67\%, 11.34\%, 12.12\%, and 13.67\% in the four scenarios. Baselines mainly directly extract chewing-related features from in-ear sound waveforms through 1D convolutional networks. 
Then, they detect chewing windows and aggregate them into eating episodes using heuristic rules, which often leads to fragmented detections and inaccurate episode boundaries. However, \systemName~explicitly models the characteristics of chewing activities and extracts representative temporal and spectral features from chewing-induced in-ear sounds. By jointly exploiting event-level and frame-level detection, \systemName~is able to achieve more accurate episode detection. 
Furthermore, the bottom two rows of Tab.~\ref{tab:eating_episode_detection} further validate the effectiveness of the proposed event-level and frame-level collaborative episode parsing strategy in Sec.~\ref{sec:mechanics_aware_diet_episode_parsing}. Removing the event-level candidate detection module leads to a noticeable performance degradation, with the average F1-score decreasing from 96.08\% to 91.23\% among four scenarios. Similarly, removing the frame-level chewing identification refinement module causes a more substantial performance drop, reducing the average F1-score to 87.58\%. 

% This result demonstrates that coarse-grained episode localization effectively suppresses irrelevant audio segments and provides reliable temporal priors for subsequent parsing.  This observation highlights the importance of fine-grained chewing analysis in distinguishing eating activities from other acoustically similar oral behaviors. 

% Compared with the single classifier, the evaluation metrics of the ensemble classifier are significantly improved by 6.21\%, 3.81, and 5.64\%, in terms of the averaged precision, averaged recall, and the averaged F1-score. Additionally, the single classifier still achieves over 90\% in terms of evaluation metrics, indicating the effectiveness of event-related
% feature extraction in Sec.~\ref{sec:event_feature_extraction}. 

\begin{figure}[t]
\centering
    \begin{minipage}[t]{1\linewidth}
        \centering
        \includegraphics[width=0.9\linewidth]{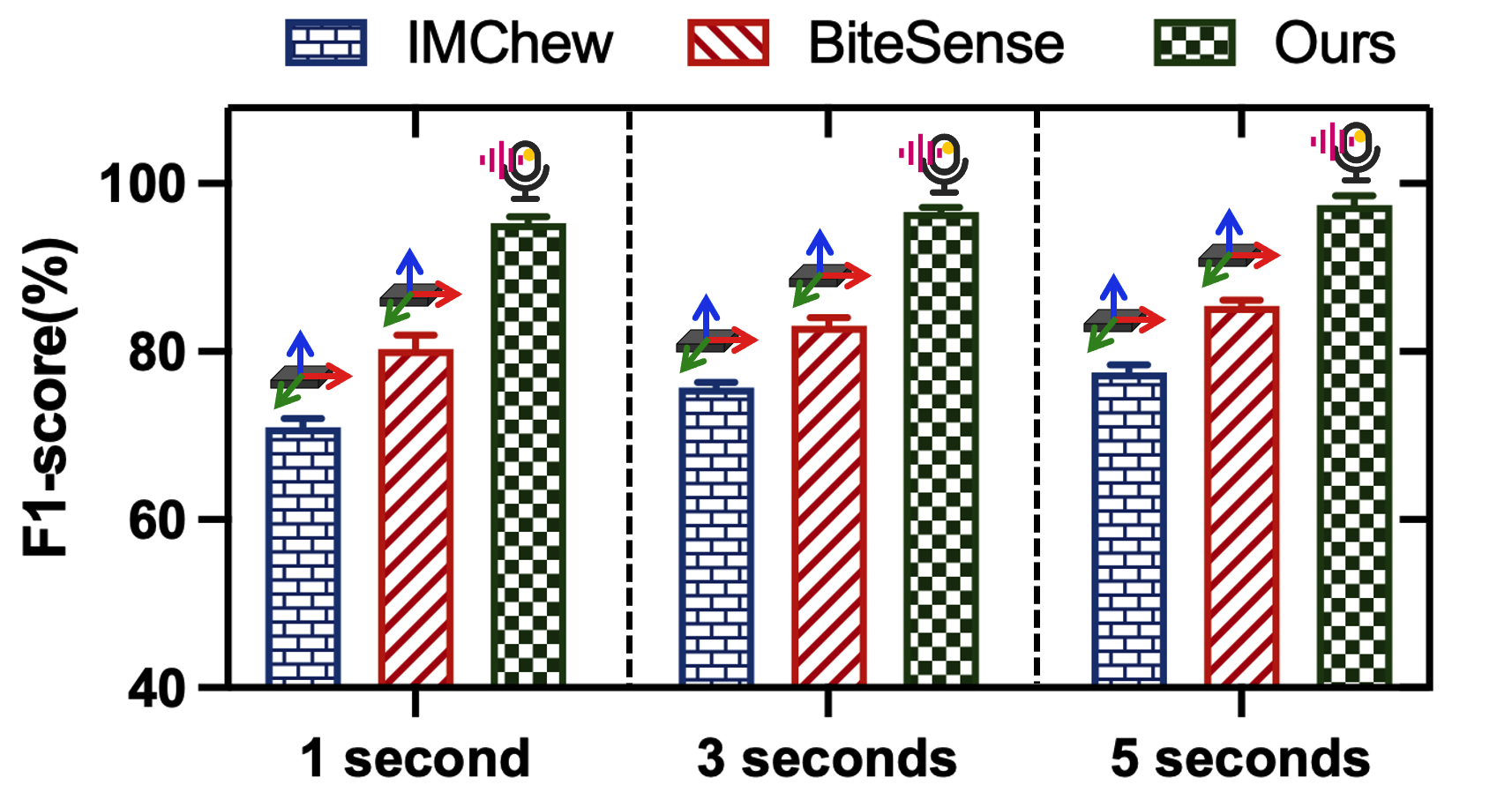} \caption{\label{fig:baseline_modality_comparison}
        \textbf{Modality comparison of eating episode detection.}}
    \end{minipage}
    \vspace{-0.5cm}
\end{figure}

\subsubsection{\textbf{Modality Comparison}} In addition to the above baselines that are also based on chewing-induced in-ear sounds, we conduct a detailed comparison between our solution and state-of-the-art solutions of other modalities in terms of eating episode detection. In particular, BiteSense~\cite{chugh2025bitesense} and IMChew~\cite{ketmalasiri2024imchew} are IMU-based earable eating activity recognition solutions that capture jaw movements during chewing. In our work, we set the IMU sensor (MPU-6050) on earphones with a sampling rate of 100 Hz to collect the inertial data. The F1-score is shown in Fig.~\ref{fig:baseline_modality_comparison}. As for 1-second input signals, our solution can achieve 23\% and 16\% improvement of F1-score compared to prevailing solutions. This is because in-ear sound has a wider bandwidth than inertial data and can therefore contain more unique information. In addition, even slight body movements also interfere with IMU-based eating activity detection. Yet, \systemName~can remove the negative impact of body movements effectively in Sec.~\ref{sec:signal_enhancement}. Lastly, the results in Fig.~\ref{fig:baseline_modality_comparison} reveal that as the input signal duration increases, \systemName~can yield better detection performance. That is because long-duration audio signals contain more temporal and spectral information.

\subsubsection{\textbf{Meal-level Evaluation}} Furthermore, we conduct the meal-level evaluation to validate the effectiveness of \systemName. A meal is defined as a continuous eating period recorded during breakfast, lunch, or dinner. A meal is considered correctly detected if at least one detected eating episode overlaps with the corresponding ground-truth meal interval for more than 10 seconds. A missed meal is counted when no detected eating episode is matched with a ground-truth meal. A false meal detection is counted when a detected eating episode does not overlap with any ground-truth meal interval. The precision, recall, and F1-score of meal-level detection are 99.23\%, 97.67\%, and 98.44\%, respectively. The above results indicate that \systemName~can enable accurate detection from fine-grained to coarse-grained levels.

% \begin{table}[t]
% \centering
% \caption{\label{tab:comparison_other_solutions} \textbf{Baseline comparison in terms of eating episode detection with F1-score.}}
%     \resizebox{0.9\linewidth}{!}{
%         \begin{tabular}{cccc}
%             \toprule
%              & \textbf{1 second} & \textbf{3 seconds} & \textbf{5 seconds}   \\ \hline
%             \textbf{IMChew} &73.15\% &  75.24\% &  74.33\%    \\
%             \textbf{BiteSense} &80.65\% &  83.79\% &  84.41\%   \\
%             \rowcolor[HTML]{E6E6FA} \textbf{\systemName} & 96.17\% & 97.04\% & 98.93\%   \\
%             \bottomrule
%         \end{tabular}
%     }
% \end{table}

\begin{figure}[t]
\centering
    \begin{minipage}[t]{1\linewidth}
        \centering
        \includegraphics[width=1\linewidth]{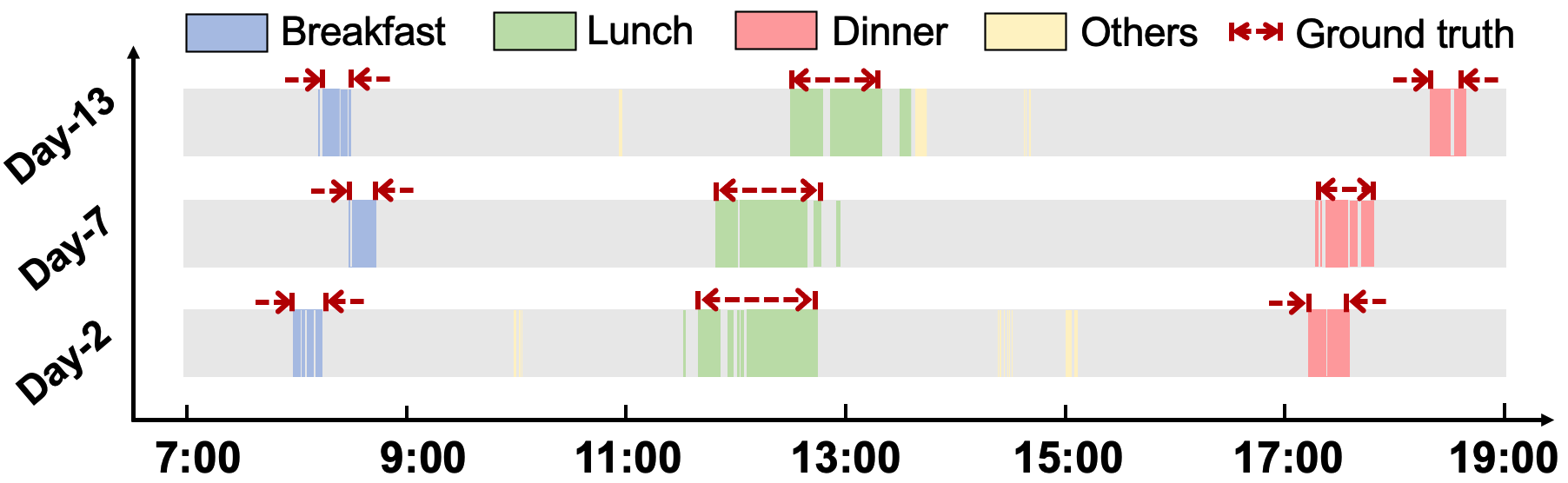} \caption{\label{fig:long_term_study}
        \textbf{Long-term evaluation of eating episode detection.}}
    \end{minipage}
\end{figure}

\subsection{\textbf{Longitudinal Study in The Wild}}
To assess the performance of long-term eating episode detection, we track a participant over a period of time. We categorize breakfast, lunch, and dinner based on eating moments. Food consumed between 7:00 and 9:00 is classified as breakfast, while that eaten between 11:00 and 13:30 is categorized as lunch, and food ingested between 17:00 and 19:00 is designated as dinner. Fig.~\ref{fig:long_term_study} illustrates the detected eating episodes and their corresponding ground-truth annotations over three days. Although slight differences exist between the detected episode boundaries and the ground truth, \systemName~successfully captures the major eating events throughout the day and accurately identifies their temporal occurrence. It indicates that \systemName~can provide diet logging for users in daily life and make up for the shortcomings of manual recording.

% Although estimated eating moments and ground truth have not yet achieved a complete match, \systemName~can still identify the eating time relatively accurately and effectively track eating activities along the time dimension. 

\begin{table*}[]
\caption{\textbf{Micro evaluation of food texture recognition and ablation study.}}
\label{tab:micro_evaluation_food_texture} 
% \renewcommand{\arraystretch}{1.3}
% \normalsize
\renewcommand{\arraystretch}{1.2}
\resizebox{0.99\linewidth}{!}{
    \begin{tabular}{ccccccccccccc}
    \toprule
    
    \multirow{2}{*}{} & \multicolumn{3}{c}{ \textbf{Scenario 1}}      & \multicolumn{3}{c}{ \textbf{Scenario 2}}       & \multicolumn{3}{c}{ \textbf{Scenario 3}}       & \multicolumn{3}{c}{ \textbf{Scenario 4}} \\ 
                      & \textbf{Precision} & \textbf{Recall} & \textbf{F1-score} & \textbf{Precision} & \textbf{Recall} & \textbf{F1-score} & \textbf{Precision} & \textbf{Recall} & \textbf{F1-score} & \textbf{Precision}     & \textbf{Recall}     & \textbf{F1-score}    \\ \midrule
    % \textbf{CNN}        & 1         & 1      &          & 1         &        &          &           &        &          &               &            &             \\
    % \textbf{BiLSTM}        &           & 1      &          &           &        &          &           &        &          &               &            &             \\ \hdashline
    \rowcolor[HTML]{E6E6FA}  \textbf{CWT-\systemName}  & 77.79\%   & 70.66\% & 74.22\% & 76.45\% & 73.68\% & 75.07\%   & 77.12\% & 74.10\%  & 75.61\%   & 74.53\%   & 75.26\%   & 74.89\%   \\
    \textbf{STFT-\systemName}  & 70.19\%   & 69.34\% & 69.77\% & 73.12\%  & 71.52\%  & 72.32\%   & 73.34\%  & 71.86\%  & 72.60\% & 71.54\%   & 69.27\%   & 70.41\%   \\
    \textbf{- w/o Attention}   &  66.04\%  & 62.31\% & 64.17\% & 65.37\% &  61.23\%   &  63.30\%  & 66.67\%  &  64.39\% &  65.53\%  &  65.23\%  & 63.72\%   &  64.48\%  \\
    \textbf{- w/o Temporal}  & 65.30\% & 60.60\% &  62.95\% & 65.78\%   & 60.71\%    & 63.25\%   
    & 66.33\%  & 63.13\%   & 64.73\%   
    & 65.12\%  &  61.29\%  & 63.21\%  \\
    \textbf{- w/o Spectral} & 35.31\%   & 32.12\%  & 33.72\% 
    & 36.67\%  & 34.56\%  & 35.62\%
    & 35.75\%  & 32.02\%  & 33.89\%   
    & 35.98\%  & 32.97\%  & 33.47\%    \\ \bottomrule
    \end{tabular}
    }
\end{table*}

% \begin{figure}[b]
%     \centering
%     \includegraphics[width=1\linewidth]{./Figures/event_detection_accuracy.png}
%     \caption{\textbf{Event detection accuracy with different classifiers.}}
%     \label{fig:event_detection_accuracy}
% \end{figure}

\subsection{\textbf{Performance of Food Texture Recognition}}

\begin{figure}[t]
\centering
    \begin{minipage}[t]{1\linewidth}
        \centering
        \includegraphics[width=0.95\linewidth]{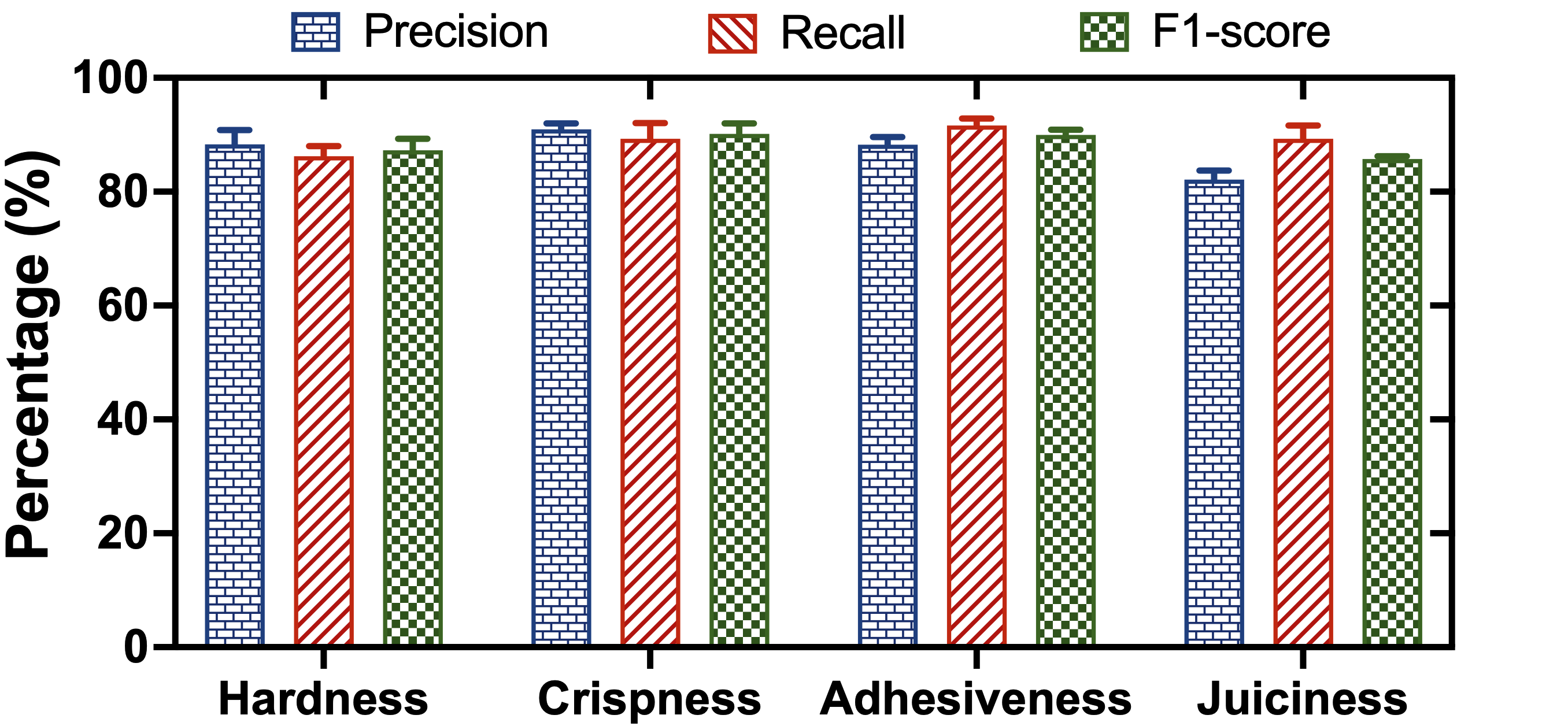} \caption{\label{fig:macro_evaluation}
        \textbf{Macro evaluation of food texture recognition.}}
    \end{minipage}
\end{figure}

\subsubsection{\textbf{Macro Metrics Evaluation}}

Firstly, we evaluate the food texture recognition performance of \systemName~by treating each food texture attribute independently and calculating the averaged precision, recall, and F1-score. Fig.~\ref{fig:macro_evaluation} demonstrates the macro evaluation in terms of food texture recognition. The averaged F1-scores across in terms of hardness, crispness, adhesiveness, and juiciness are 87.35\%, 90.17\%, 90.32\%, and 85.72\%, respectively, indicating that \systemName~can effectively recognize food textures. We can notice that \systemName~experiences a certain performance degradation when identifying the food texture of juiciness. Unlike hardness, crispness, and adhesiveness, which directly affect chewing mechanics and generate distinctive acoustic responses, juiciness is primarily associated with moisture release during mastication, exhibiting weaker acoustic discriminability. Furthermore, among foods involved in the experiments, juiciness frequently coexists with other texture attributes, making it easy to confuse with other texture attributes.

% Furthermore, in the foods involved in the experiment, juiciness often coexists with other texture properties, making it easy to confuse with other texture properties, resulting in lower accuracy.

% accurate classification of this texture attribute more challenging.

% That is because juicy foods are more likely to be confused with other texture attributes, resulting in low precision. 

\subsubsection{\textbf{Micro Metrics Evaluation}}
Furthermore, we also conduct a micro evaluation to validate the effectiveness of \systemName~in terms of food texture recognition. Different from the macro evaluation, the micro evaluation considers all labels of the sample as a whole. The predicted values need to match all labels to be considered a positive prediction. 
As shown in Tab.~\ref{tab:micro_evaluation_food_texture}, the averaged precision, the averaged recall, and the averaged F1-score are 76.47$\pm$1.41\%, 73.43$\pm$1.96\%, and 74.95$\pm$0.57\%, across four scenarios, respectively. Compared with the single-texture classification task, the performance of multi-texture classification degrades. That is because the characteristics of multiple textures may overlap, making it difficult to accurately identify the boundaries. For instance, the hardness attribute and the juiciness attribute share a similar spectral structure among low-frequency components.

% multi-texture classification involves distinguishing between multiple textures within the same image or scene, which increases the complexity of the task. 

% The presence of multiple textures can create confusion for classification algorithms, as they may struggle to accurately identify the boundaries and features of each texture due to overlapping characteristics.
% % In addition, leveraging, indicating that CWT can be 

\subsubsection{\textbf{Ablation Study}}
To explore the contribution of each module in food texture recognition, we conduct an ablation study. As shown in Tab.~\ref{tab:micro_evaluation_food_texture}, ``w/o Attention", ``w/o Temporal", and ``w/o Spectral" represent \systemName~works without the assistance of the attention module, temporal feature extraction module, and spectral feature extraction module, respectively. From the comparison of the results, we can clearly find that each module in the food texture recognition model is essential. Especially, without the contribution of the spectral feature extraction module, the performance of \systemName~drops by nearly 40\% in terms of F1-score. That is because the frequency domain information contains richer fine-grained information than the time domain, leading to a greater contribution to the overall performance of food texture recognition.
Regarding the effectiveness of CWT representations, \systemName~with CWT representation outperforms \systemName~with STFT representation by nearly 5\%. This is because the CWT representation shows great capability in high-resolution time-frequency structure representation and thus can offer fine-grained information to food texture recognition.

\begin{figure}[t]
\centering
    \begin{minipage}[t]{1\linewidth}
        \centering
        \includegraphics[width=0.98\linewidth]{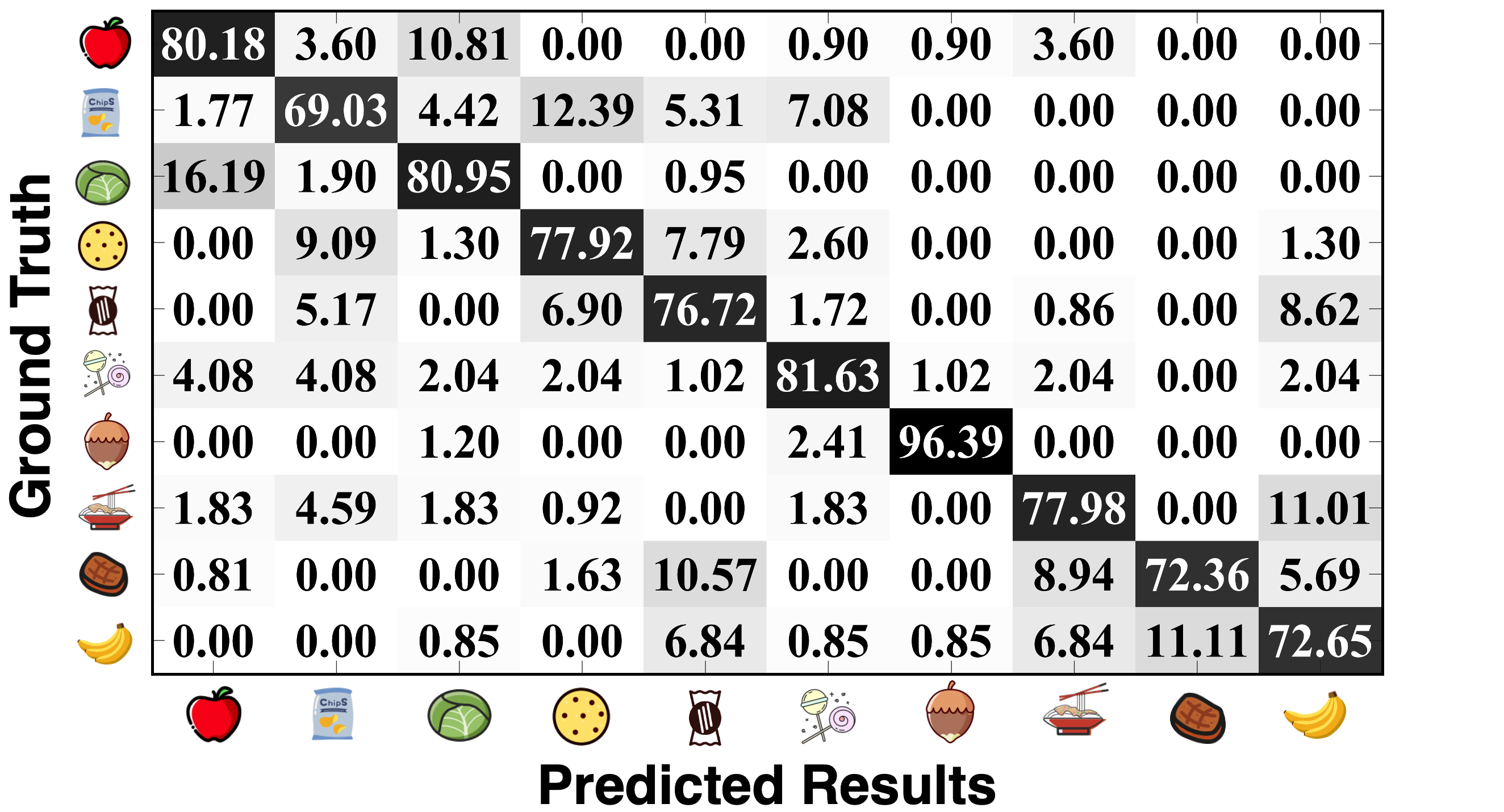} \caption{\label{fig:confusion_matrics}
        \textbf{Confusion matrix of food type recognition.}}
    \end{minipage}
\end{figure}

% In addition, \systemName~with the CWT as input significantly outperforms that with the STFT as input.
% \begin{itemize}
%     \item ``w/o Attention":
%     \item ``w/o Temporal":
%     \item ``w/o Spectral":
% \end{itemize}

\subsection{\textbf{Performance of Food Type Recognition}} \label{sec:performance_food_type}
The confusion matrix of food type recognition is shown in Fig.~\ref{fig:confusion_matrics}, where the element of $(i, j)$ represents the percentage of samples with the ground truth of the $i$-th class but are classified as the $j$-th class. \systemName~can achieve an averaged accuracy of 78\% across 10 food types. Specifically, for food types with unique textures such as nuts, apples, and lettuce, \systemName~can effectively distinguish them with high accuracy. However, if several food types share similar texture attributes, the recognition accuracy degrades. For instance, while potato chips and cookies exhibit similar crispness, cookies possess a higher hardness attribute than potato chips. Likewise, banana is occasionally recognized as beef or noodles due to the shared adhesiveness-related chewing patterns. These results and key observations suggest that food texture plays a dominant role in shaping chewing-induced in-ear sounds and directly affects food type recognition performance.

% which limits the ability of \systemName~to effectively differentiate between them. 

% \textcolor{red}{Why}

% \systemName~can achieve an averaged accuracy of 78\% across 10 food types. We can clearly observe that the recognition accuracy degrades when it comes to distinguishing foods with similar texture attributes, such as chips and cookies.

% \begin{figure*}[t]
% \centering
%     \begin{minipage}[t]{0.48\linewidth}
%         \centering
%         \includegraphics[width=0.85\linewidth]{./Figures/impact_user_group_2.png} \caption{\label{fig:impact_user_group}
%         \textbf{Impact of user group on food type recognition.}}
%     \end{minipage} \hfill 
%     \begin{minipage}[t]{0.48\linewidth}
%         \centering
%         \includegraphics[width=0.95\linewidth]{./Figures/impact_hardware_type.png} \caption{\label{fig:impact_hardware_type}
%         \textbf{Impact of hardware on food type recognition.}}
%     \end{minipage} 

% \end{figure*}

\begin{figure}[t]
\centering
    \begin{minipage}[t]{1\linewidth}
        \centering
        \includegraphics[width=0.8\linewidth]{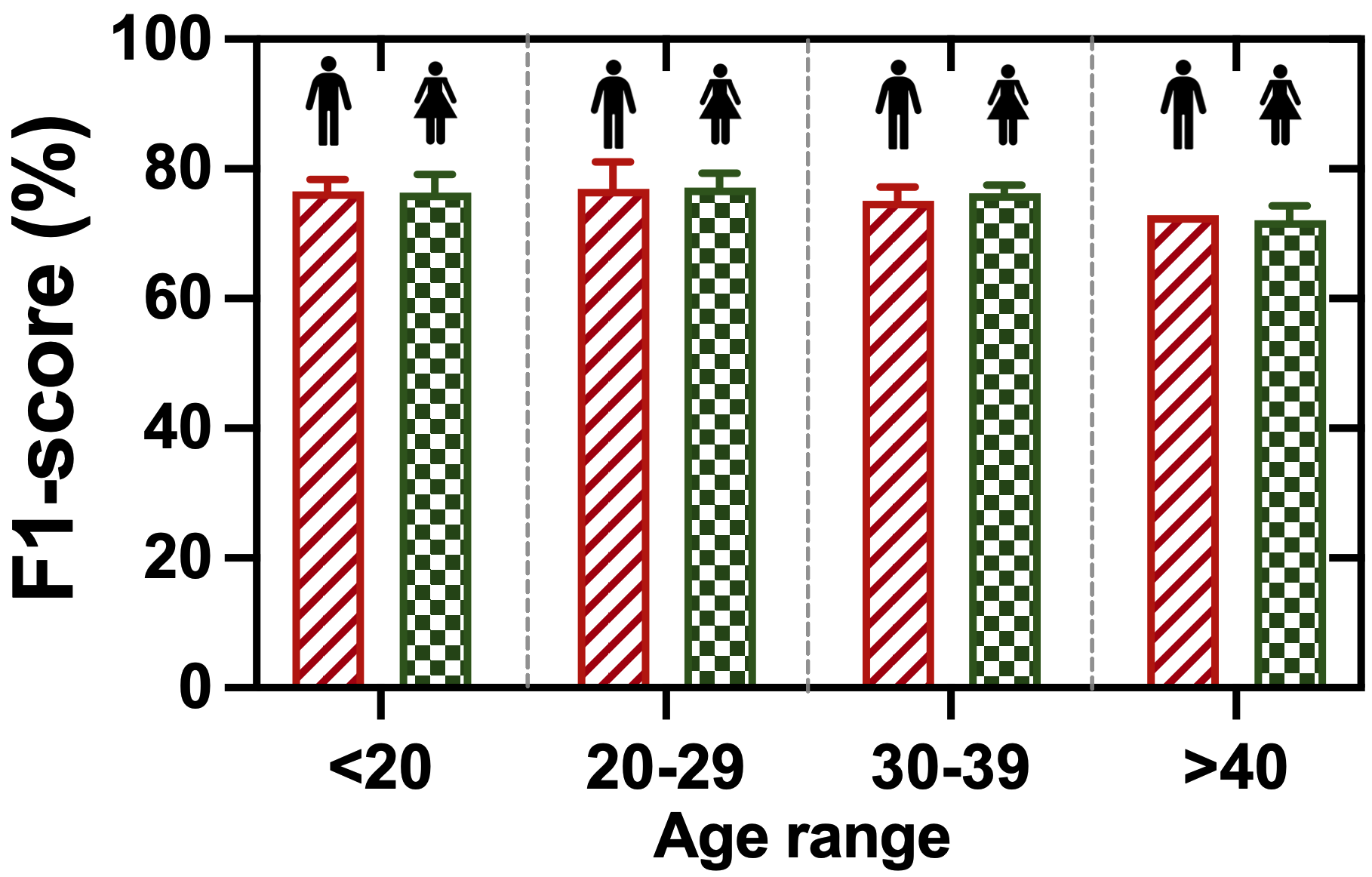} \caption{\label{fig:impact_user_group}
        \textbf{Impact of user group on food type recognition.}}
    \end{minipage} \hfill 
    % \begin{minipage}[t]{0.48\linewidth}
    %     \centering
    %     \includegraphics[width=0.95\linewidth]{./Figures/impact_hardware_type.png} \caption{\label{fig:impact_hardware_type}
    %     \textbf{Impact of hardware on food type recognition.}}
    % \end{minipage} 
\end{figure}

\begin{figure}[t]
\centering
    \begin{minipage}[t]{1\linewidth}
        \centering
        \includegraphics[width=0.85\linewidth]{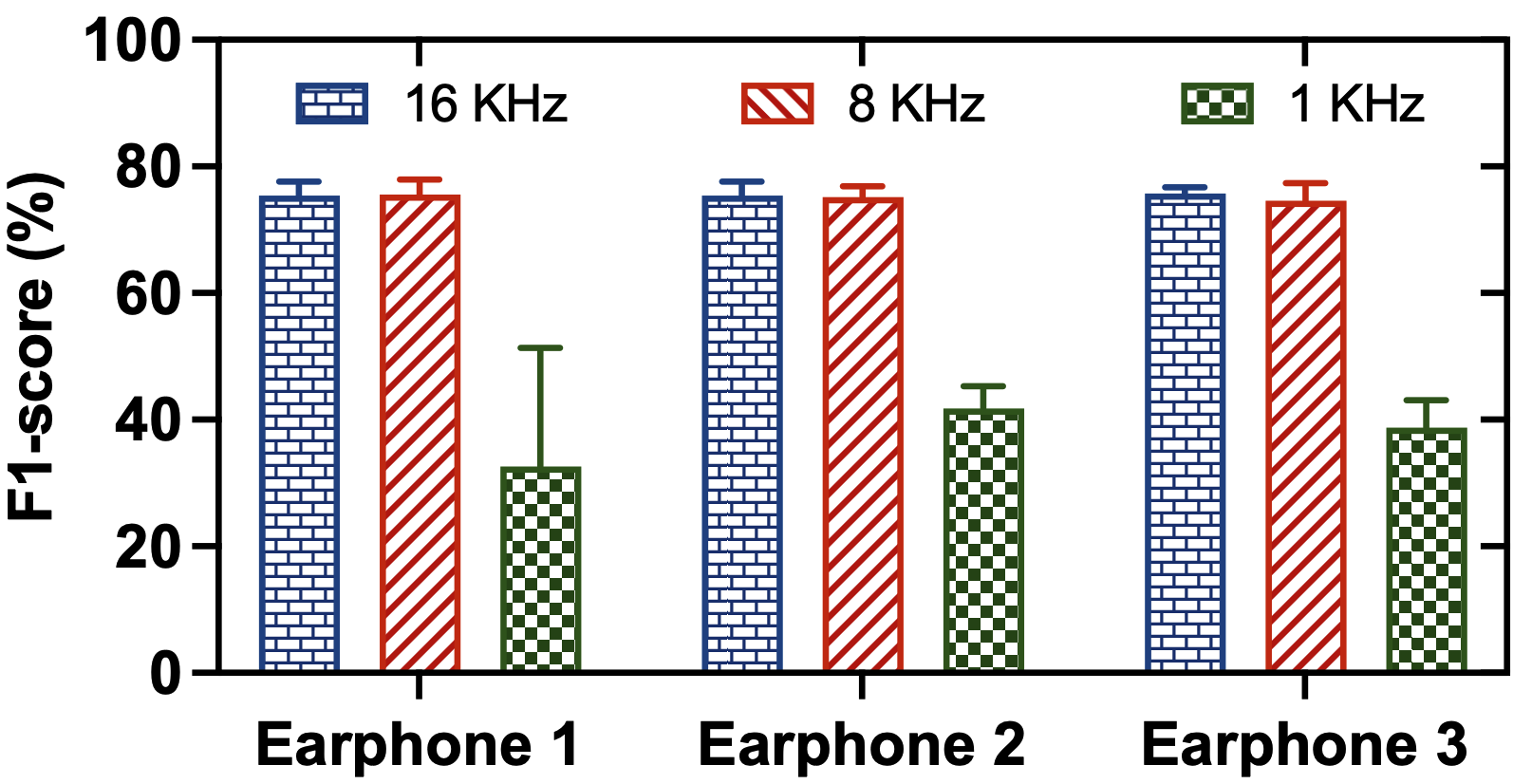} \caption{\label{fig:impact_hardware_type}
        \textbf{Impact of hardware on food type recognition.}}
    \end{minipage} 
\end{figure}

\begin{table}[t]
\centering
\caption{\label{tab:cross_subject_validation} \textbf{F1-score of the cross-subject evaluation}.}
\renewcommand{\arraystretch}{1.2}
\resizebox{1\linewidth}{!}{
\begin{tabular}{cccc}
\toprule
    & \multicolumn{1}{c}{\begin{tabular}[c]{@{}c@{}}\textbf{Eating Episode} \\ \textbf{Detection}\end{tabular}} & \multicolumn{1}{c}{\begin{tabular}[c]{@{}c@{}}\textbf{Micro Texture} \\ \textbf{Recognition}\end{tabular}} &   \multicolumn{1}{c}{\begin{tabular}[c]{@{}c@{}}\textbf{Food Type} \\ \textbf{Recognition}\end{tabular}} \\ \hline

\textbf{Scenario 1} & 96.32\% ($\pm$ 1.51) & 75.48\% ($\pm$ 2.72) & 79.87\% ($\pm$ 1.71 ) \\

\textbf{Scenario 2} & 96.68\% ($\pm$ 0.76) & 73.85\% ($\pm$ 3.64) & 77.63\% ($\pm$ 3.48 ) \\

\textbf{Scenario 3} & 95.93\% ($\pm$ 1.27) & 74.71\% ($\pm$ 3.99) & 78.33\% ($\pm$ 4.44 ) \\

\textbf{Scenario 4} & 95.79\% ($\pm$ 0.53) & 72.04\% ($\pm$ 2.50) & 77.46\% ($\pm$ 3.27 ) \\

\bottomrule
\end{tabular}
}
\end{table}

\subsection{\textbf{Cross-Subject Generalization Evaluation}}
To rigorously evaluate the cross-subject generalization capability of \systemName, we further conducted a subject-independent evaluation using leave-one-group-out cross-validation. Specifically, we randomly divide the 32 participants into 8 groups, with 4 participants in each group. In each fold, data from one user group is reserved exclusively for testing, while data from the remaining user groups is used for training. Tab.~\ref{tab:cross_subject_validation}~presents the cross-subject evaluation results. Among the eating episode detection, micro texture recognition, and food type recognition tasks, cross-subject evaluation results only show a slight difference from the above subject-dependent evaluation results. That is because \systemName~is designed to extract discriminative and generalized features from chewing-induced sounds for eating detection and food recognition. The above comparable evaluation results under this rigorous evaluation protocol confirm the robustness and cross-subject generalization capability of \systemName.

\subsection{\textbf{Robustness Study}}
% \subsubsection{\textbf{Impact of External Noise}}

% As demonstrated in Tab.~\ref{tab:eating_episode_detection} and Tab.~\ref{tab:micro_evaluation_food_texture}, four scenarios with different noise levels (Scenario-1: 36 dB, Scenario-2: 43 dB, Scenario-3: 55 dB, and Scenario-4: 64 dB) are involved in the robustness evaluation. \systemName~achieve comparable performance across different scenarios in terms of eating episode detection and food texture recognition, demonstrating a strong noise resistance capability. It also indicates that the designed denoising approach in Sec.~\ref{sec:signal_enhancement} can effectively remove external noises. In addition, in-ear chewing sounds propagate along the bone conduction channel, which is non-sensitive to ambient noise. 

% Such strong noise resistance capability mainly benefits from two aspects. First, \systemName~capture in-ear sounds propagating along the bone conduction channel that is non-sensitive to ambient noise. Second, 

\subsubsection{\textbf{Impact of Motion Artifacts}}
In most cases, individuals do not engage in significant physical activity (like walking and running) while eating. Indeed, light-amplitude body movements involving the head and hands are frequently observed during eating. In the above experiments, we did not restrict any body movements of participants. Experimental results in Tab.~\ref{tab:eating_episode_detection} and Tab.~\ref{tab:micro_evaluation_food_texture} indicate that our system demonstrates robustness against motion artifacts. That is because the frequency components of in-ear sounds induced by most body movements are distributed below 100 Hz, exhibiting a distinguished pattern with in-ear chewing sounds.

\subsubsection{\textbf{Impact of User Group}}
Since the eating patterns and body geometry vary from person to person, individual variability interferes with the performance of \systemName. 
In total, there are 32 participants with different age ranges and BMI profiles involved in our evaluation. Based on their age ranges and genders, we evaluate the food type recognition performance to explore the impact of the user group. Fig.~\ref{fig:impact_user_group} shows the F1-score of food type recognition. Small variances between recognition accuracy of different user groups indicate that \systemName~yields generalization among different user groups.

% As shown in Fig.~\ref{fig:impact_user_group}, we can observe that \systemName~yields generalization among different user groups.

\subsubsection{\textbf{Impact of Hardware Type}}
In the data collection, three types of earphones with various microphone types (equipped with different types of microphones) are leveraged to record in-ear audio signals. Fig.~\ref{fig:impact_hardware_type} demonstrates food type recognition performance of different earphone types. The performance differences between the different earphone types are small, indicating that \systemName~yields a generalized performance among different hardware types. Additionally, a decrease in the audio sampling rate results in a gradual decline in recognition performance. That is because the significant components of chewing-induced in-ear audio signals are mainly distributed below 1 KHz. Thus, downsampling raw audio signals to 8 KHz can balance computing efficiency and recognition accuracy.

% \begin{figure}[t]
% \centering
%     \begin{minipage}[t]{1\linewidth}
%         \centering
%         \includegraphics[width=1\linewidth]{./Figures/music_impact_1.png} \caption{\label{fig:music_impact}
%         \textbf{CWT representation of music playback and target in-ear chewing sounds.}}
%     \end{minipage}
% \end{figure} 

\begin{figure}[t]
\centering
    \begin{minipage}[t]{1\linewidth}
        \subfloat[\textbf{F1-score of food texture recognition.}]{\includegraphics[width=0.47\textwidth]{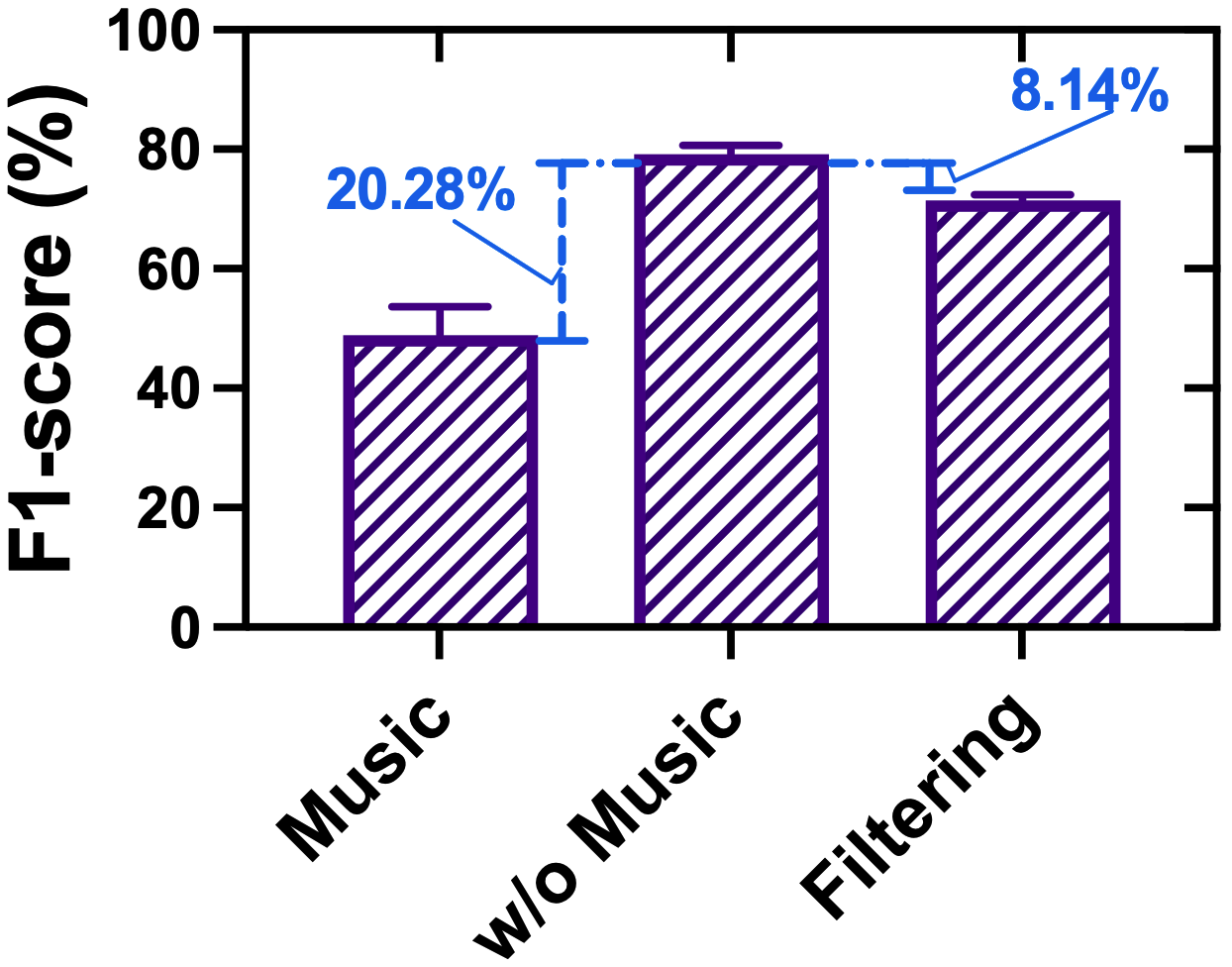}
    \label{fig:music_f1}} \hfill
     \subfloat[\textbf{F1-score of food type recognition.}]{\includegraphics[width=0.47\textwidth]{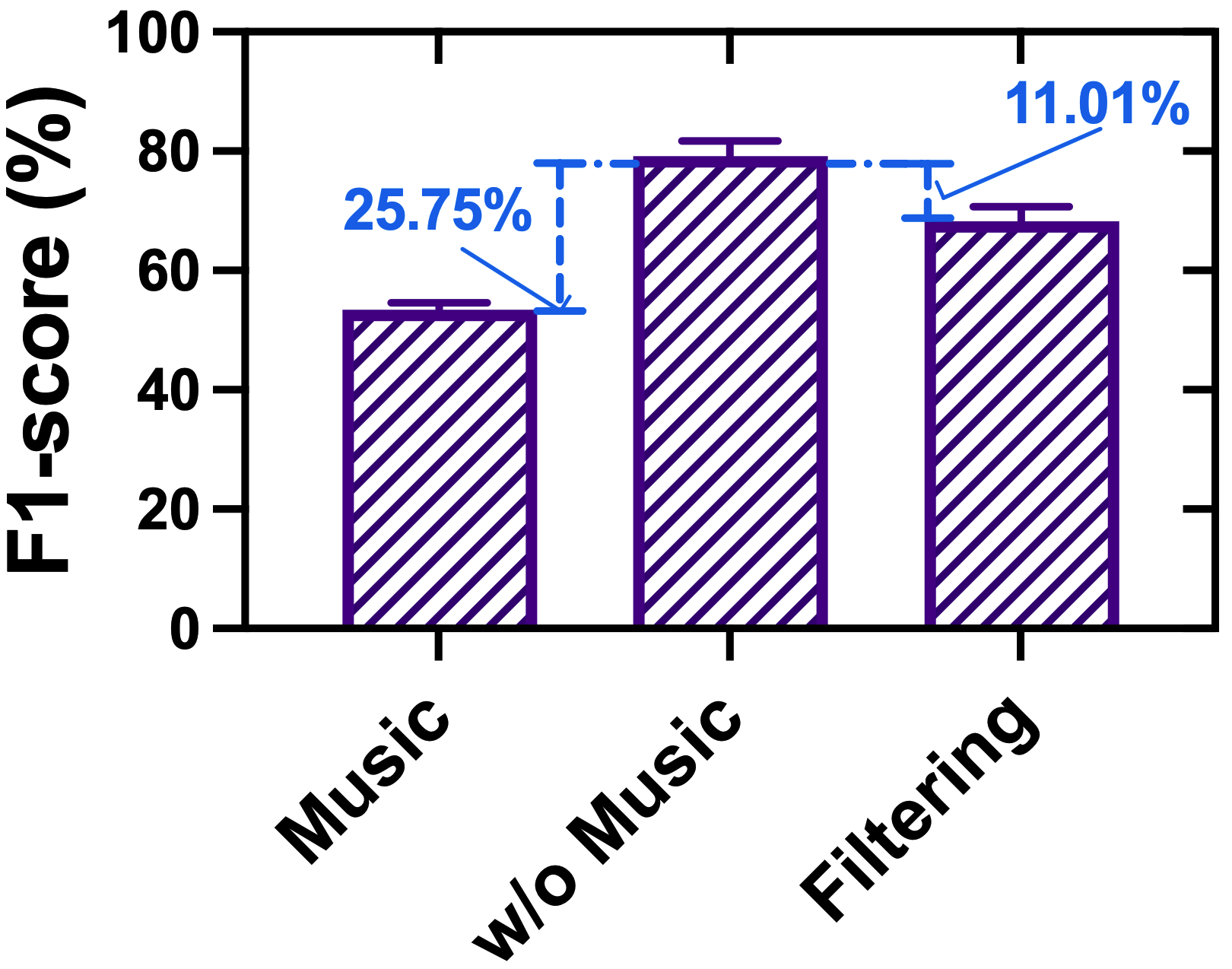}
    \label{fig:music_acc}}
    \caption{\label{fig:music_impact_acc_f1_score}\textbf{Impact of music playback from earphones.}} 
    \end{minipage}
    \vspace{-2mm}
\end{figure}

\begin{figure}[t]
    \centering
    \begin{minipage}[t]{1\linewidth}
        \centering
        \includegraphics[width=0.85\linewidth]{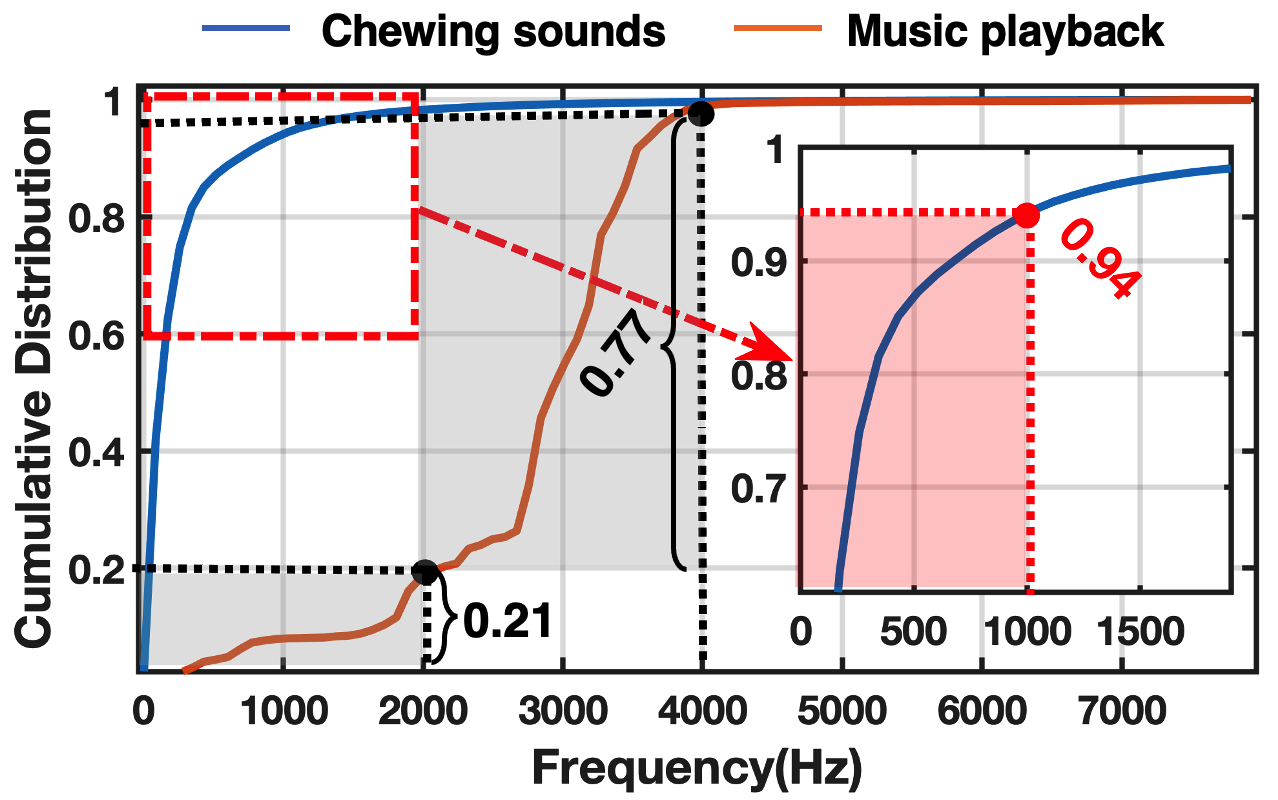}
    \caption{\textbf{Cumulative spectral distribution comparison.}}
    \label{fig:cumulative_distribution}
    \end{minipage}
    \vspace{-0.5cm}
\end{figure}

\subsubsection{\textbf{Impact of Sound Playback}}\label{sec:impact_sound_playback}
\systemName~relies on inward-facing microphones on earables to collect chewing-induced sounds for diet monitoring. In practice, individuals often play music, listen to podcasts, or make phone calls through the same device. Any audio playback from earphones' speakers mixes with in-ear chewing sounds, potentially reducing SNR or even triggering false chewing detection. Next, we record in-ear chewing sounds while simultaneously playing a music song (``My All") using the hardware prototype. The evaluation results of micro food texture recognition and food type recognition are shown in Fig.~\ref{fig:music_impact_acc_f1_score}. We can find that music playback interference causes \systemName~to decrease its performance in food texture recognition and food type recognition by nearly 20\% and 25\%, respectively. 

Furthermore, we conduct a quantitative cumulative spectral power distribution comparison between chewing sounds and music playback, as shown in Fig.~\ref{fig:cumulative_distribution}. It can be observed that chewing sounds exhibit a highly concentrated spectral distribution, with approximately 94\% of the signal energy located below 1 kHz. In contrast, music signals possess a much broader frequency spectrum, where only about 21\% of the energy is distributed below 2 KHz, and nearly 77\% of the energy is distributed from 2 KHz to 4 KHz. Thus, based on the difference in spectral power distribution, we leverage a low-pass filter to remove the interference components above 2000 Hz. After filtering, the performance of \systemName~improves noticeably in both tasks, as shown in Fig.~\ref{fig:music_impact_acc_f1_score}. Although filtering cannot completely eliminate the performance degradation due to the remaining low-frequency overlap, it significantly improves robustness under music playback conditions. Sec.~\ref{sec:discussion_noise_replayed} will continue to discuss potential solutions to this issue.

% Fig.~\ref{fig:music_impact} intuitively demonstrates the negative impact of sound playback on in-ear chewing sounds. Furthermore, 

% Yet, the acoustic echo cancellation technology has been widely used in earphones to eliminate the sound playback interference~\cite{storn1996echo}. 

% The playback audio can be viewed as an echo signal re-entering the ear canal and being captured alongside chewing-induced sounds. Existing Acoustic Echo Cancellation (AEC) techniques have been widely studied, which can be leveraged to address the sound playback interference. Yet, this echo is shaped by the unique and dynamic acoustic path of the ear canal, which is further influenced by jaw movement and earbud fit. Such conditions make the echo time-varying and potentially nonlinear, complicating its removal. 

% Specifically, the first earphone 1 and Earphone are commercial earphones, Earphone

% As shown in Fig.~\ref{fig:impact_hardware_type}, \systemName~among different hardware types.

% \subsection{\textbf{Deployability Study}}
% time delay and memory consumption.

% Mobile and Laptop CPU, Server GPU.

\begin{figure}[h]
\centering
    \begin{minipage}[t]{1\linewidth}
        \centering
        \includegraphics[width=0.8\linewidth]{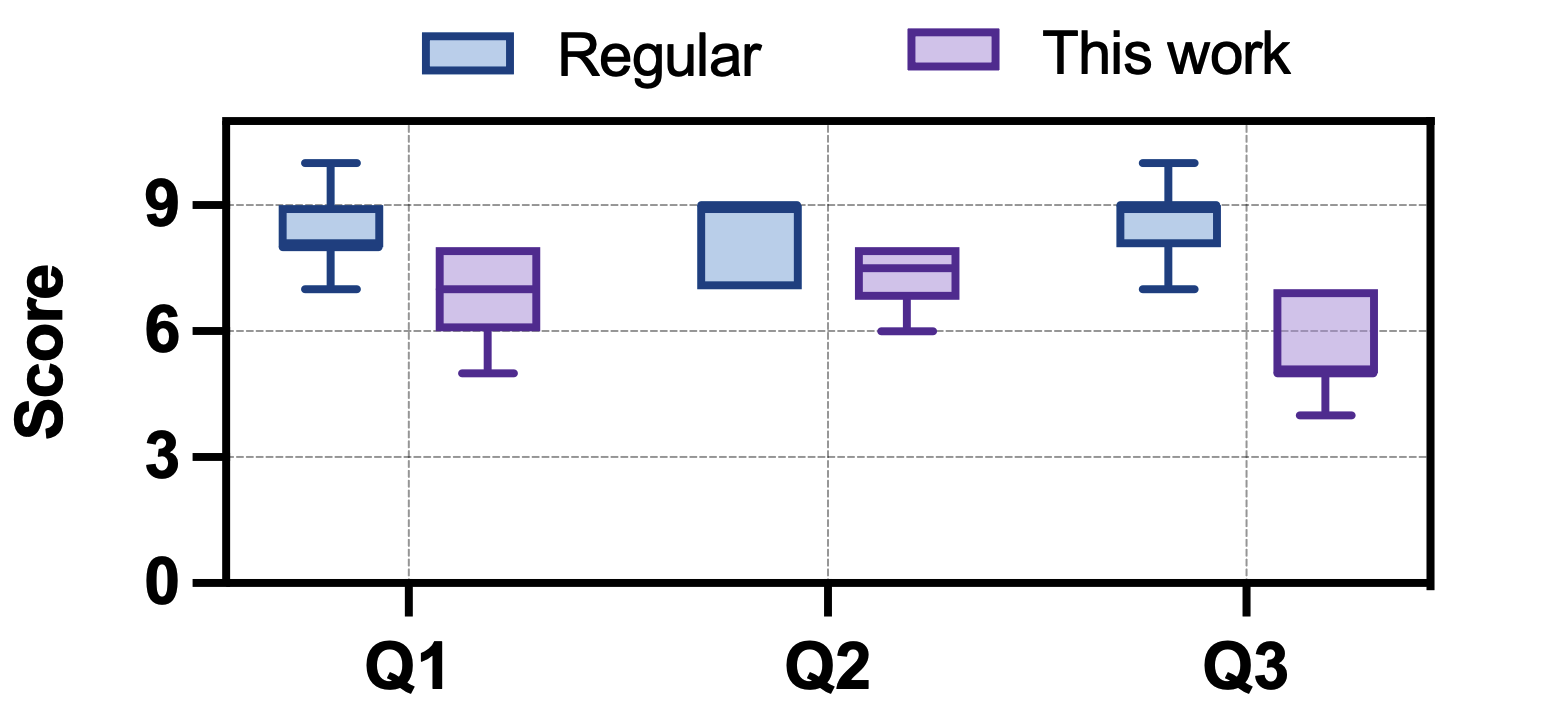} \caption{\label{fig:user_experience_study_comparison}
        \textbf{User experience study.}}
    \end{minipage}
    \vspace{-0.5cm}
\end{figure}

\subsection{\textbf{User Experience Study}}
To evaluate the user experience of \systemName, we developed a 10-point scale questionnaire comprising three items. \textit{Q1}: How comfortable did you feel wearing our prototype while eating? \textit{Q2}: How convenient do you find it to carry our prototype in your daily life? \textit{Q3}: To what extent did our solution interfere with your eating activities? All participants are involved in this user experience study. To offer experience reference, we also require participants to wear regular earphones while eating. The subjective scores of user experience are shown in Fig.~\ref{fig:user_experience_study_comparison}. Compared with regular earphones, \systemName~can bring an excellent experience to users in terms of \textit{Q1} and \textit{Q2}, demonstrating the comfort and portability of our solution. To avoid interference between the earphones’ built-in speaker playback and the chewing sounds inside the ear canal, the playback function is disabled during eating. Consequently, for \textit{Q3}, our solution receives significantly lower scores than regular earphones.

% In order to make our hardware prototype fit the ear canal of participants and improve user experience, we provide different types of silicone ear tips for participants to choose from. 

% As shown in Fig.~\ref{fig:user_experience_study_comparison}, we can clearly find that \systemName~can bring an excellent experience to users in terms of acceptability, comfort, and portability. 

\subsection{\textbf{System-level Deployability Evaluation}}

\begin{table}[t]
\centering
\caption{\label{tab:runtime_analysis} \textbf{Runtime latency analysis (seconds)}.}
\renewcommand{\arraystretch}{1.2}
\resizebox{1\linewidth}{!}{
\begin{tabular}{cccc}
\toprule
    & \textbf{Module 1+2} & \textbf{Module 3} & \textbf{Overall}      \\ \hline
    
% \includegraphics[width=0.5cm]
%    {./Figures/laptop.png} \textbf{Server (GPU)} & 1.86 ($\pm$ 0.06) & 0.07 ($\pm$ 0.02) & 3.64 ($\pm$ 0.22)\\

\includegraphics[width=0.38cm]
   {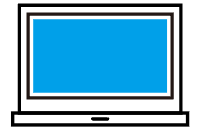} \textbf{Laptop (CPU)} & 0.53 ($\pm$ 0.07) & 0.25 ($\pm$ 0.02) & 0.97 ($\pm$ 0.15)\\

\includegraphics[width=0.28cm]
   {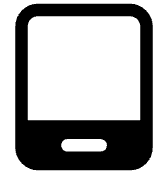} \textbf{Mobile Phone} & 1.23 ($\pm$ 0.01) & 1.18 ($\pm$ 0.28) & 1.63 ($\pm$ 0.57)\\

\includegraphics[width=0.25cm]
   {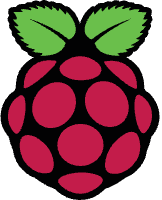} \textbf{Raspberry Pi} & 2.48 ($\pm$ 0.04) & 6.12 ($\pm$ 0.01) & 9.76 ($\pm$ 0.45)\\
\bottomrule
\end{tabular}
}
\end{table}

\begin{table}[t]
\centering
\caption{\label{tab:energy_consumption} \textbf{Energy consumption analysis (J)}.}
\renewcommand{\arraystretch}{1.2}
\resizebox{1\linewidth}{!}{
\begin{tabular}{cccc}
\toprule
    & \textbf{Module 1+2} & \textbf{Module 3} & \textbf{Overall}      \\ \hline
    
\includegraphics[width=0.28cm]
   {./Figures/mobilephone.png} \textbf{Mobile Phone} & 3.78 ($\pm$ 0.12) & 7.21 ($\pm$ 0.36) & 13.66 ($\pm$ 2.19)\\

\includegraphics[width=0.25cm]
   {./Figures/raspi.png} \textbf{Raspberry Pi} & 8.23 ($\pm$ 1.73) & 32.38 ($\pm$ 3.20) & 43.11 ($\pm$ 2.16)\\
\bottomrule
\end{tabular}
}
\end{table}

To explore the deployability of \systemName, we conduct a system-level evaluation on three types of computing platforms, \ie, a laptop (Macbook Pro), a mobile phone (Redmi Note 12 Turbo), and a Raspberry Pi 5. Since the earphones are connected to the computing platform via a wired connection in our hardware prototype, we mainly focus on the overhead of signal processing and model inference. Module 1, Module 2, and Module 3 represent the audio signal enhancement module (Sec.~\ref{sec:signal_enhancement}), the diet episode parsing module (Sec.~\ref{sec:mechanics_aware_diet_episode_parsing}), and the dietary analysis and logging module (Sec.~\ref{sec:dietary_analysis_logging}), respectively. 
For most chewing episodes, the duration is usually no more than 10 seconds. Thus, we leverage 10-second in-ear audio signals as input to calculate the runtime latency and energy consumption of each module of \systemName. 

The designed deep learning model occupies most of the memory. On the laptop, the memory usage is about 4531.54 MiB. Thus, the model is first converted to ONNX and then deployed on the mobile phone and the Raspberry Pi. After that, the model occupies 909.02 MiB of memory and requires 716 MiB of memory for inference. Tab.~\ref{tab:runtime_analysis} shows the runtime latency of each module in \systemName. We can find that even on the resource-constrained Raspberry Pi, the runtime latency of the entire pipeline is still below 10 seconds. Furthermore, energy consumption is evaluated on the mobile phone and the Raspberry Pi, because they represent resource-constrained deployment platforms for wearable dietary monitoring. However, the laptop typically offers abundant computational resources and stable power supplies, making energy consumption less critical. Tab.~\ref{tab:energy_consumption} gives the energy consumption results. The overall energy consumption of \systemName~is 13.66 J on the mobile phone and 43.11 J on the Raspberry Pi. 
However, based on the measurements, the energy consumption of in-ear sound acquisition is about 3.71 J on the mobile phone and 8.23 J on the Raspberry Pi. Thus, one potential energy-saving deployment solution for the future is to have terminal devices only responsible for data collection and some low-power preprocessing, while offloading energy-intensive computing tasks to nearby edge servers. 

\subsection{\textbf{Clinical Validation}}

Beyond dietary monitoring, we further explore the potential of \systemName~for oral function assessment in real clinical environments. In total, the clinical validation involves 51 patients with masticatory disorders (such as masseter muscle dysfunction and dysphagia) who visited the outpatient department of Nanjing Medical University Affiliated Hospital\footnote{All data collection procedures are approved by the Internal Review Board (IRB) of Nanjing Medical University Affiliated Hospital.}.

\begin{figure}[t]
    \centering
    \includegraphics[width=1\linewidth]{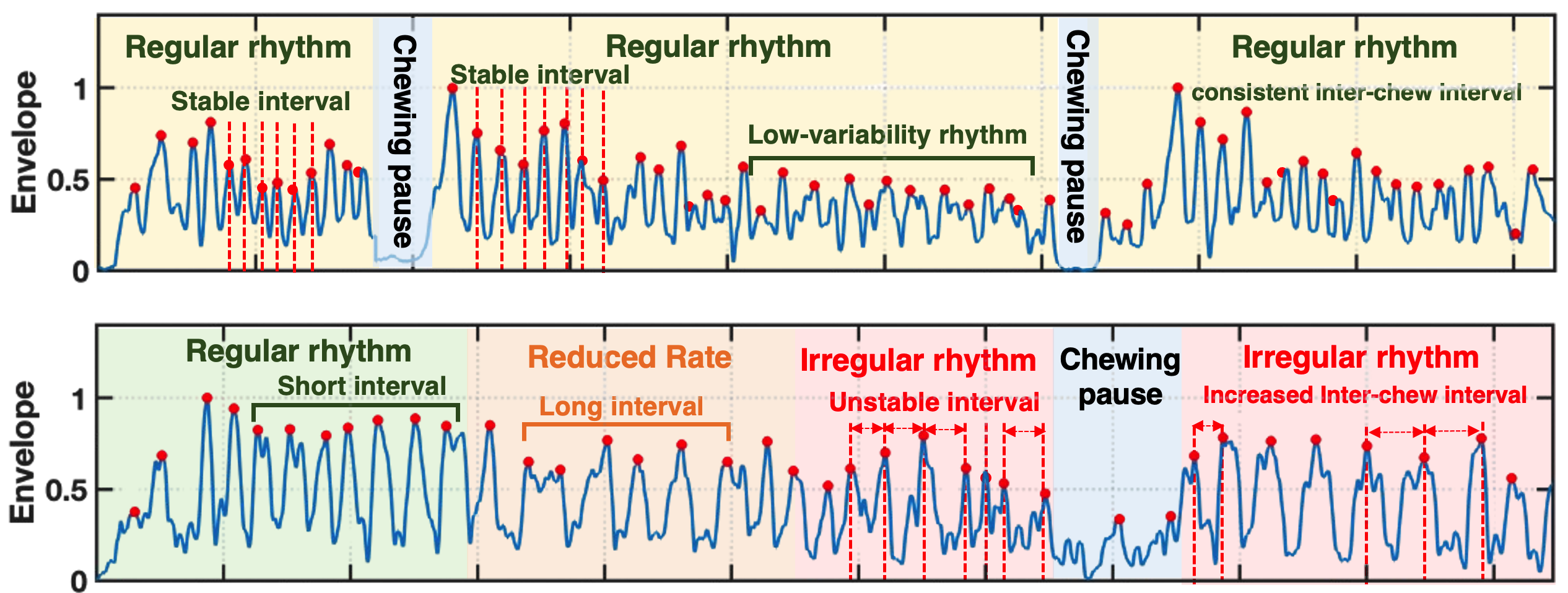}
    \caption{\textbf{Chewing patterns comparison between a normal person and a patient.}}
    \label{fig:clinical_study}
\end{figure}

\begin{figure}[t]
    \centering
    \begin{minipage}[t]{1\linewidth}
        \centering
        \includegraphics[width=0.8\linewidth]{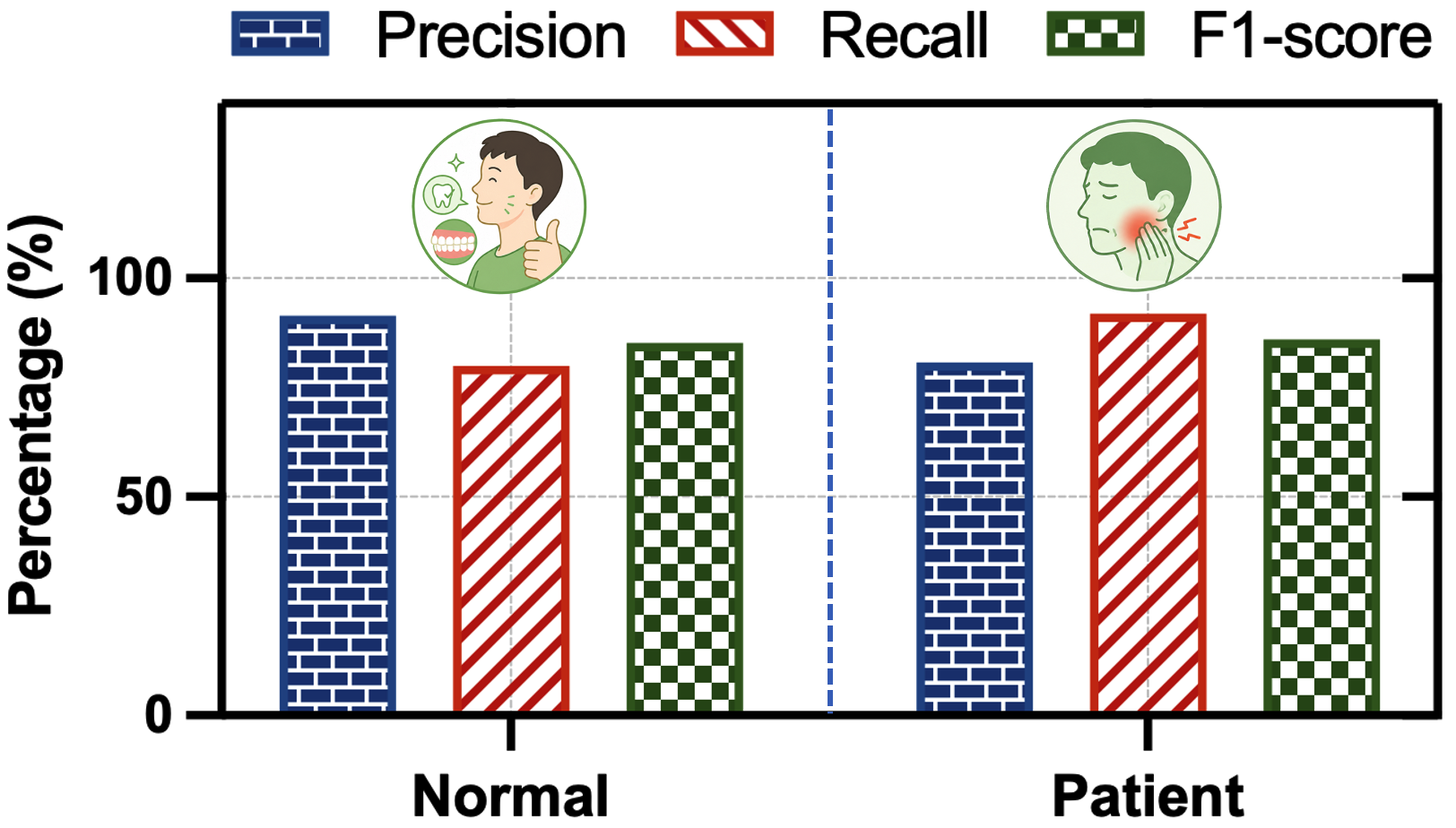}
    \caption{\textbf{Chewing patterns comparison between a normal person and a patient.}}
    \label{fig:clinical_validation_result}
    \end{minipage}
    \vspace{-0.5cm}
\end{figure}

These masticatory disorders often alter chewing biomechanics and oral motor behaviors, which are directly reflected in chewing-induced in-ear sounds. As shown in Fig.~\ref{fig:clinical_study}, compared with normal chewing activities, patients with masticatory disorders typically demonstrate inconsistent chewing rhythm, irregular chewing-swallowing coordination, and prolonged chewing duration. Since \systemName~can explicitly model and track the entire eating episodes, the chewing behavioral patterns and dietary information derived by \systemName~can provide a rich representation of masticatory disorders. Thus, we extract the above representative chewing behavioral patterns (\eg, chewing episodes, chewing rhythm, burst density, and temporal consistency) from clinical data and use a Support Vector Machine model for masticatory disorder identification. As shown in Fig.~\ref{fig:clinical_validation_result}, \systemName~identifies normal and abnormal chewing patterns and further recognizes the masticatory disorders with an averaged precision of 80.77\%, an averaged recall of 91.88\%, and an averaged F1-score of 85.97\%. The above results demonstrate the potential of \systemName~as a non-invasive and continuous tool for masticatory disorder and oral function monitoring in clinical conditions. In the future, we will further investigate its feasibility and applicability in large-scale clinical studies, promoting the use of ear-worn sensing technologies for non-invasive healthcare applications.

\section{\textbf{Discussion}}

\subsection{\textbf{Unseen Food Types and Textures}}

Although \systemName~achieves high accuracy on our limited dataset compared to other state-of-the-art solutions~\cite{chugh2025bitesense, shin2022mydj}, real-world diets encompass hundreds or even thousands of food types, many of which exhibit texture properties not represented in our current work. This distribution gap makes the problem inherently open-set, where previously unseen food types or novel texture profiles may lead to degraded recognition performance. Addressing such an issue requires methods that can generalize beyond the training set, such as zero-shot learning, domain generalization, or online adaptation strategies. Future research should also consider incremental model updating in deployment, enabling \systemName~to continuously learn from new dietary contexts without catastrophic forgetting, thus improving robustness to the vast diversity of real-world food textures and types.
Additionally, we will investigate incorporating objective physical measurements (\eg, texture analyzer measurements or fracture-force measurements) as complementary texture proxies to further improve the granularity of food texture annotations. 

\subsection{\textbf{Recognition of Mixed Foods}}
\systemName~is designed to distinguish individual food types under semi-controlled scenarios. Daily eating habits often involve mixed-food consumption, where users ingest multiple food items within a short time window or simultaneously in a single bite. In such cases, the chewing-induced audio signals become a composite of multiple texture attributes, leading to overlapping spectral and temporal patterns. This superposition effect poses a significant challenge for \systemName. Potential strategies include multi-modality fusion techniques to capture food intake information from multiple perspectives. Future work should also consider data collection protocols that systematically include mixed-food samples, improving robustness in real-world dietary monitoring.

\begin{figure}[t]
\centering
    \begin{minipage}[t]{1\linewidth}
        \centering
        \includegraphics[width=0.9\linewidth]{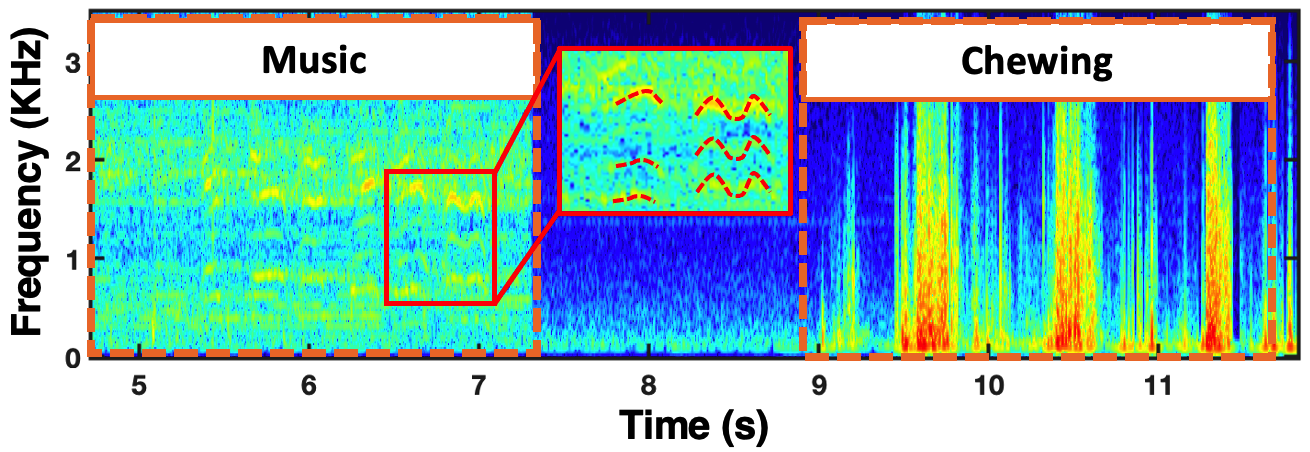} \caption{\label{fig:music_chewing_stft}
        \textbf{Time-frequency representation comparison.}}
    \end{minipage}
    \vspace{-0.5cm}
\end{figure}

\subsection{\textbf{Sound Playback from Earphones}} \label{sec:discussion_noise_replayed}

As analyzed in Sec.~\ref{sec:impact_sound_playback}, the audio playback from earphones' speakers significantly affects the detection performance of \systemName. The playback audio can be viewed as an echo signal re-entering the ear canal and being captured alongside chewing-induced sounds. The difference in physical generation mechanisms of chewing sounds and music playback results in their unique time-frequency characteristics. In addition to the difference in spectral energy distributions, Fig.~\ref{fig:music_chewing_stft} shows that music playback and chewing sounds exhibit fundamentally different time-frequency characteristics. While music signals present continuous harmonic and formant structures, chewing sounds are dominated by intermittent burst-like events with distinct temporal dynamics. The substantial structural discrepancy between the two sources indicates that they are not fully overlapped in the time-frequency domain, making source separation feasible. Inspired by the prior literature~\cite{han2024earspeech}, deep learning-based signal enhancement can be introduced to eliminate the interference induced by audio playback of earphones. We will continue to focus on this common challenge of ear-worn sensing in our future work. 

\section{\textbf{Conclusion}}

In this work, we presented \systemName, a single-modality, multi-dimensional dietary monitoring system leveraging inward-facing microphones on earables to capture chewing-induced in-ear sounds. By integrating carefully designed signal processing with a multi-task learning framework, \systemName~enables accurate eating episode detection, food texture recognition, and food type classification without the need for multiple sensors or intrusive hardware modifications. Comprehensive experiments with 32 participants across diverse real-world dining scenarios demonstrate high robustness and effectiveness. \systemName~maintains strong performance in both short-term and longitudinal evaluations, highlighting its potential for continuous, fine-grained diet logging in daily life.

% \clearpage
\bibliographystyle{IEEEtran}
\bibliography{IEEEabrv,mybibfile}

\end{document}